\documentclass[12pt]{article}
\usepackage{epsfig}
\usepackage{cite}
\usepackage[figuresright]{rotating}
\usepackage{lscape}
\def\dfrac#1#2{{\displaystyle{#1\over#2}}}
 
\begin{document}
  \thispagestyle{empty}
 \begin{flushright}
        ATL-COM-INDET-2005-009
\end{flushright}\bigskip\bigskip\bigskip\bigskip\bigskip
 \begin{center}
   {\huge\bf 
     A Method to check the Connectivity for the ATLAS TRT Detector
   }
 \end{center}\bigskip\bigskip
 \begin{center}{{\large 
       N.~Ghodbane~$^a$, X.~Pons~$^a$, O.~M.~Rohne~$^b$\\
               }
\bigskip
     {\normalsize 
       $^a$European Laboratory for Particle Physics (CERN), 
                 Geneva, Switzerland.\\
       $^b$ Department of Physics and Astronomy, University of 
                 Pennsylvania, Philadelphia, Pennsylvania, USA.}
     }\end{center}\bigskip\bigskip
 \bigskip\begin{center}{\large  Abstract}\end{center}
\begin{center}\begin{minipage}{5.2truein}
We report on a technique developed to diagnostic for non working readout channels of the Transition Radiation Tracker (TRT), one of the three inner detectors
 of the ATLAS experiment. From a detailled study, we show that  99.6 \% of the readout channels of the TRT endcap-detector are perfectly operational.

\end{minipage}\end{center}
\newpage

\section{Introduction}
\label{sect-intro}
The ATLAS Transition Radiation Tracker (TRT) \cite{ATLASTRT}, part of
the ATLAS inner detector, is designed to provide electron
identification and charged particle track reconstruction at the nominal design
luminosity of the CERN Large Hadron Collider (LHC). The initial TRT consists
of two parts, a barrel section comprising three cylindrical rings each 
containing 32 identical modules made of axially aligned straws (2$\times$52544 readout
channels) 
and two end-cap TRT parts, made each of two sets of identical
 and independent eight-plane wheels with radially oriented straws (12 A-type
 eight-plane wheels,
 8 B-type) for a total number of readout channels equal to
  245760 straws.

\begin{figure}[htb]
\begin{center}
  \epsfig{file=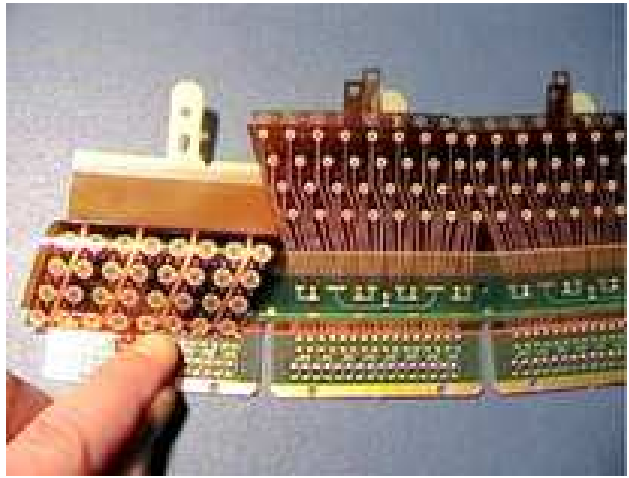,width=.5\textwidth}
\caption{Photograph of a  Wheel End-cap Board (WEB) of the ATLAS TRT End-cap detector.}
\label{fig:WEB}
\end{center}
\end{figure}

The main interface of each of these wheels to the outside world consists of 2~$\times$~32 flex-rigid printed  circuit boards
called Wheel End-cap Board or WEB as shown in Fig. \ref{fig:WEB}, used to connect the straws
to the high-voltage supply and to collect the signal from the wire to the front-end electronics.

\begin{figure}[htb]
\begin{center}
  \epsfig{file=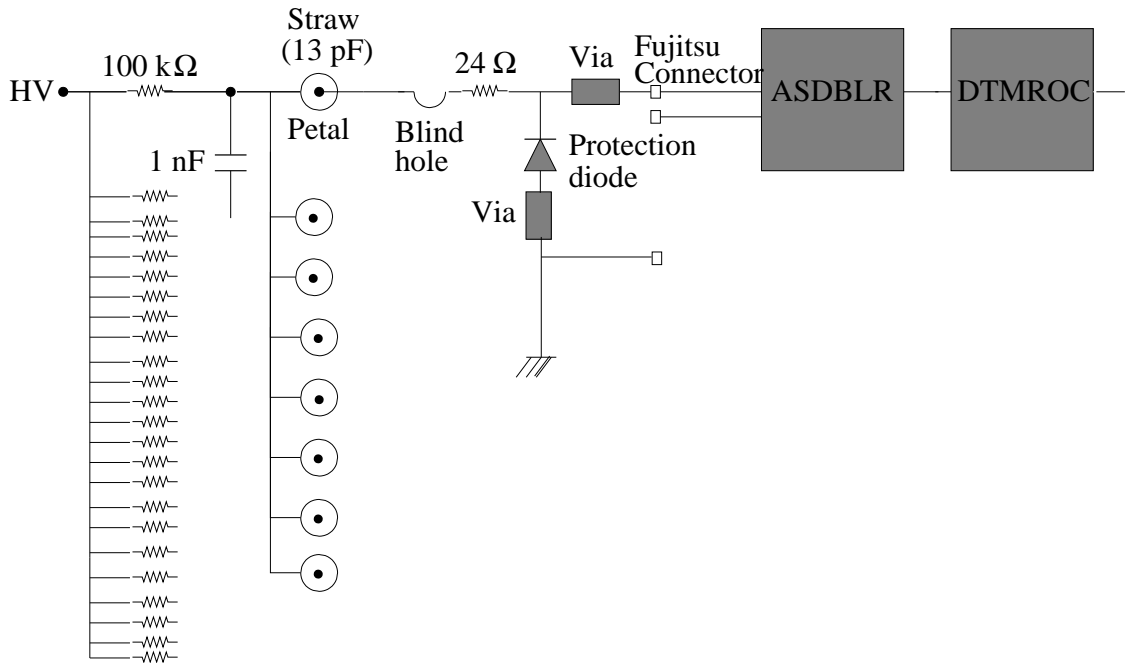,width=0.9\textwidth}
\caption{Schematic view of a TRT end-cap readout chain.}
\label{fig:StrawElectronicChannel}
\end{center}
\end{figure}

The analogue signal collected from the anode through the WEB is transmitted to the 
on-detector-front-end electronics boards, which comprise:
\begin{itemize}
\item[-] the TRT Analog Shaper Discriminator Base Line Restorer chip
  (ASDBLR) \cite{ASDLR}, which integrates 8 identical channels on a 3.3~$\times$~3.6~mm~$^{2}$ die. Each
  channel features a preamplifier connected to the anode wire with an
  intrinsic noise of 0.3 fC and a dynamic range which exceeds 100 fC, a shaper
  for the ion tail cancellation ($\tau \simeq 7$ ns), a baseline restorer and
 two discriminators  with adjustable thresholds. A low-level threshold (typically $\simeq$~250 eV) is
  used to provide a precise  measurement of the drift time for charged
  particles and  a high-level threshold (typically $\simeq$~6-7 keV) is used to identify the absorbed transition
  radiation (TR) photons.

\item[-] the status of the low-level discriminator is then fed into the Drift-Time 
 Measurement Read-Out Chip (DTMROC) \cite{DTMROC}, which samples the signal eight times per
  bunch crossing through a Delay Lock Loop (DLL), thereby  providing drift time measurements with a 25/8 ns time bin
  size. The result of the high-level threshold is latched and as a
  result the DTMROC produces a total of 9 data bits per channel foreach bunch crossing. A total of three bunch crossings are read out for each straw 
 ( 27 bits total over 75 ns).

\end{itemize}
A schematic view of one channel is presented on Fig. \ref{fig:StrawElectronicChannel}, with some specific details of the WEB connection to the anode wires which are relevant to the results presented in this note (blind holes and vias).

\begin{figure}[htb]
\begin{center}
  \epsfig{file=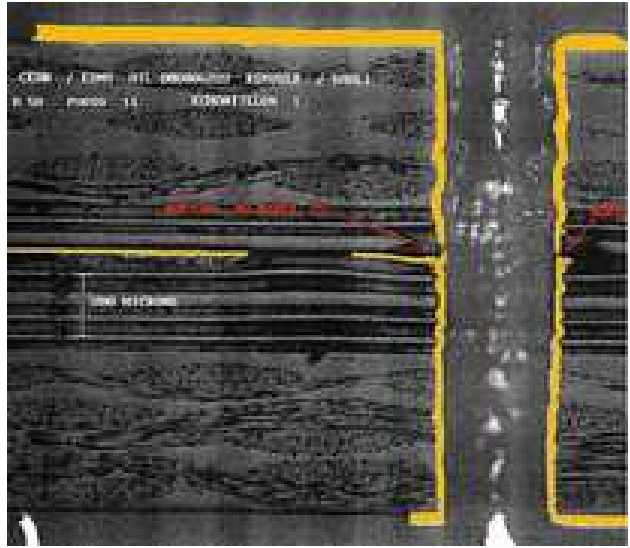,width=.5\textwidth}
\caption{Crack due to a bad metallisation of one signal via of a TRT end-cap WEB.}
\label{fig:ViaProblems}
\end{center}
\end{figure}

As an integral part of the quality assurance process implemented during and after assembly of the TRT
detector, the different wheels of the TRT end-caps have to
fulfill several stringent acceptance criteria \cite{ACCEPTANCE} and are submitted to a serie of 
well-defined tests, such as high-voltage tests, wire-tension
measurements, gas-gain uniformity measurements, etc...\\
During these so-called {\textit{acceptance tests}} (AT), 
straws found to be out of specifications are disconnected from their high
voltage group by un-soldering the 24$\Omega$ protection resistor (Fig. \ref{fig:StrawElectronicChannel}).
Careful studies have shown that most of the problematic channels are due to metallisation cracks in the signal vias 
or  (more rarely) blind holes of the WEBs, as shown by the photograph in Fig. \ref{fig:ViaProblems}.\\
After repair of these channels, the total number of disconnected channels after the acceptance tests is below 0.2\% for the first set of end-cap wheels (122880 channels).

All relevant repair information, together with a full list of disconnected and anomalous channels is provided as part of an electronic passport, which is stored in an ORACLE-based database foreach eight-plane TRT wheel.

After completion of all the Acceptance tests, the eight-plane wheels are stacked horizontally and are then made available 
for installation of the front-end electronic boards.\\
The integration sequence comprises:

\begin{itemize}
\item[1)] a qualification of the front-end boards based on stringent selection criterias 
(all channels operational and their electronic response within specifications).
\item[2)] a noise scan over the 8-bits low-level threshold with a step of one, taking 500 events per scan point.
\item[3)] a built in test pulse scan over the low-level threshold with a step of one, taking 250 events per scan point.
\item[4)] a built in test pulse scan over the high-level threshold with a step of one, taking 250 events per scan point.
\item[5)] a run using the accumulation mode of the DTMROC chip at one set of thresholds (120 DAC $\simeq$ 300 eV for the tracking threshold and 45 DAC $\simeq$ 1.2 keV ), with 50 event at each of the different gating time windows.
\end{itemize}
Tests 2), 3) and 4) are repeated when the electronics is connected to the detector.\\
The results of each of these tests are stored in a powerful MySQL-based database and retrieved for analysis with a PHP-based web interface.

To check for connectivity of the different readout channels, the latest off- and on-detector noise scans are used together with the accumulation mode scan. The ditribution of the noise rate differences is fitted to a Gaussian curve of mean $\mu$ and RMS $\sigma$ and channels with values below $\mu - 3\times \sigma$ are studied in more details.\\
First, the electronic board is inspected for any faulty component. For instance, the Fujitsu female connectors are carefully checked to have an impedance about 28 k$\Omega$ from one input (active or dummy) to ground. Rare are the electronic boards which fail this test, since stringent quality requirements are applied to select boards to be mounted on the detector.\\
Then, using a battery-operated pF-meter, the different Fujitsu male connectors on the detector are probed looking at some abnormal capacitance from active pin to ground. This method, combined with the value obtained with the rate integral technique, permits to locate the problem on the detector side, at the via close to the active pin of the Fujitsu, or at the blind hole or directly at the straw as depicted on Fig. \ref{fig:StrawElectronicChannel}. Abnormal channels show usually a capacitance value lower by more than 5 pF than the average of the corresponding WEB.\\
Finally a last technique uses the special "diode check" function of  a volt-meter. A small DC current is injected between the diode ground and the signal output and then the forward voltage drop of the diode is measured. An abnormal value, with respect to the Fujitsu average, is directly linked to 
a problem of one of the two vias as shown on Fig.  \ref{fig:StrawElectronicChannel}.

During the implementation of this procedure, robust and redundant techniques have been developed to identify problematic channels, which may have appeared since the acceptance testing of the wheels or the qualification of the front-end electronic boards.

\section{The Noise Rate Integral Technique}
\label{ITMethod}

\begin{figure}[ht]
\begin{center}
  \epsfig{file=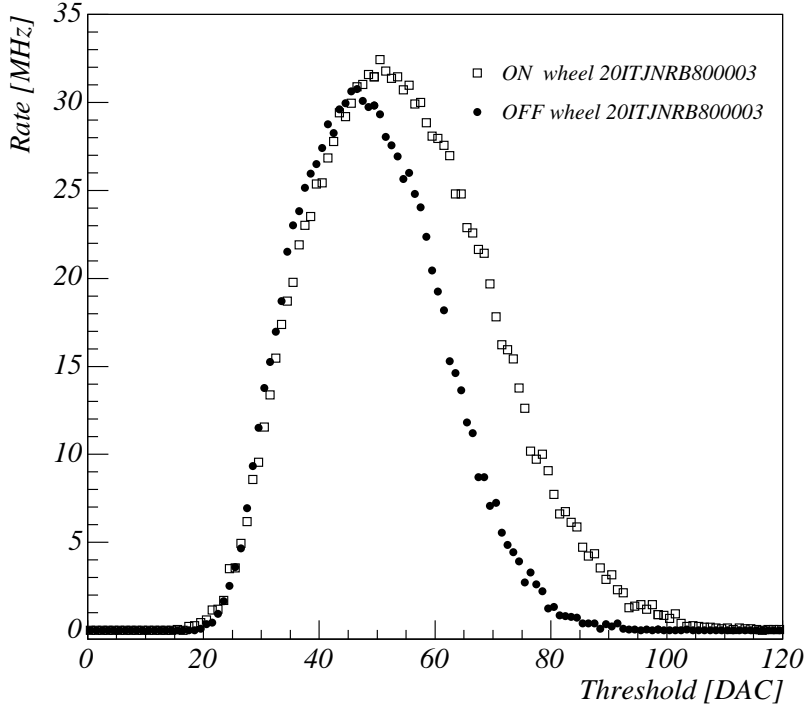,width=.7\textwidth}
\caption{Noise rate as a function of the low-level threshold DAC setting for a TRT end-cap board, when measured 
off-detector (full circles) and on-detector (open squares). 
For an operational setting of close to 200 eV ($\simeq$ 100-120 DAC counts), the noise counting rate per straw is about 300 kHz.}
\label{fig:RateDistribution}
\end{center}
\end{figure}

The first method used to characterise and quantify the board performance is the
noise rate shown in Fig. \ref{fig:RateDistribution} as a function of the low-level threshold applied to the ASDBLR chip.
 This rate is
defined as the number of transitions from low to high discriminator level output 
for a given DAC setting of the low-level threshold. For small discriminator values of the DAC
setting,
the output level is almost always high for all eight time bins and the number of
 transitions therefore drops to zero for low values of the low-level threshold.\\
For higher values of
the DAC setting, the number of transitions from low  to high increases
and reaches a maximum in the region between 40-50 DAC counts (near the zero
threshold of the comparator of the ASDBLR chip).
 The typical value of the
maximum corresponds to the maximum bandwidth of the ASDBLR ($\simeq$ 35 MHz).
For even larger DAC settings, the number of transitions decreases and reaches the design specifications (a rate of about 300 kHz) 
for low-threshold values of about $\simeq$ 100-120 DAC counts, which corresponds to a low-threshold close to 200 eV.

\begin{figure}[ht!]
\begin{center}
\begin{tabular}{cc}
\epsfig{file=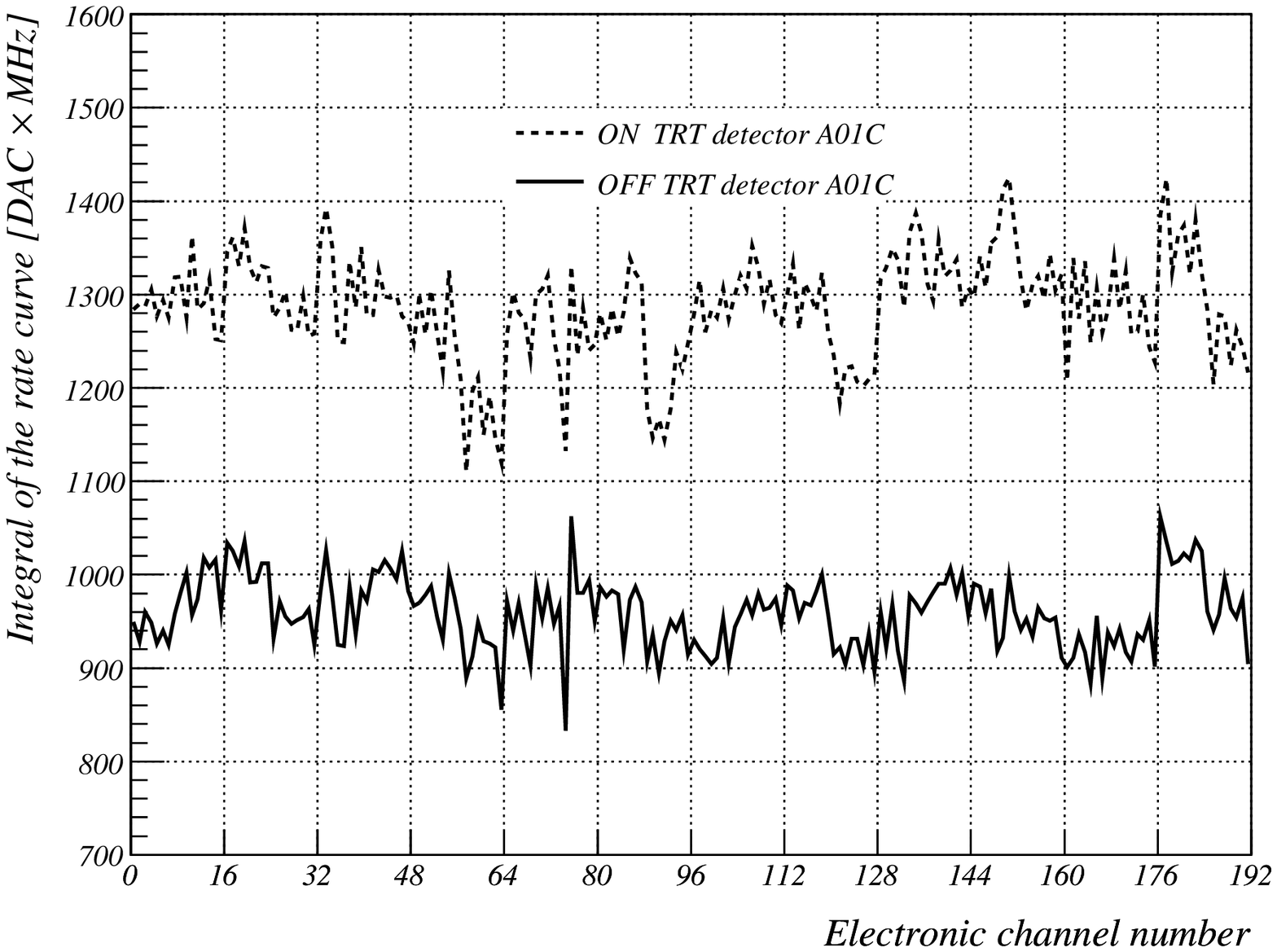,width=.55\textwidth} &\epsfig{file=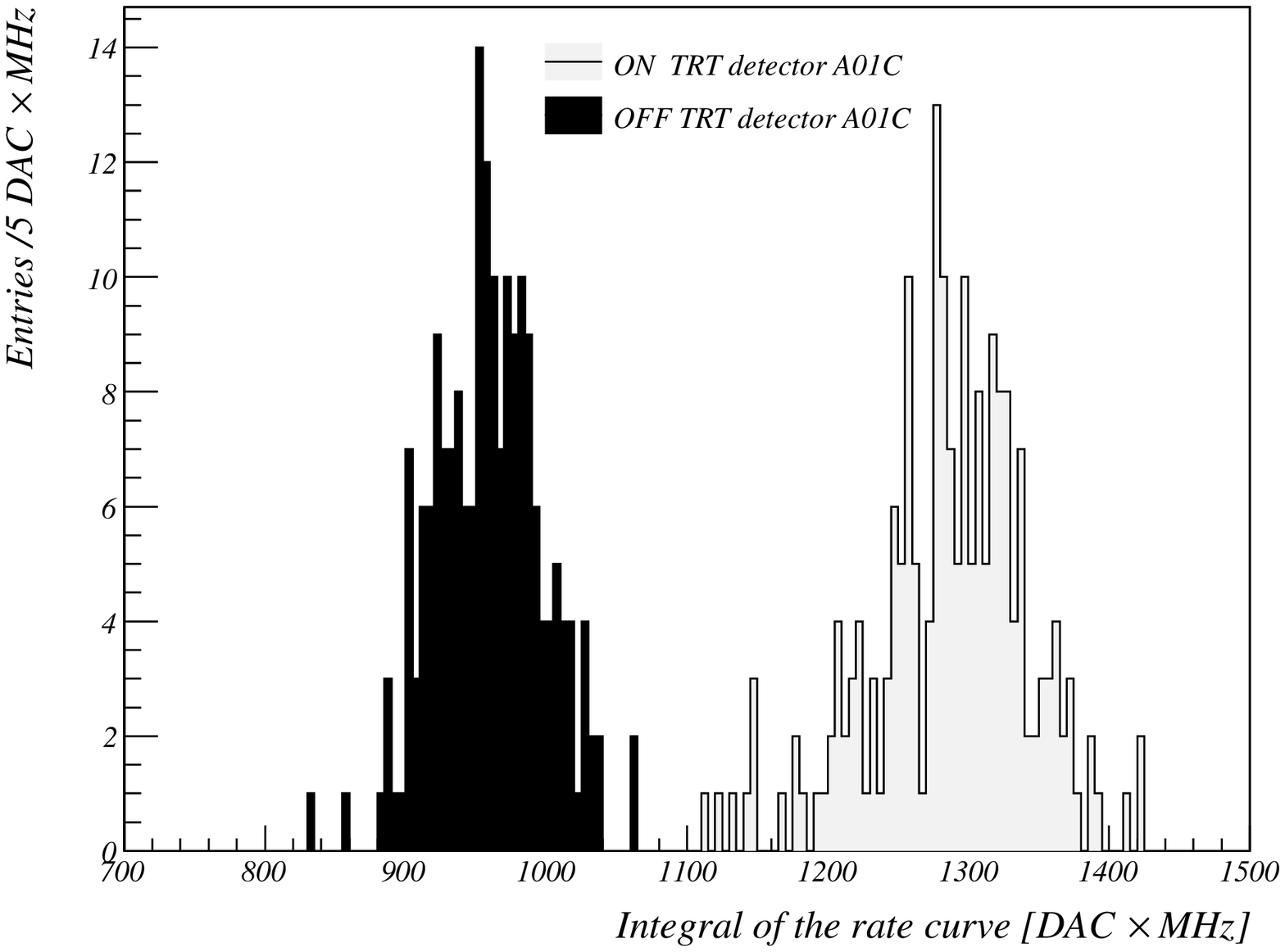,width=.55\textwidth}\\
\end{tabular}
\caption{Left: integral of the noise rate curve for one module (192 channels) on- (dashed line) and off- a TRT endcap-detector. Right: distribution of the integrals of the rate curves for one TRT endcap WEB (192 channels) off (black histogram) and on detector.}
\label{fig:IT333231}
\end{center}
\end{figure}

As show in Fig. \ref{fig:RateDistribution}, once the front-end electronics board is connected to the TRT wheel, 
the input load increases to that of the WEB plus the straw itself and therefore the total Equivalent Noise Charge
(ENC) is larger. This results then in a broader noise rate curve, which can be used to identify disconnected channels.\\
Several quantities, based on the noise rate curve shown in Fig. \ref{fig:RateDistribution}, have been investigated:
\begin{itemize}
\item[-] the $\simeq$2 \% occupancy level (summed over three bunch crossing) corresponding to a rate of 300 kHz;
\item[-] within the approximation of a Gaussian distribution for the noise rate curve, the $\sigma$ obtained from a Gaussian fit to 
the distributions of Fig. \ref{fig:RateDistribution};
\item[-] the integral of the noise rate curves can be computed and compared as shown in Fig. \ref{fig:IT333231}.
\end{itemize}

The integral noise rate turns out to be the most sensitive and reliable measurement to identify problematic channels of the type described in
the introduction, i.e. channels with metallisation problems causing poor connectivity between the ASDBLR chip and the anode wire.

\begin{figure}[ht!]
\begin{center}
\begin{tabular}{cc}
  \epsfig{file=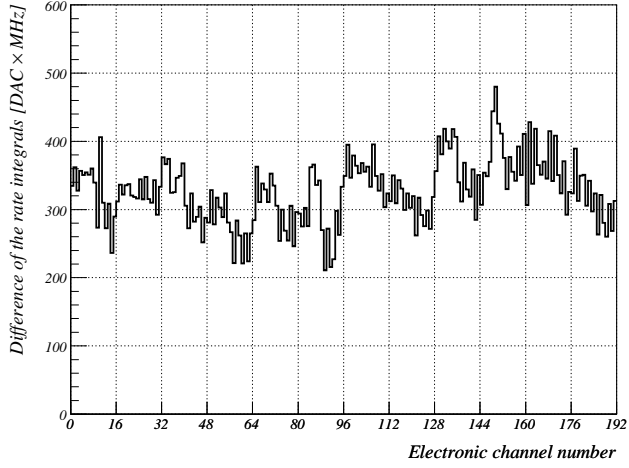,width=.55\textwidth} &   \epsfig{file=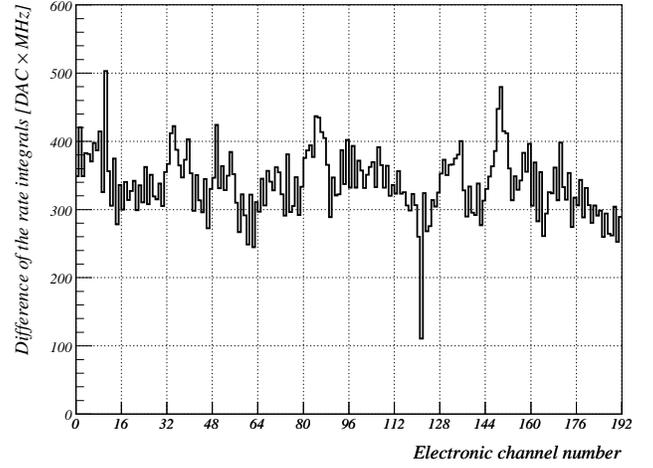,width=.55\textwidth} \\
  \epsfig{file=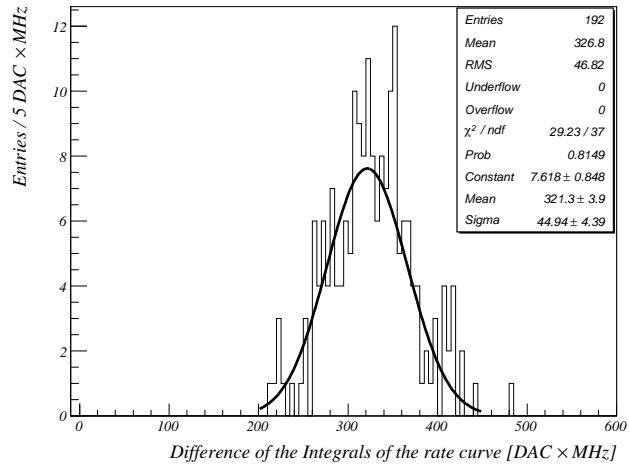,width=.55\textwidth} &   \epsfig{file=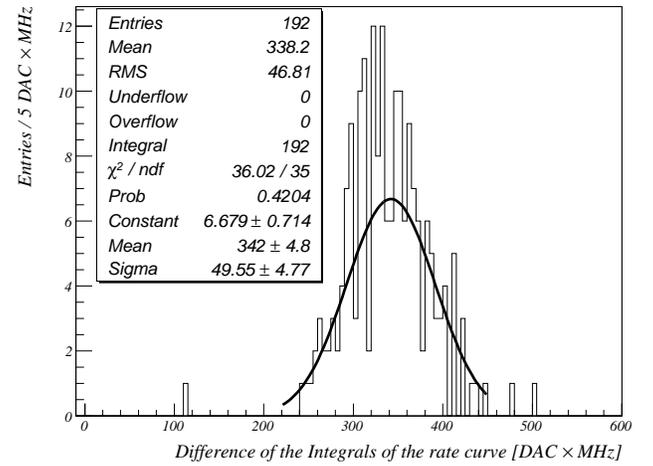,width=.55\textwidth} \\
\end{tabular}
\caption{Top: Distribution of the difference between the noise rate integral on-detector and off-detector as a function of channel number for a well-behaved WEB (left) and for a WEB with one channel displaying metallisation problems not identified by the acceptance tests (right). Bottom: distribution of the noise rate integral difference for the same WEBs as above.}
\label{fig:IT333231Diff}
\end{center}
\end{figure}

Fig. \ref{fig:IT333231} clearly demonstrates the good discrimination potential of the integral noise rate, since a separation of about 300 counts (DAC$\times$MHz) is observed on average between the off- and on-detector measurements.\\
This is illustrated on Fig. \ref{fig:IT333231Diff} which shows the difference between the two noise rate integrals
for a well-behaved WEB and for a WEB with one channel displaying metallisation problems not seen during the acceptance testing.
For the well-behaved WEB, the distribution is centered around 320 DAC$\times$MHz counts with a Gaussian RMS of about 50 DAC$\times$MHz counts.
Problematic channels appear at much lower values and front-end board behavior have been removed by fitting each distribution to find its mean and RMS.

The mean value of the distributions shown in Fig. \ref{fig:IT333231Diff} are also quite sensitive to the two analogue voltages used to power the ASDBLR chips. A small shift of one of these voltages from its nominal setting at $\pm 3$ Volts  will result in a shift of the noise rate integral difference distribution. A careful study of the possible systematic effects due to these analogue voltages was performed by changing the voltage settings by $\pm$0.1 Volts for the noise scan measurements both on- and off-detector. These result in large shifts of the noise rate integral difference distributions from 250 to 390 DAC $\times$ MHz.
To minimise such effects, at each measurement on- and off-detector, the applied analogue voltages  were tuned to be within$\pm 0.02$ Volts of the nominal  $\pm 3$ Volts.\\
The selection criteria defining channels as problematic from these measurements were chosen as follows: the distribution of the noise rate integral differences was fitted to a Gaussian curve of mean $\mu$ and RMS $\sigma$ and channels with values below $\mu - 3\times\sigma$ were flagged as problematic.\\
The noise scan rate integral technique was then combined to another independant method described in the next section to fully characterise the problematic channels.

%
\section{The Accumulation Mode Technique}
\label{DCAnalysis}
The DTMROC chip features a so-called {\it{accumulation mode}} for the high-level threshold hits.
By enabling this mode, the front-end electronics can be operated in such a way that, once the high-level threshold bit has been set, the bit remains set until the relevant
DTMROC configuration register bit is cleared.
\begin{figure}[ht]
\begin{center}
\epsfig{file=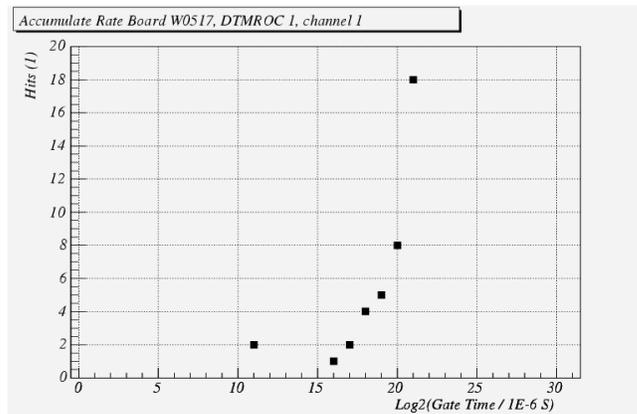, width=.55\textwidth}
\caption{Total number of events with a high-level threshold bit using the accumulation mode for a total of 50 L1A software triggers.}
\label{fig:DarkCurrentHighBit}
\end{center}
\end{figure}

A technique to check for the connectivity features the accumulation mode. The TRT detector is filled with Argon-CO2 gas mixture (70:30) and a nominal high voltage of 1350 Volts is applied between the anode and the cathode of each of the straws. Under these conditions, micro-discharges occur near the straws or on the WEBs, the signal is then collected, amplified and shaped by the ASDBLR which ternary output is then fed into the DTMROC for digitization. As it is shown on Fig. \ref{fig:StrawElectronicChannel}, a problem in the electronic chain, like a bad metalization, or a problem with one of the electronic components will directly appear, since for this particular channel, the high-level threshold accumulation state will not be reached.\\
The accumulation mode technique implements the following algorithm:
\begin{itemize}
\item[1)] work at two well defined values for the low- and high-level thresholds: 120 DAC $\simeq$ 300 eV for the tracking threshold and 45 DAC $\simeq$ 1.2 keV for the TR threshold.
\item[2)] set the DTMROC chip in the accumulation mode.
\item[3)] gate for a given time $\tau_i$ ( from some $\mu$s to a few ms).
\item[4)] issue a software trigger, read the event and decode the corresponding three bunch crossing high-level threshold bit information. Check whether some of the 3 high-level bits are set. 
\item[5)] repeat steps 2) to 4) for a given number of events ($N \simeq 50$ events). Call the total number of events with response over threshold $n_i$. 

\item[f)] increase the gating time $\tau_i$, by typically a factor 2, and repeat steps 2) to 5).
\end{itemize}
Fig. \ref{fig:DarkCurrentHighBit} shows typical values for $n_i$ for different gating times $\tau_i$. As expected, $n_i$, the number of events for which the accumulation mode is reached, increases as the gating time $\tau_i$ is longer. \\
The accumulation rate, or dark current rate, is then defined as:
\begin{eqnarray}
r_{k^{'}} = \dfrac{ \Sigma_{i} n_i }{ \Sigma_i \tau_i \times \dfrac{ 1 - exp \left( - r_k \times \tau_i \right)}{r_k \times \tau_i }} 
\label{dcformulae}
\end{eqnarray}

\begin{figure}[hb!]
\begin{center}
\begin{tabular}{cc}
\epsfig{file=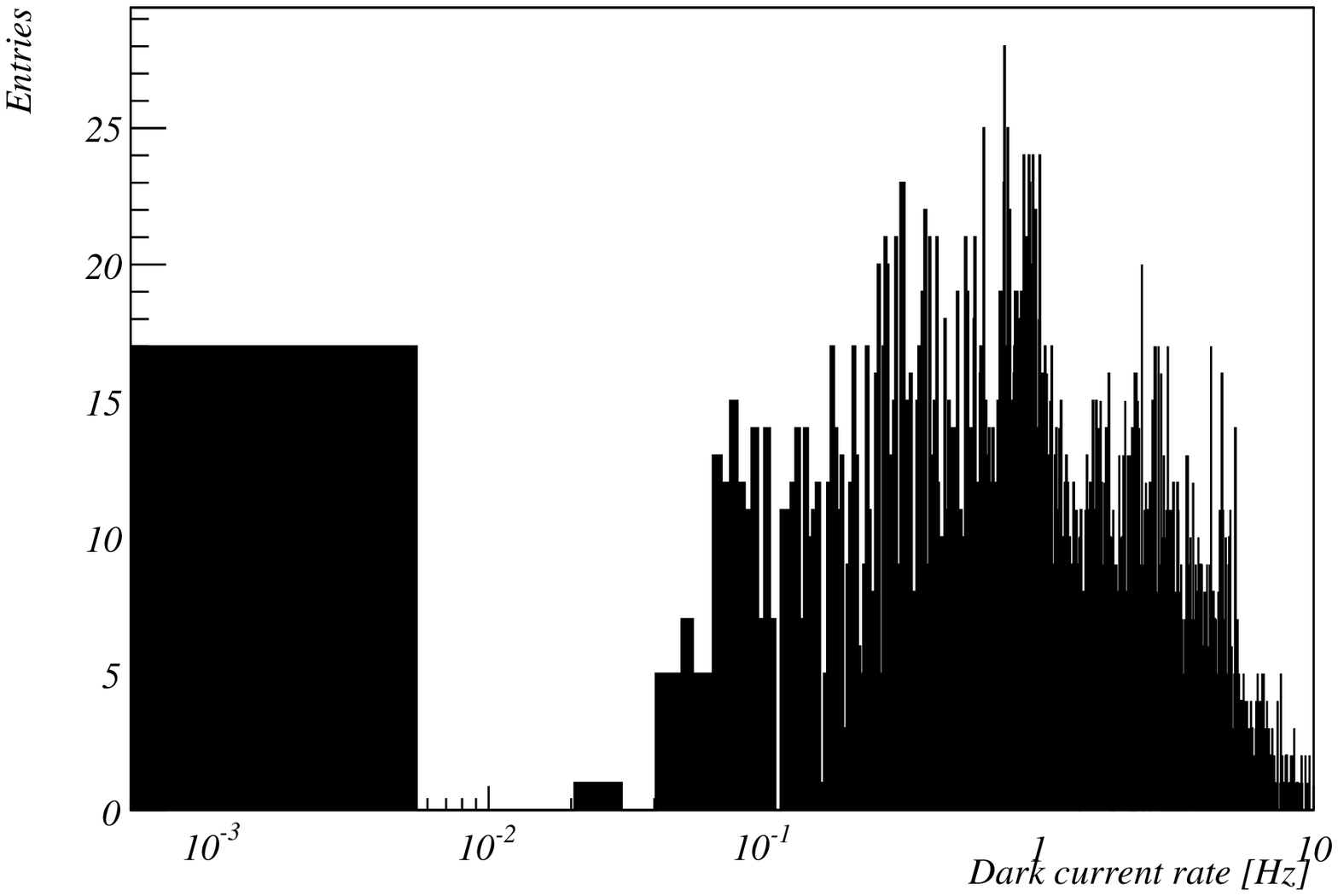, width=.45\textwidth} &
\epsfig{file=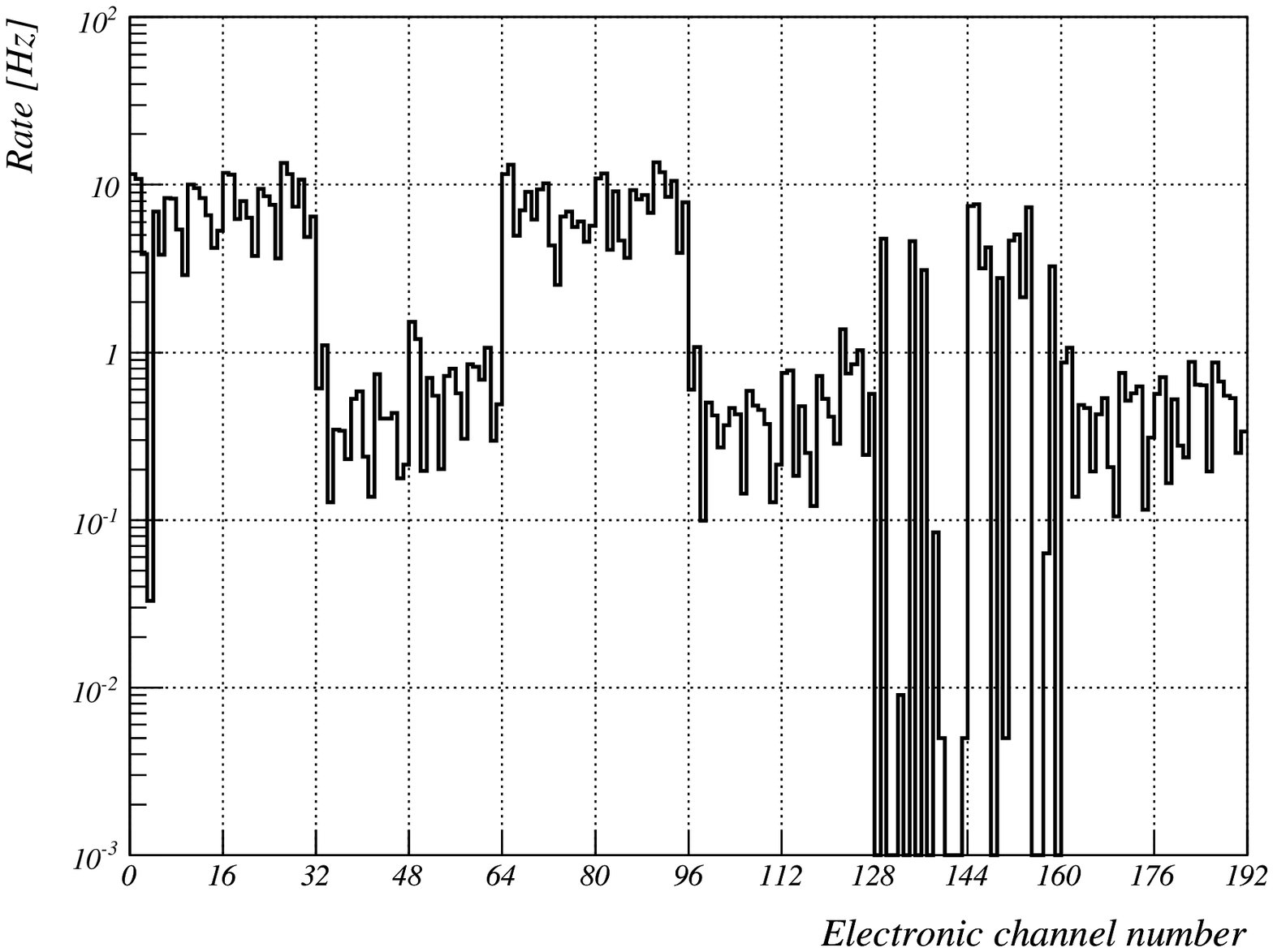,width=.45\textwidth} \\
\end{tabular}
\caption{Rate distribution for one TRT endcap WEB, using the DTMROC accumulation mode. Problematic channels have a very low activity, i.e. accumulation mode rates below 10 mHz.}
\label{fig:DarkCurrent}
\end{center}
\end{figure}

A distribution of the rate is shown on Fig. \ref{fig:DarkCurrent}. Typically, working channels have a rate value greater than 0.1 Hz,
 whereas problematic ones have values below 0.01 Hz. 
For some TRT endcap-wheels, the accumulation rate distribution shows clearly a larger activity on one of the lowest of the two 4-Planes wheels, whereas for other ones a rate of about 1 Hz is measured across all the readout channels.\\
The Accumulation mode technique, combined with the noise scan method offers then a powerful method to check for additional problematic channels in the TRT detector.

%
\section{Analysis and Results Interpretation}
\label{ITAnalysis}

\begin{figure}[ht]
\begin{center}
\epsfig{file=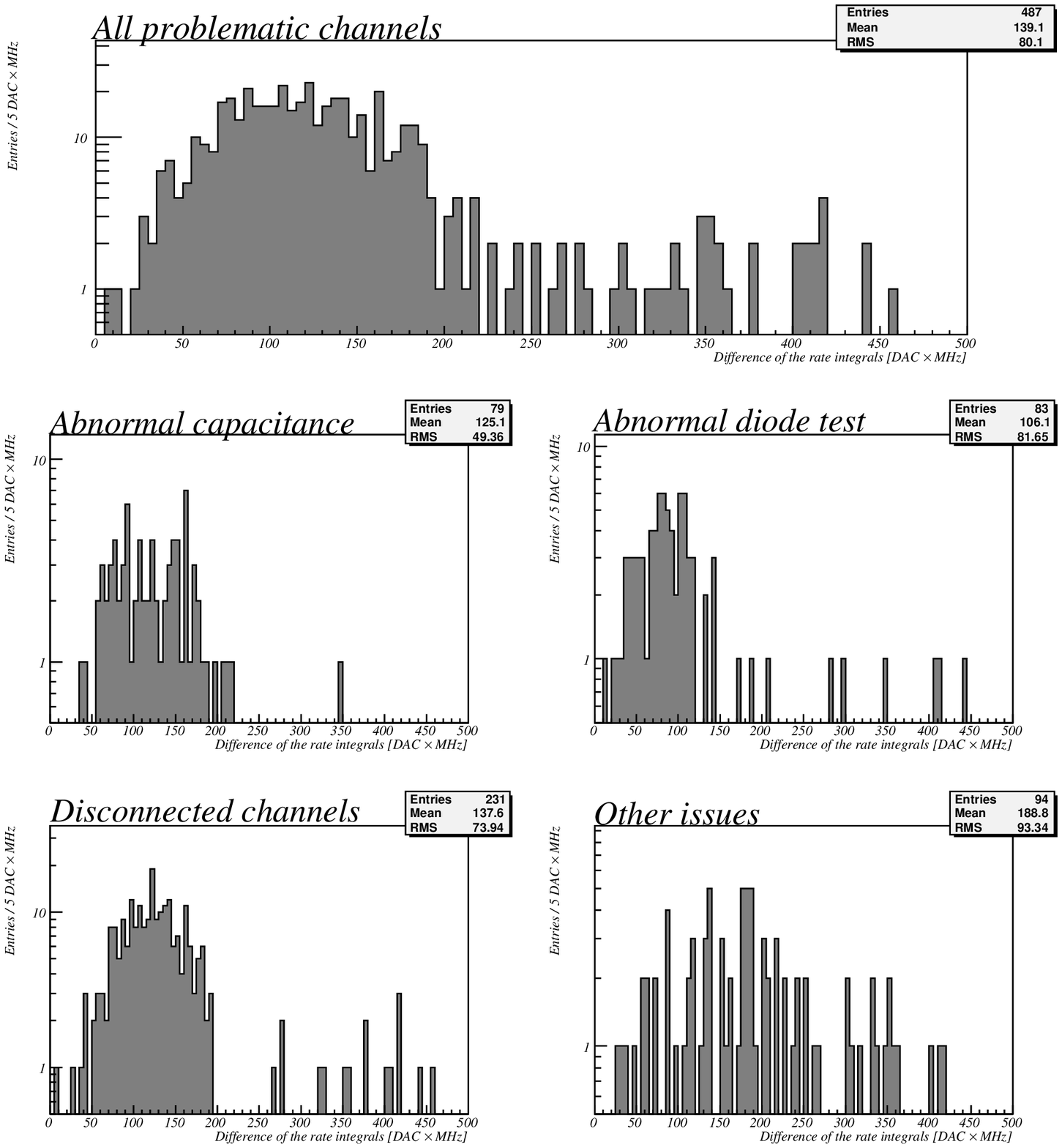,width=1.1\textwidth}
\caption{Rate integral distribution of the different problematic channels for A-type and B-type TRT wheels. }
\label{fig:BadChannelsRI}
\end{center}
\end{figure}

\begin{figure}[ht]
\begin{center}
\epsfig{file=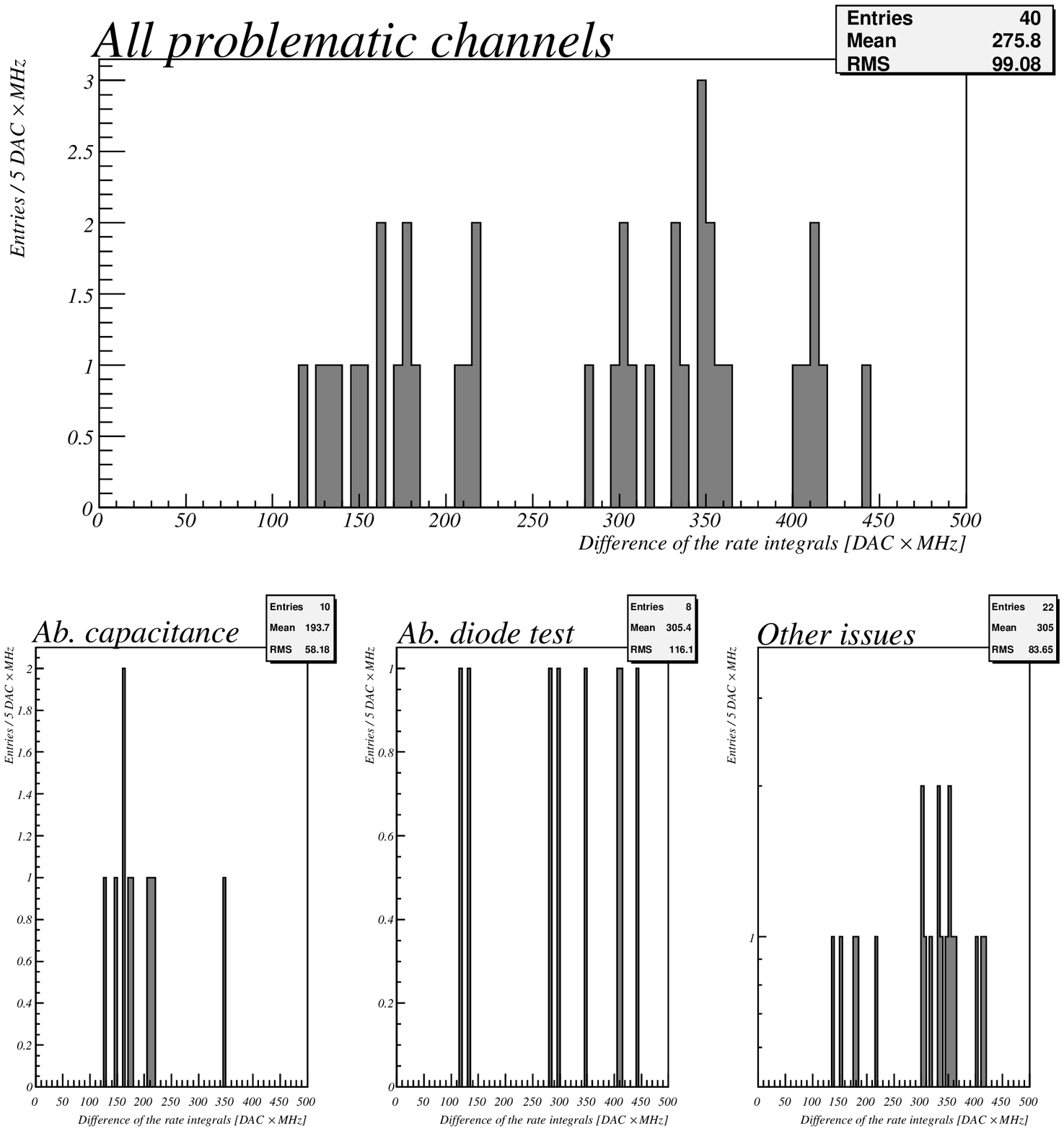,width=1.1\textwidth}
\caption{Rate integral distribution of the different problematic channels failing the Noise Scan technique for A-type and B-type TRT wheels. }
\label{fig:BadChannelsRINotSeenByRI}
\end{center}
\end{figure}

\begin{figure}[ht]
\begin{center}
\epsfig{file=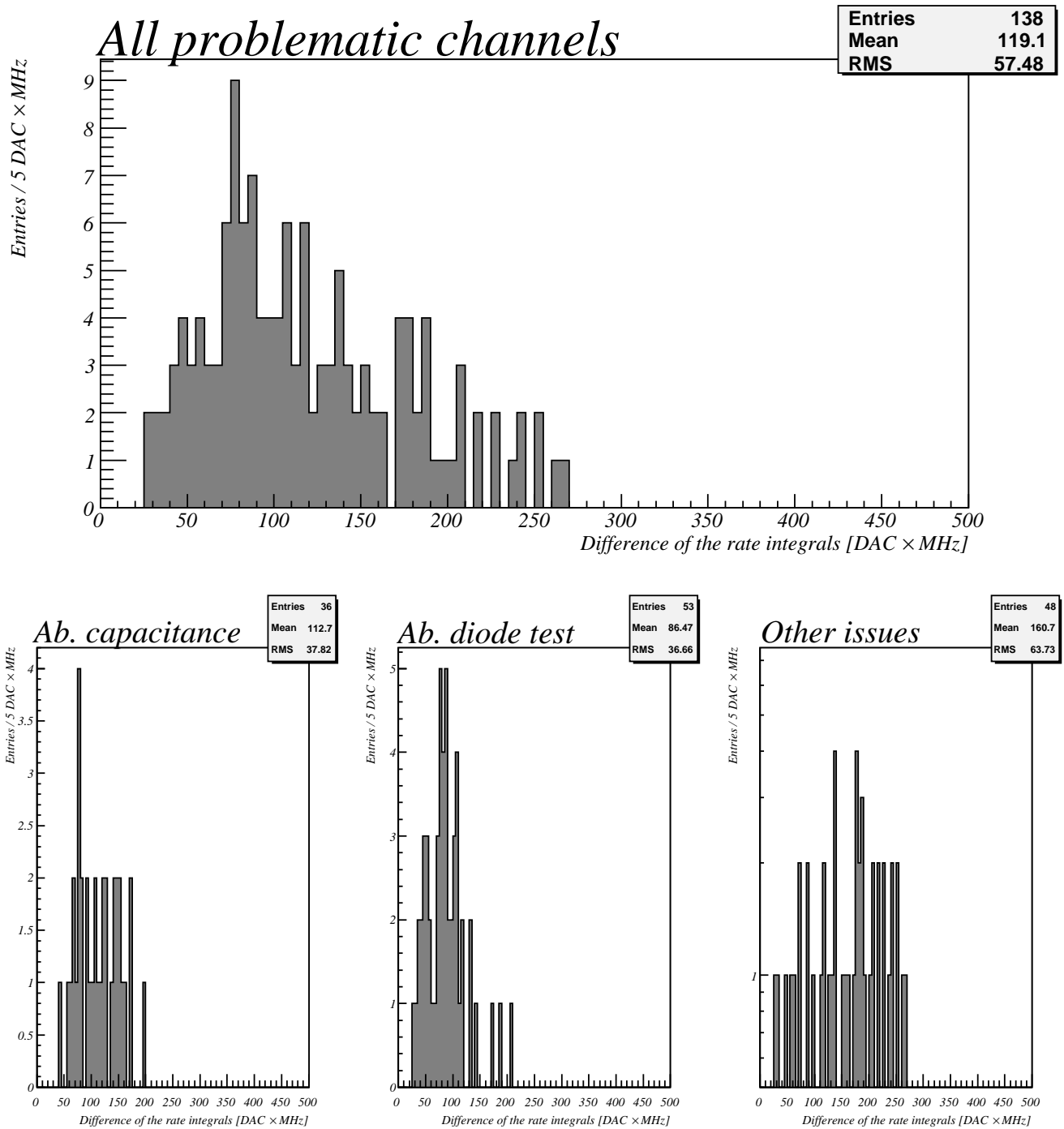,width=1.1\textwidth}
\caption{Rate integral distribution of the different problematic channels failing the Accumulation mode technique for A-type and B-type TRT wheels. }
\label{fig:BadChannelsRINotSeenByDC}
\end{center}
\end{figure}

\begin{figure}[ht]
\begin{center}
\epsfig{file=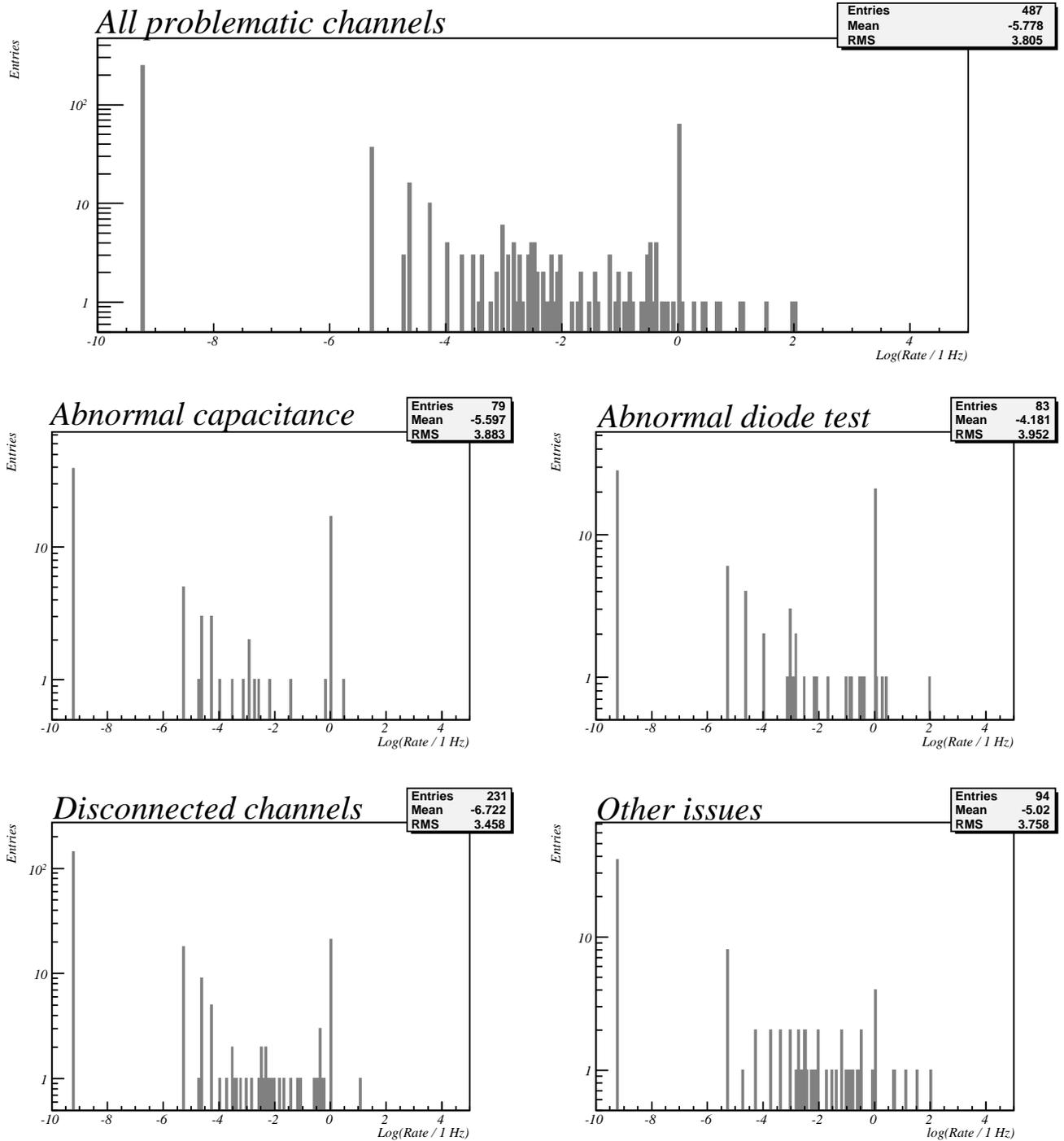,width=1.1\textwidth}
\caption{Dark current rate distribution of the different problematic channels for A-type and B-type TRT wheels. }
\label{fig:BadChannelsDC}
\end{center}
\end{figure}

\begin{figure}[ht]
\begin{center}
\epsfig{file=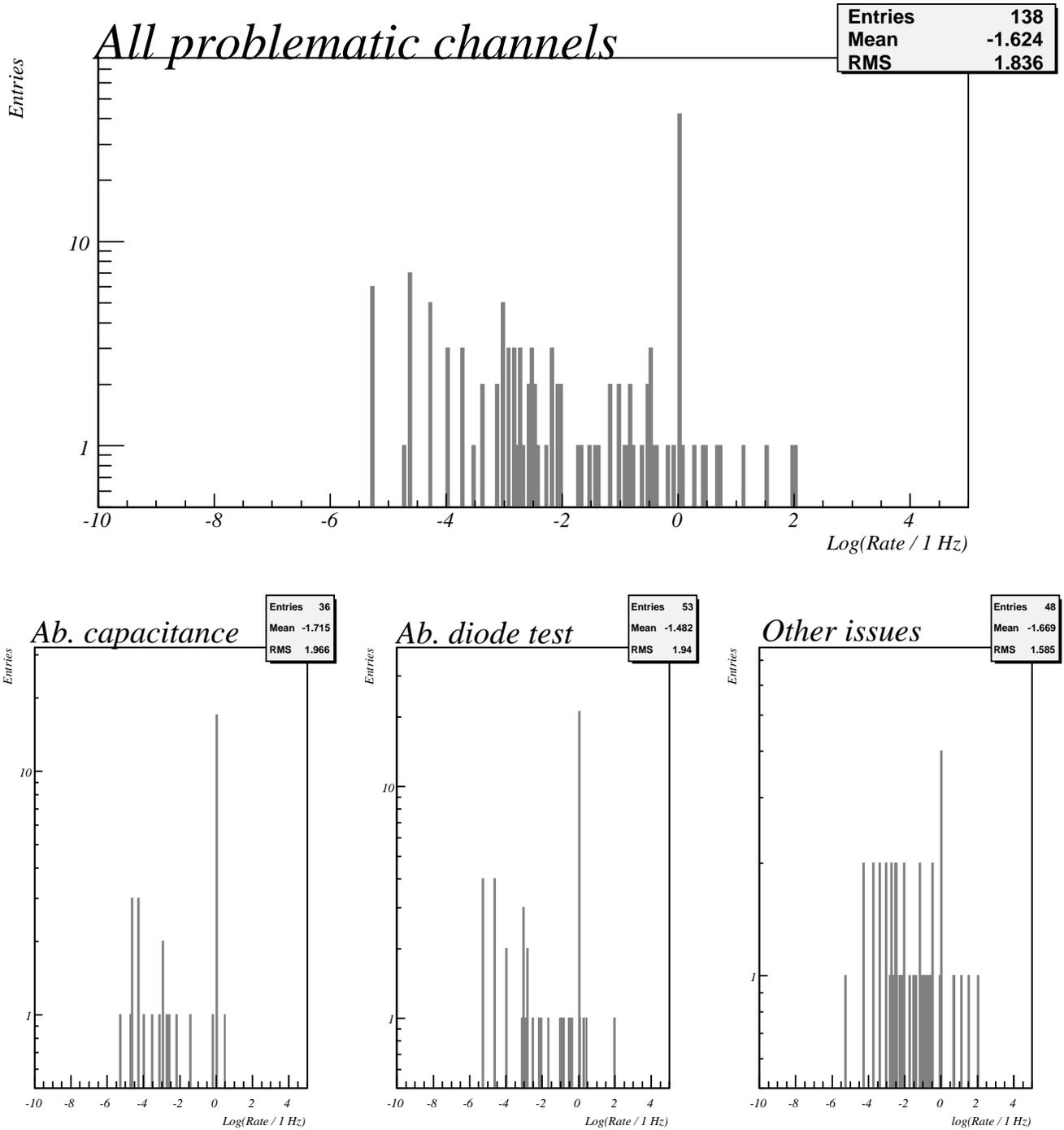,width=1.1\textwidth}
\caption{Dark current distribution of the different problematic channels failing the Accumulation mode technique for A-type and B-type TRT wheels. }
\label{fig:BadChannelsDCNotSeenByDC}
\end{center}
\end{figure}

\begin{figure}[ht]
\begin{center}
\epsfig{file=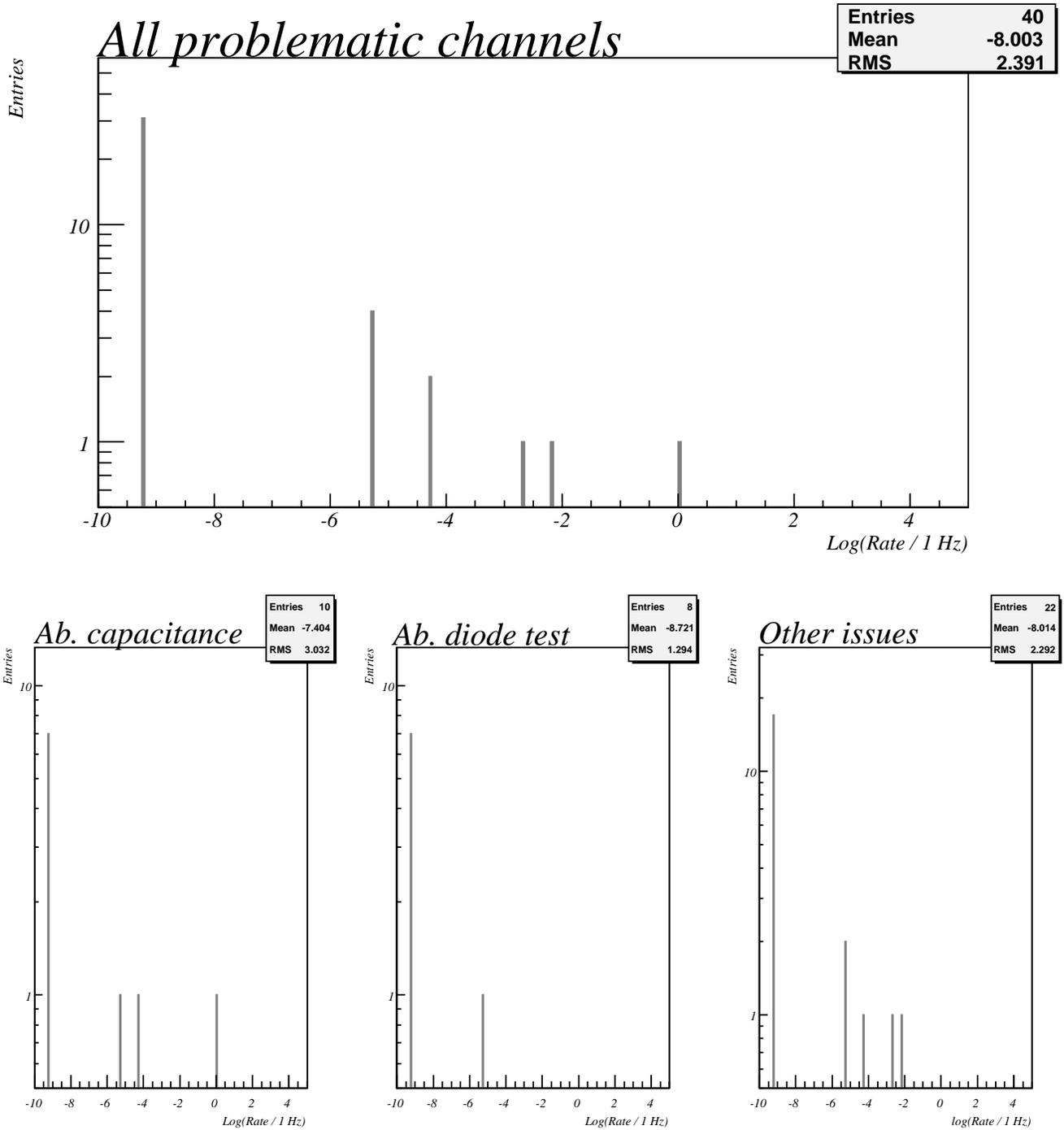,width=1.1\textwidth}
\caption{Dark current distribution of the different problematic channels failing the Noise Scan technique for A-type and B-type TRT wheels. }
\label{fig:BadChannelsDCNotSeenByRI}
\end{center}
\end{figure}

The connectivity check method based on the noise scan rate integral and the accumulation mode techniques 
has been successfully applied during the integration of the TRT endcap-detector.\\
 Finally, only 487 readout channels are found not to be responding.
These channels comprise 231 channels for which the protection resistors were unsoldered after they failed the serie of acceptance tests. 
Typically, these channels are characterised by a noise scan rate integral difference value about 130 DAC$\times$MHz counts, much lower than the WEB average ( below three $\sigma$ of the mean value $\mu$). The total distribution of the rate integral is presented on Fig. \ref{fig:BadChannelsRI}. In addition, their corresponding accumulation mode rate is about 10 times lower than the WEB average, below 0.01 Hz 
and reaches for most of these channels a typical value of 0 Hz (see Fig. \ref{fig:BadChannelsDC}).\\
 These 231 disconnected channels represent for both A-and B-type TRT endcap-detectors about 45 \% of the total number of problematic channels found at the end of the integration. At the same time, they represent a very small fraction of the total number of readout channels of the TRT endcap-detector. Only 0.19 \% of the total number of channels have been disconnected after acceptance tests.\\
The tests used to check for connectivity, during the integration of the TRT endcap-detector report a total of 256 new problematic channels, 127 for A-type and 129 for B-type wheels. \\
A non negligeable fraction of these channels, are ones for which several repairs have been done during the acceptance tests. As it is shown on Tab. \ref{tab:BadChannels}, 48 of these 256 channels have been repaired during acceptance tests and even stronger correlation exists between these channels and bad WEBs, i.e. WEBs for which some repairs were done.\\
Another interesting remark concerns the fraction of repaired channels: A-type WEBs need more repairs than B-type ones (see Tab. \ref{tab:BadChannels2}). For several wheels, like A02C for instance, all the repaired channels break again with time and thus fail the integration tests.\\
The other 208 new bad channels, perfectly operational after the completion of the acceptance tests, are largely due to WEB via issues.
Several explanations have been suggested to understand the apparition of these new faulty channels. For a small fraction, they appeared during their transportation from the acceptance test stand to the integration test area. For others, they simply have been not seen by the serie of acceptance tests, but are more sensitive to the two methods implemented during the integration. For a last fraction, a possible explanation is metallisation issues evolving with time.\\
The efficiency of the two methods described in the previous section is clearly established. it enables to find 256 new problematic channels, most of them being reported non working by both methods. For instance, only 40 channels are not found by the noise scan based technique (see Fig. \ref{fig:BadChannelsRI}, \ref{fig:BadChannelsRINotSeenByRI}, \ref{fig:BadChannelsRINotSeenByDC}, \ref{fig:BadChannelsDC}, \ref{fig:BadChannelsDCNotSeenByDC}  and \ref{fig:BadChannelsDCNotSeenByRI}). \\
At the end of the integration, the fraction of non-working readout channels represent only 0.4 \% of the 122880 of the Stack C.\\

\section{Conclusions}

\begin{figure}[ht]
\begin{center}
\begin{tabular}{cc}
\includegraphics[width=9.0cm,angle=0]{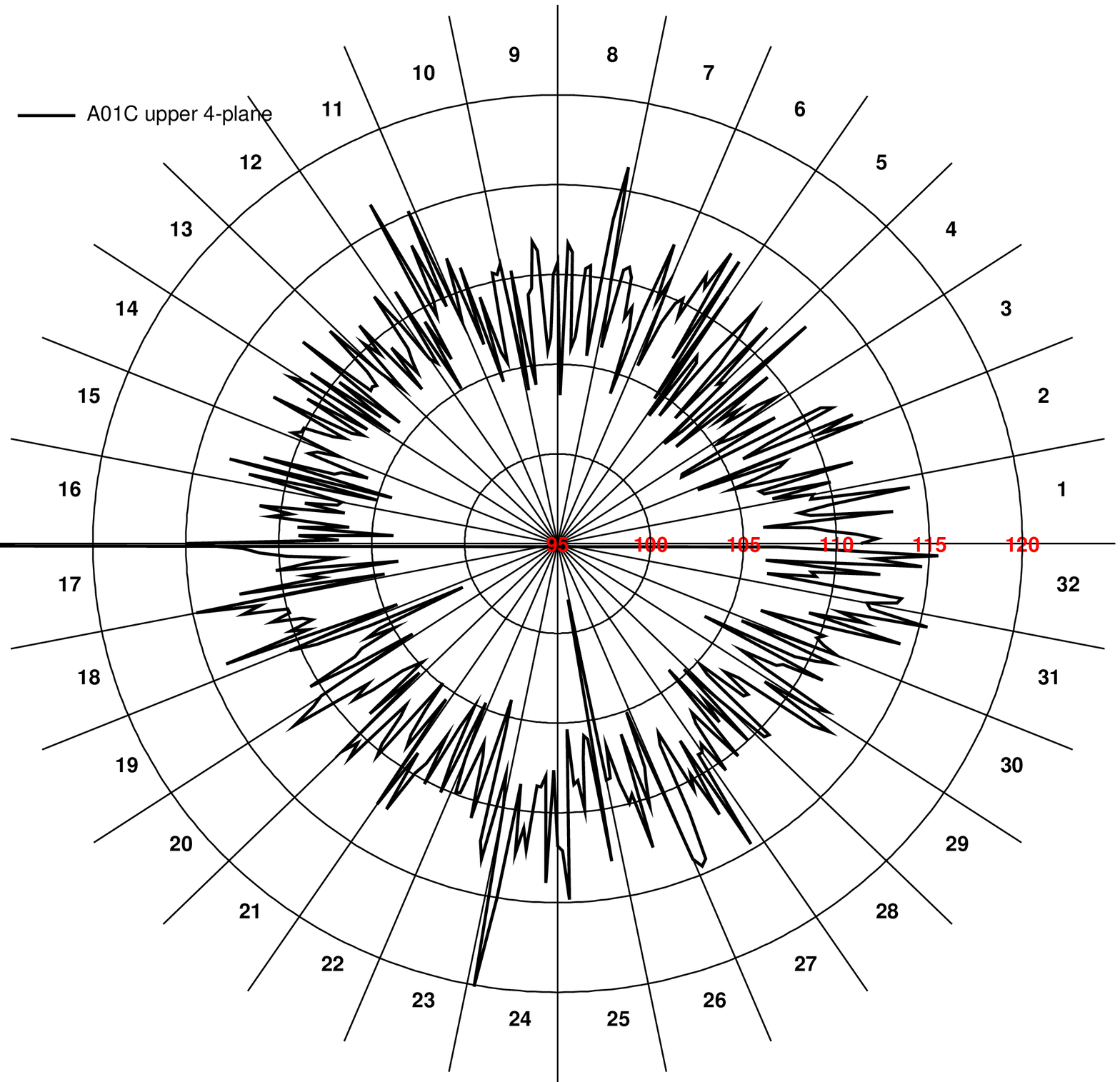}&
\includegraphics[width=9.0cm,angle=0]{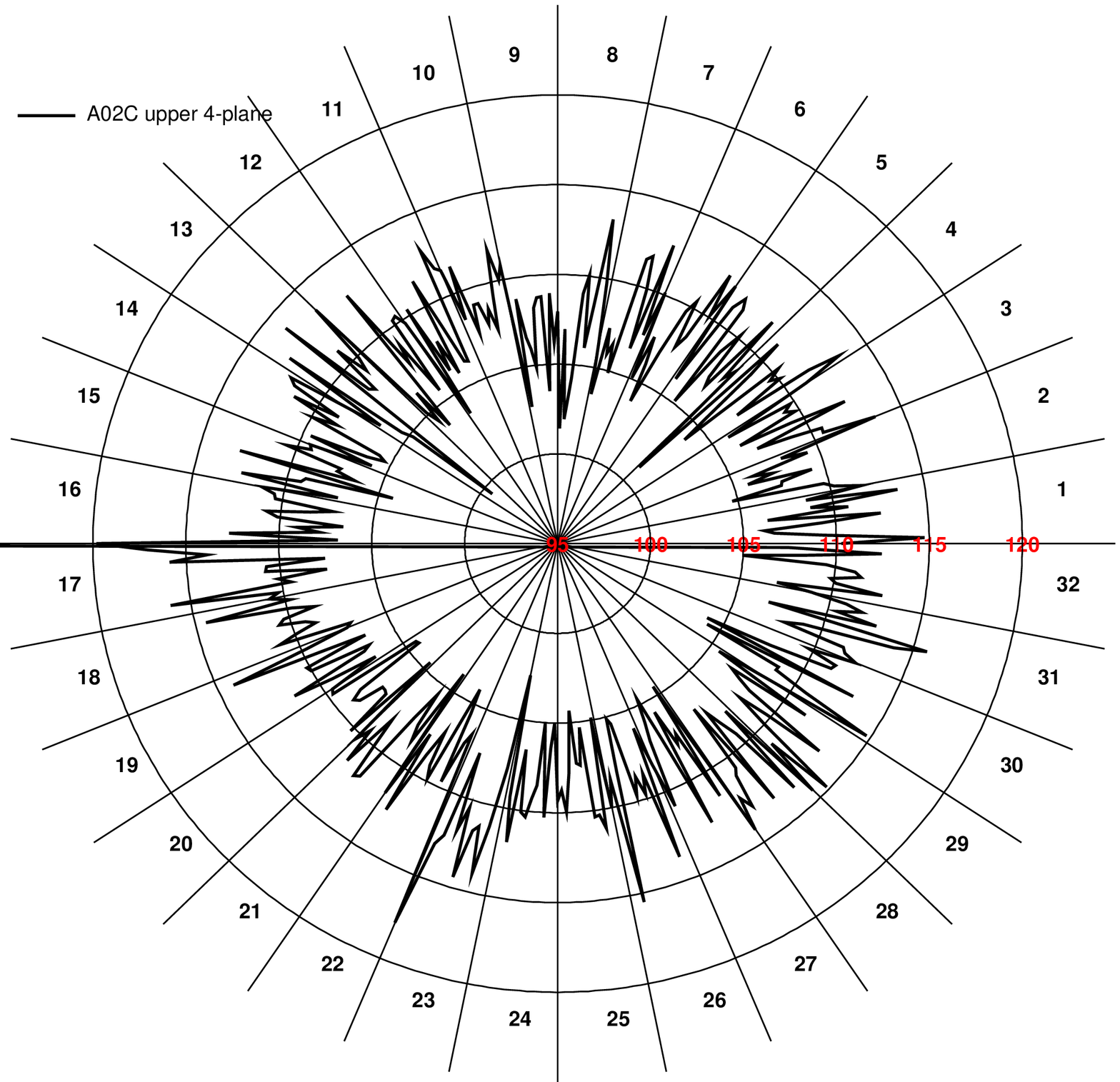}\\

\includegraphics[width=9.0cm,angle=0]{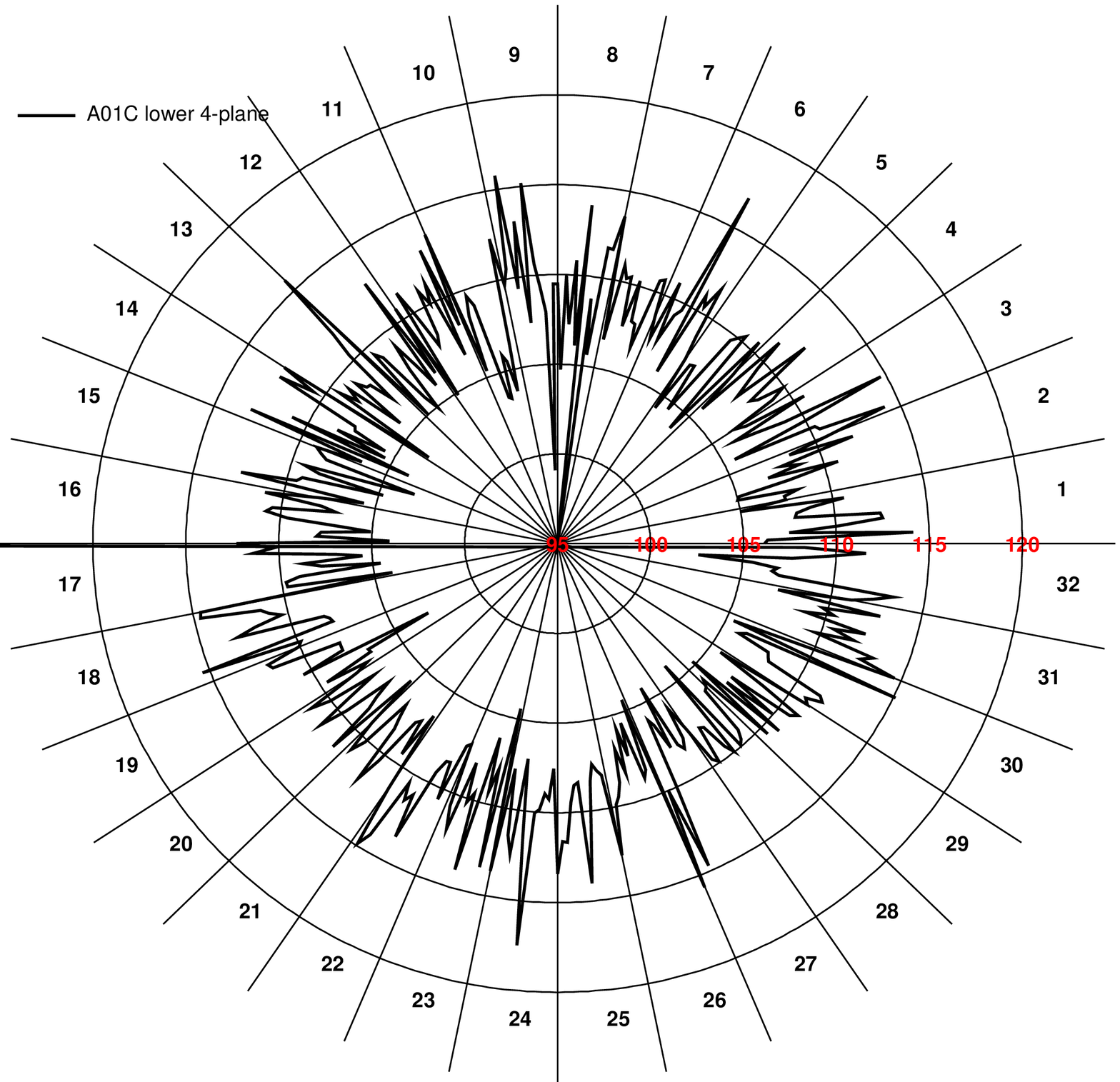}&
\includegraphics[width=9.0cm,angle=0]{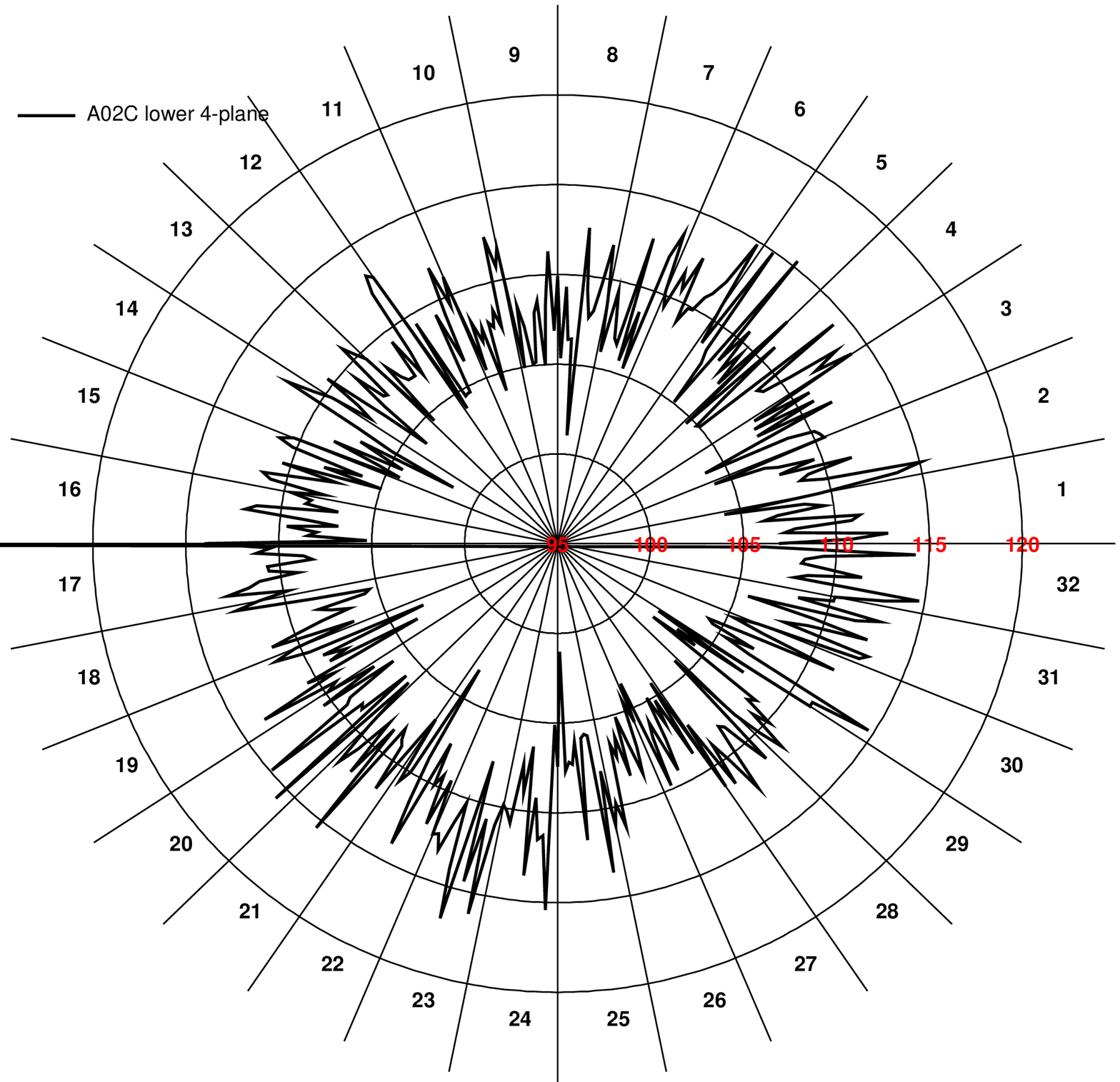}\\
\end{tabular}
\caption{Average per ASDBLR 300 kHz DAC low-threshold for the two upper and lower 4-planes wheels A01C and A02C. }
\label{fig:300kHzA1A2}
\end{center}
\end{figure}

\begin{figure}[ht]
\begin{center}
\begin{tabular}{cc}
\includegraphics[width=9.0cm,angle=0]{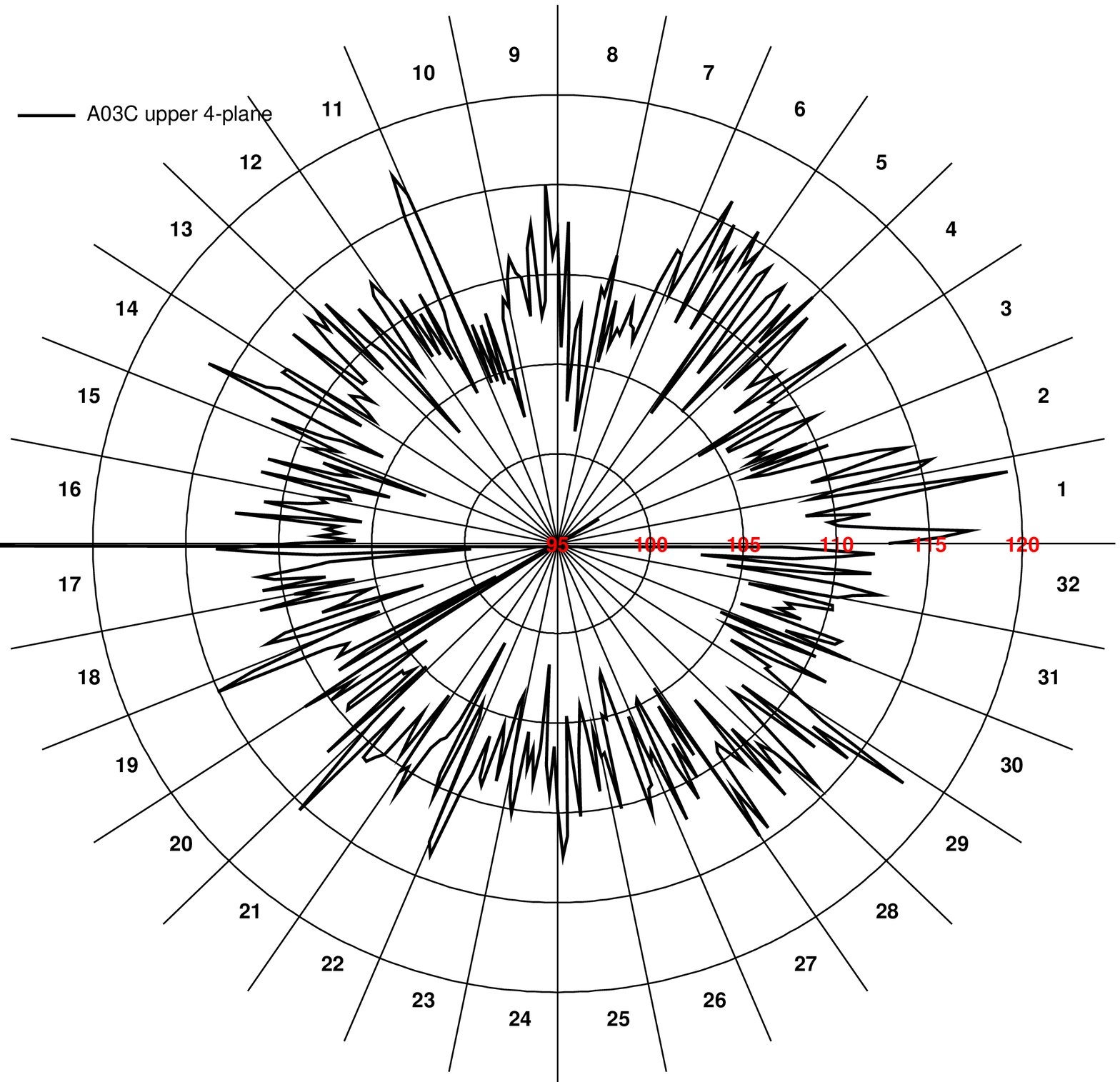}&
\includegraphics[width=9.0cm,angle=0]{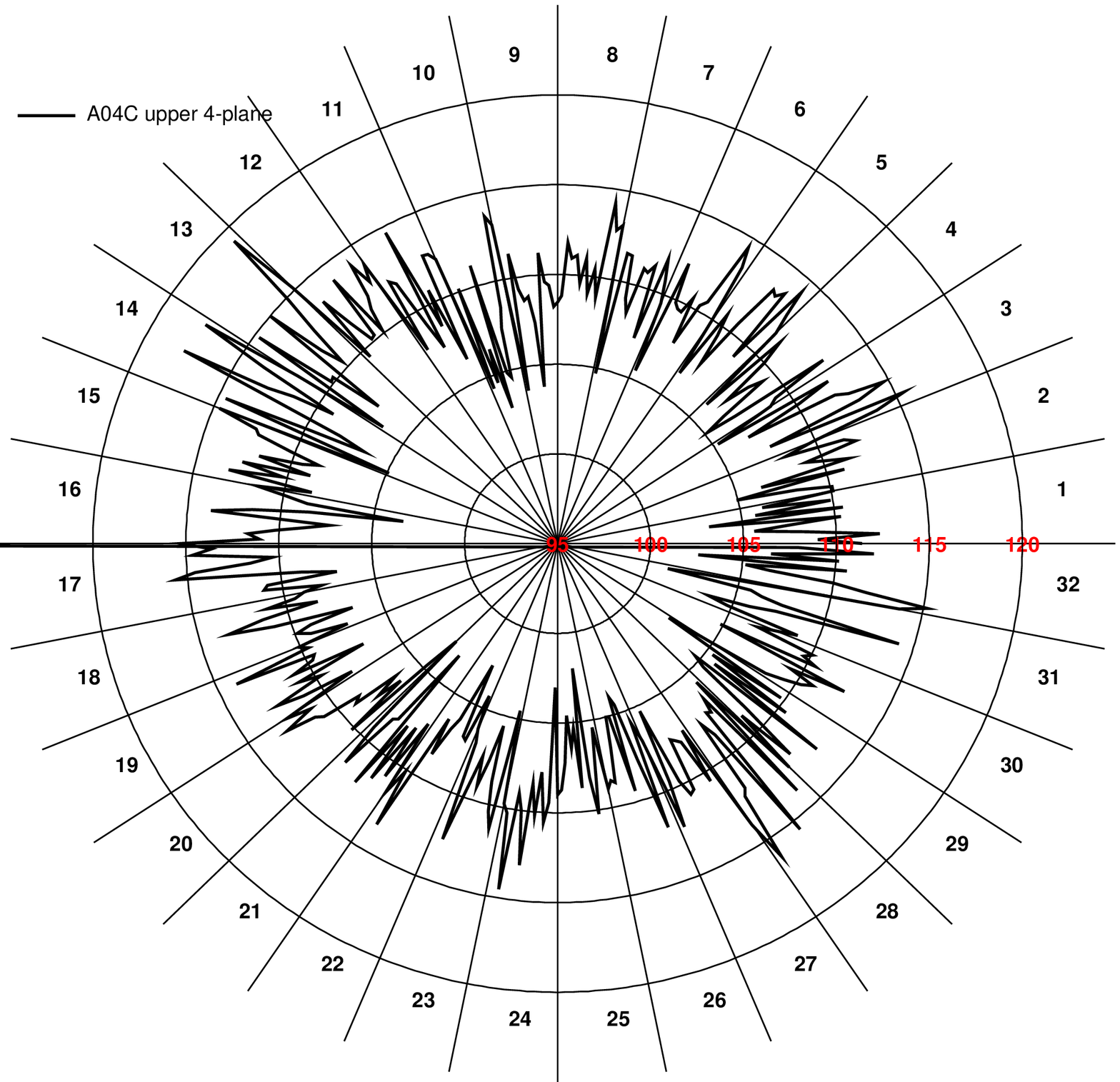}\\

\includegraphics[width=9.0cm,angle=0]{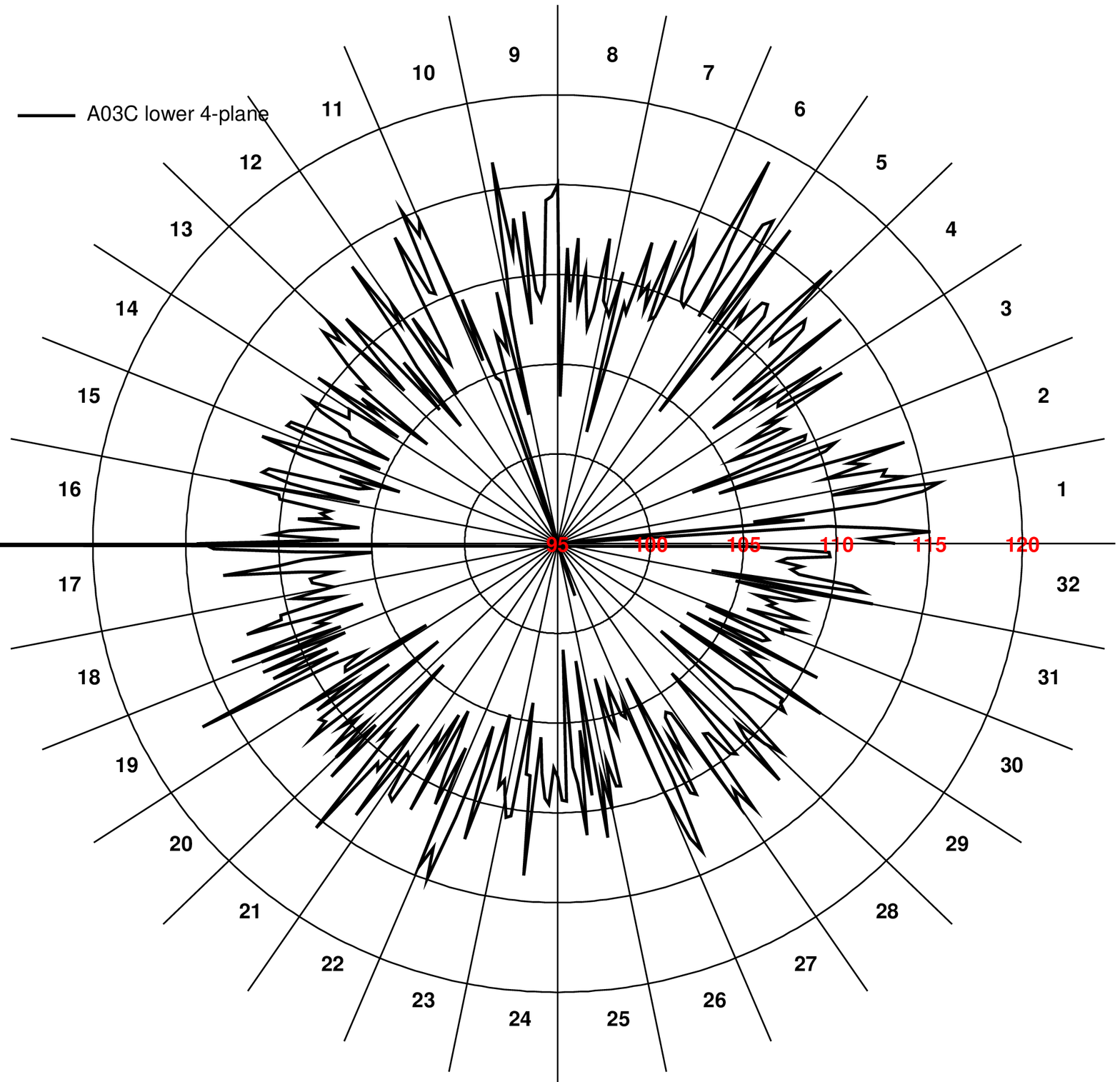}&
\includegraphics[width=9.0cm,angle=0]{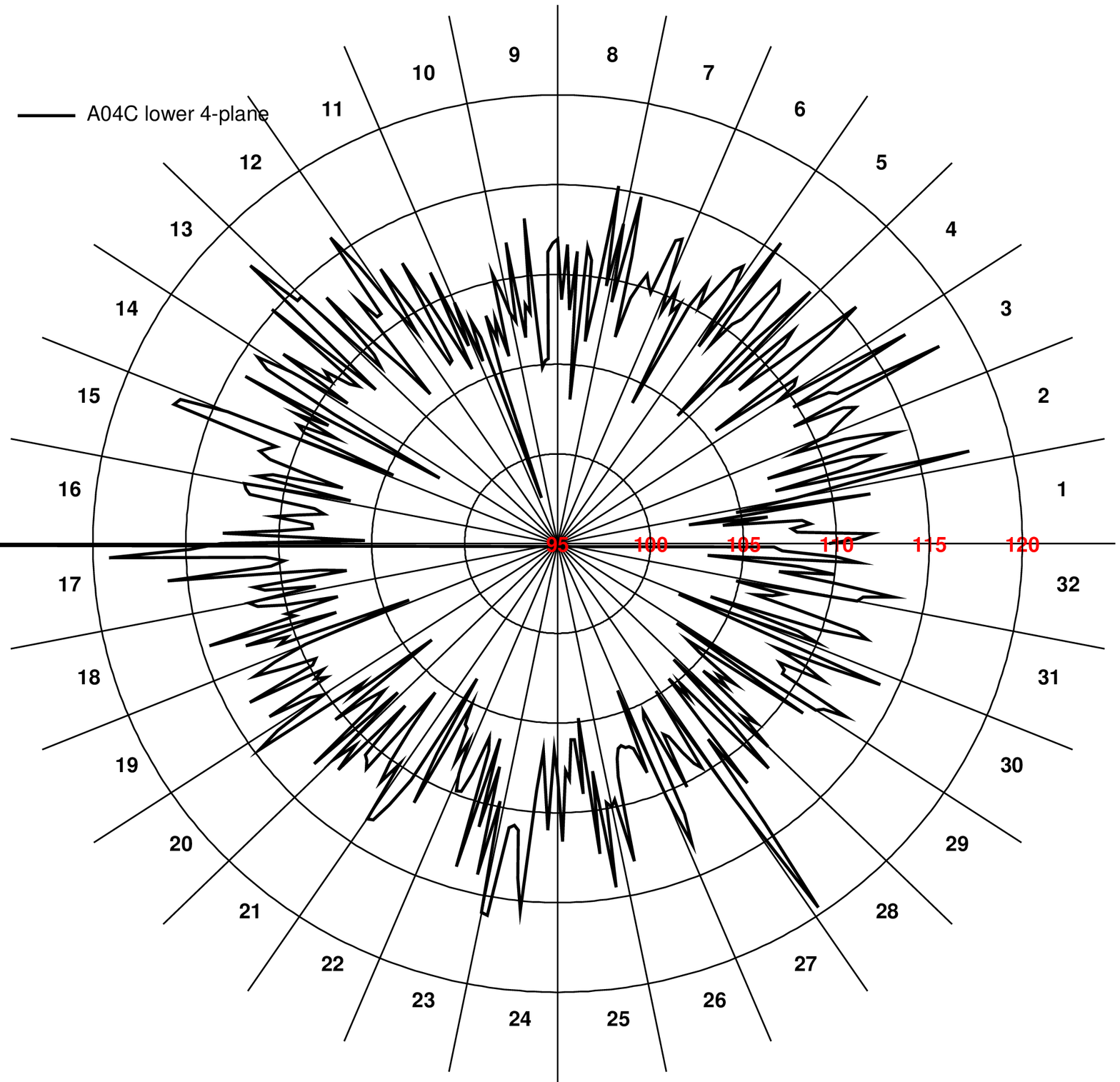}\\
\end{tabular}
\caption{Average per ASDBLR 300 kHz DAC low-threshold for the two upper and lower 4-planes wheels A03C and A04C. }
\label{fig:300kHzA3A4}
\end{center}
\end{figure}

\begin{figure}[ht]
\begin{center}
\begin{tabular}{cc}
\includegraphics[width=9.0cm,angle=0]{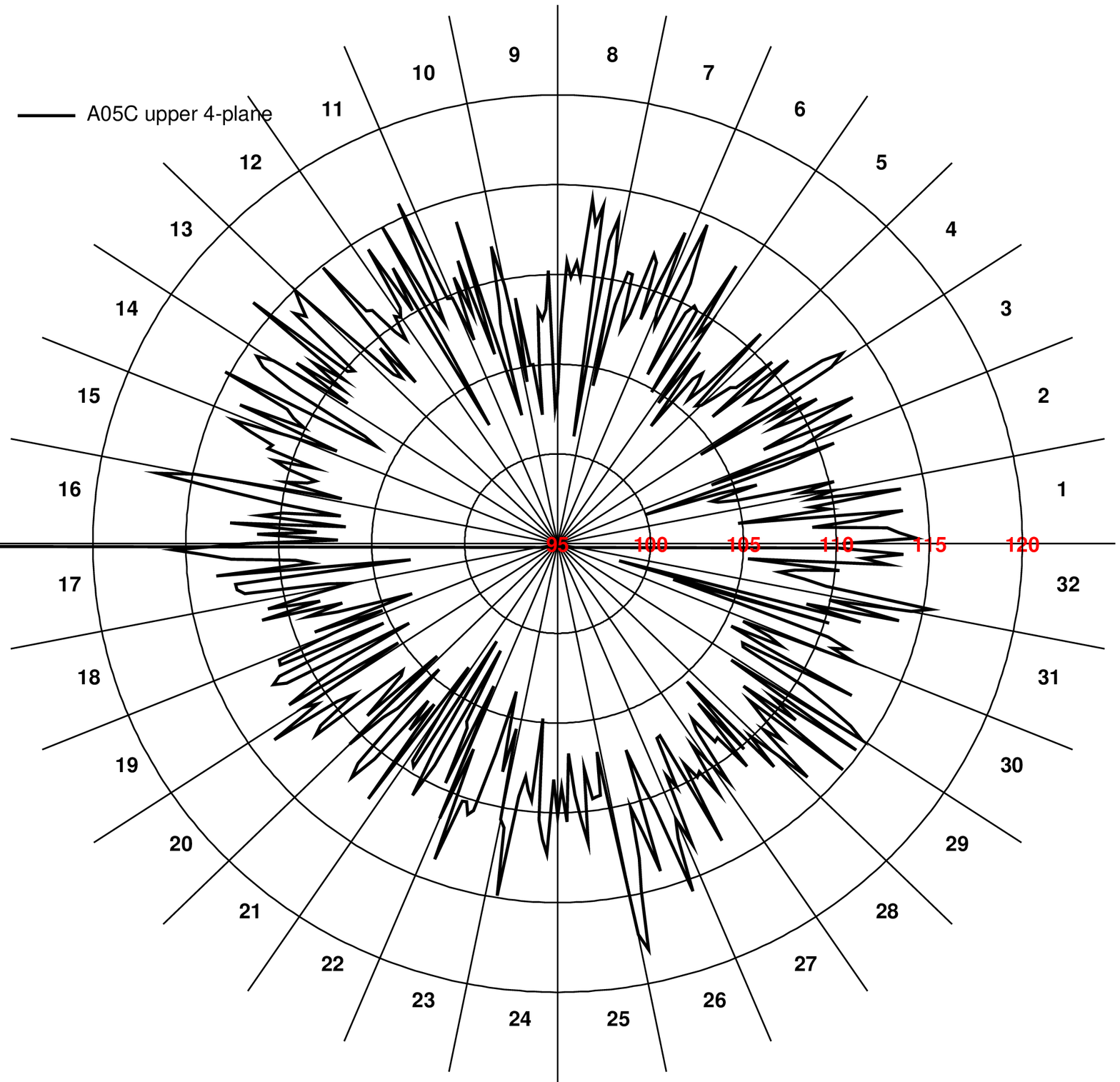}&
\includegraphics[width=9.0cm,angle=0]{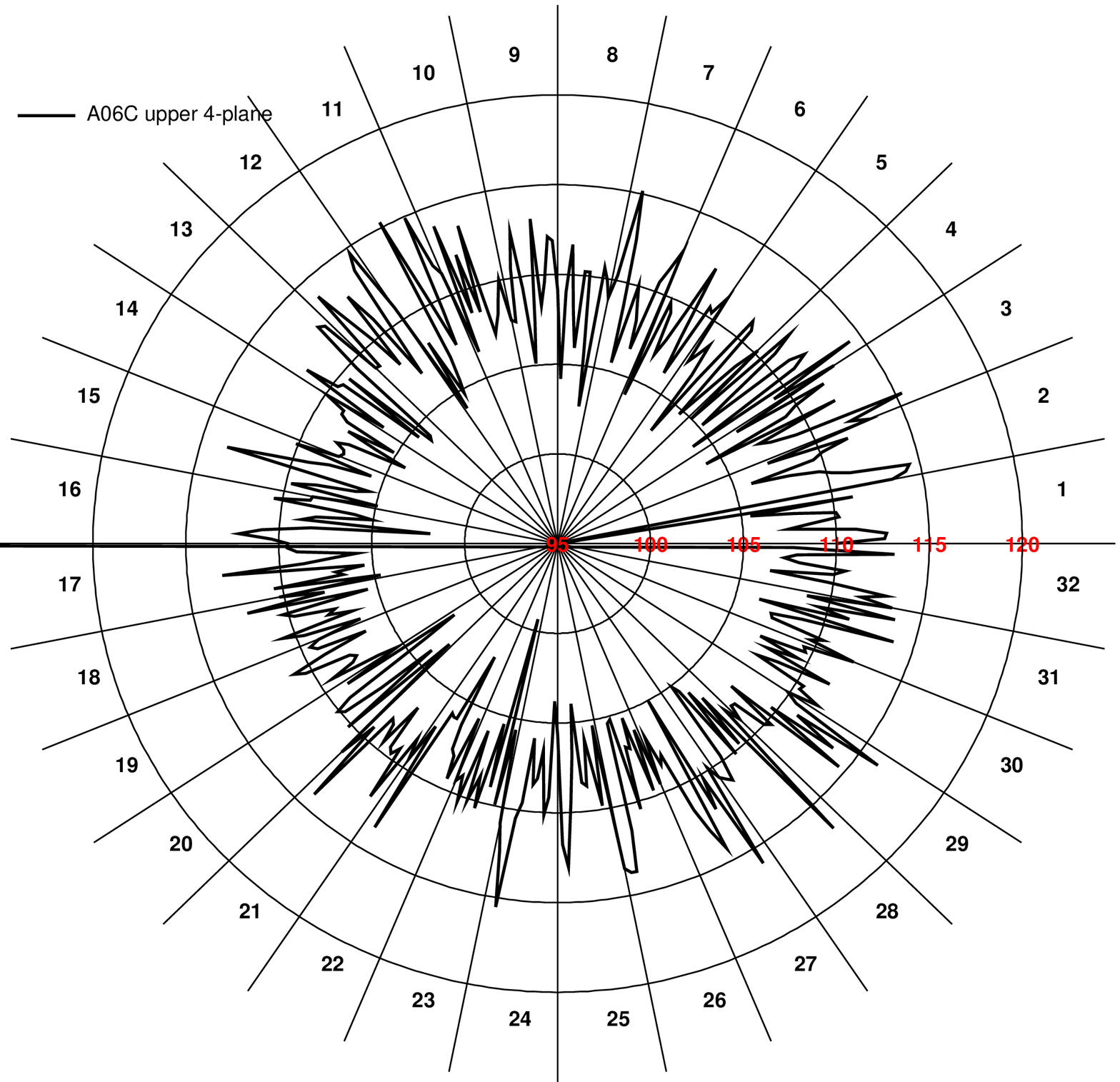}\\

\includegraphics[width=9.0cm,angle=0]{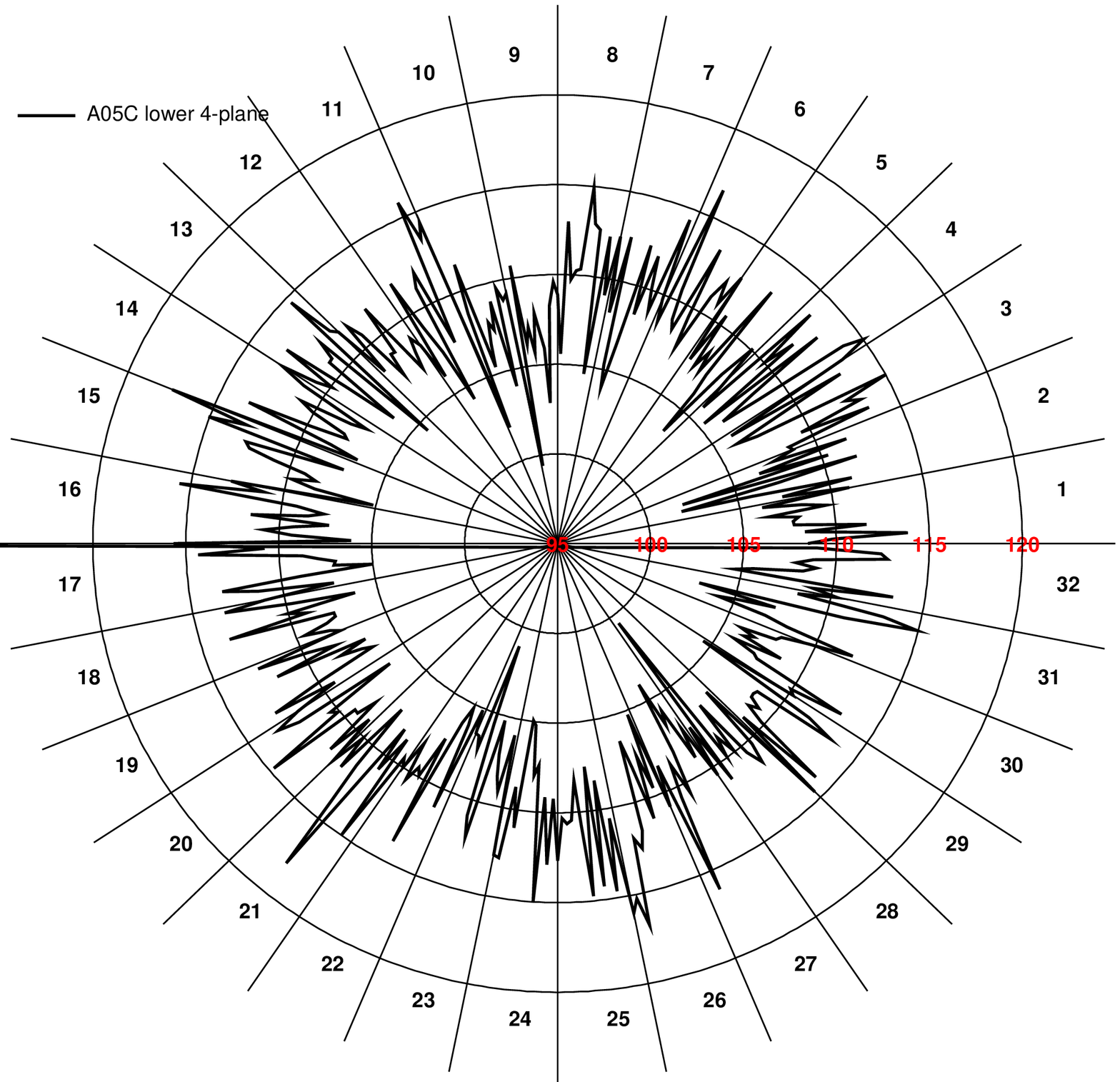}&
\includegraphics[width=9.0cm,angle=0]{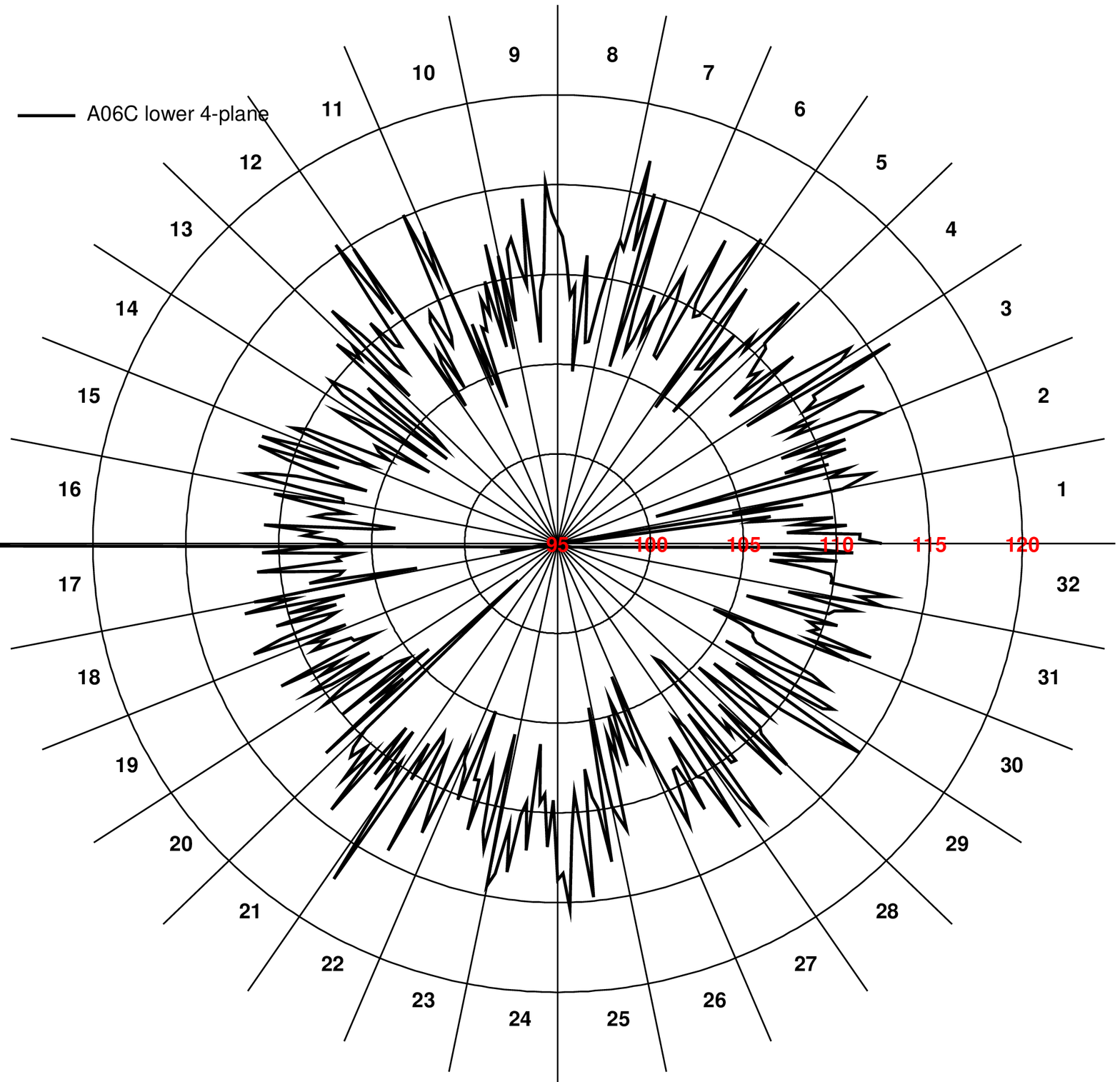}\\
\end{tabular}
\caption{Average per ASDBLR 300 kHz DAC low-threshold for the two upper and lower 4-planes wheels A05C and A06C. }
\label{fig:300kHzA5A6}
\end{center}
\end{figure}

\begin{figure}[ht]
\begin{center}
\begin{tabular}{cc}
\includegraphics[width=9.0cm,angle=0]{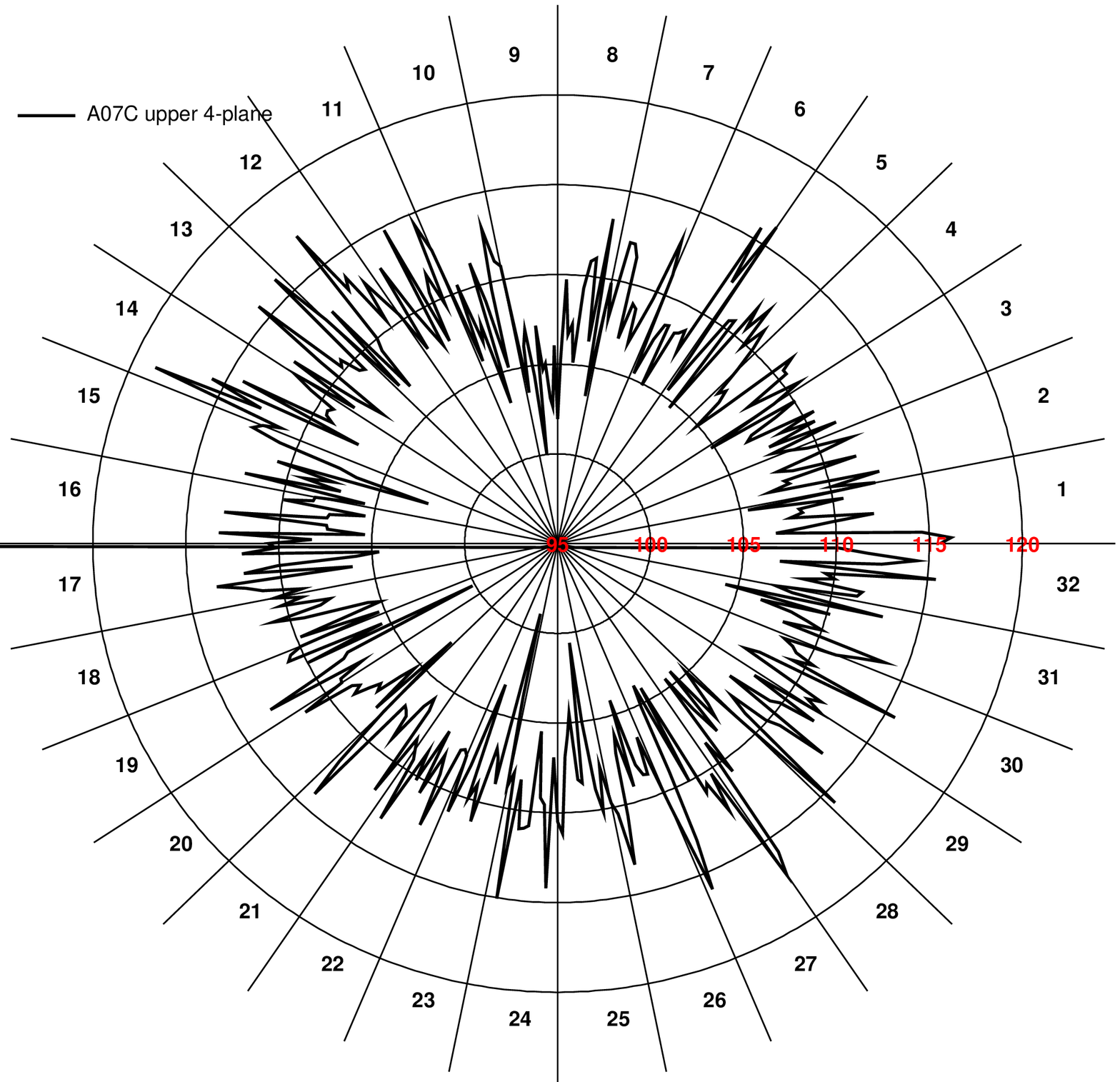}&
\includegraphics[width=9.0cm,angle=0]{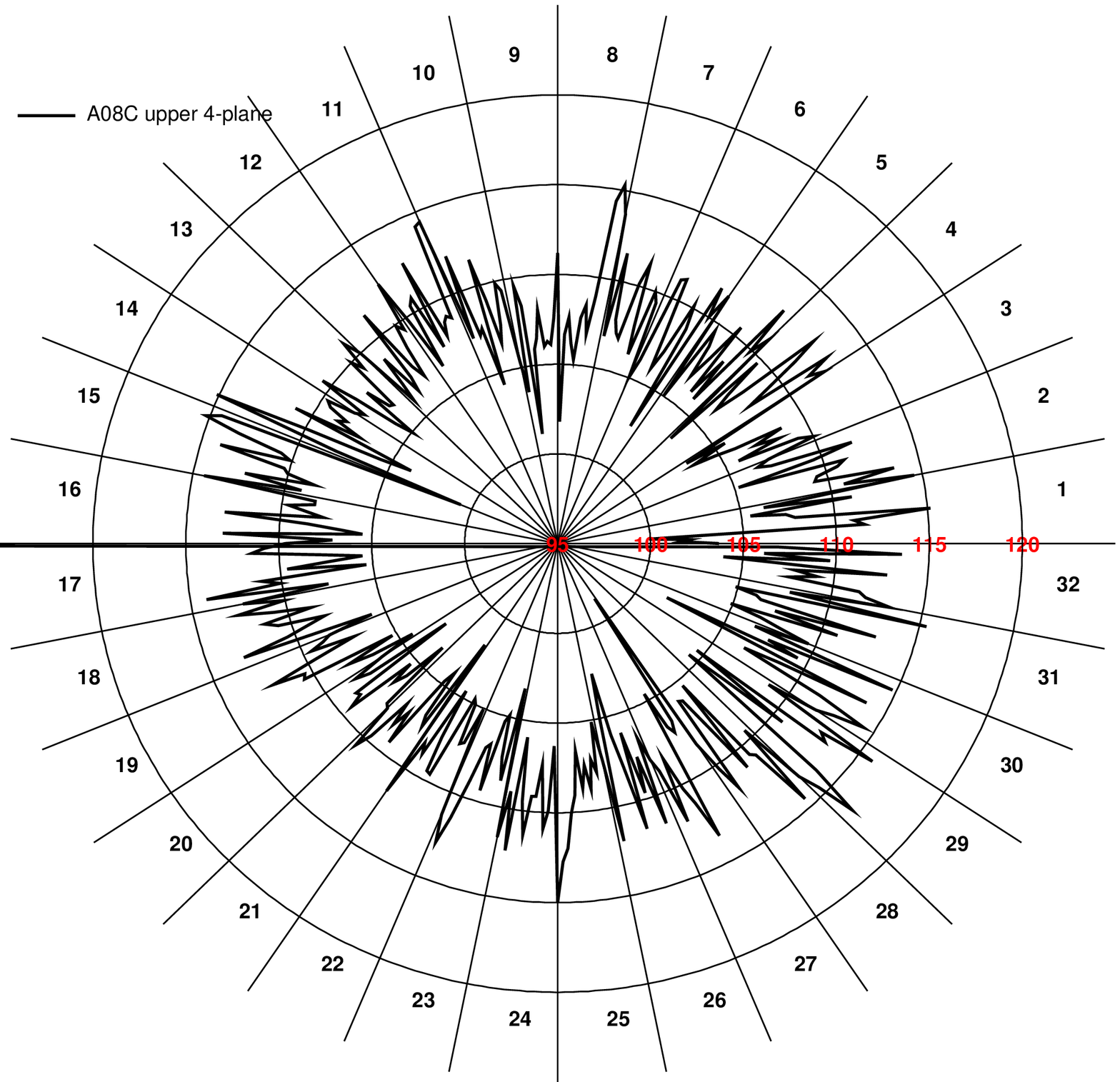}\\

\includegraphics[width=9.0cm,angle=0]{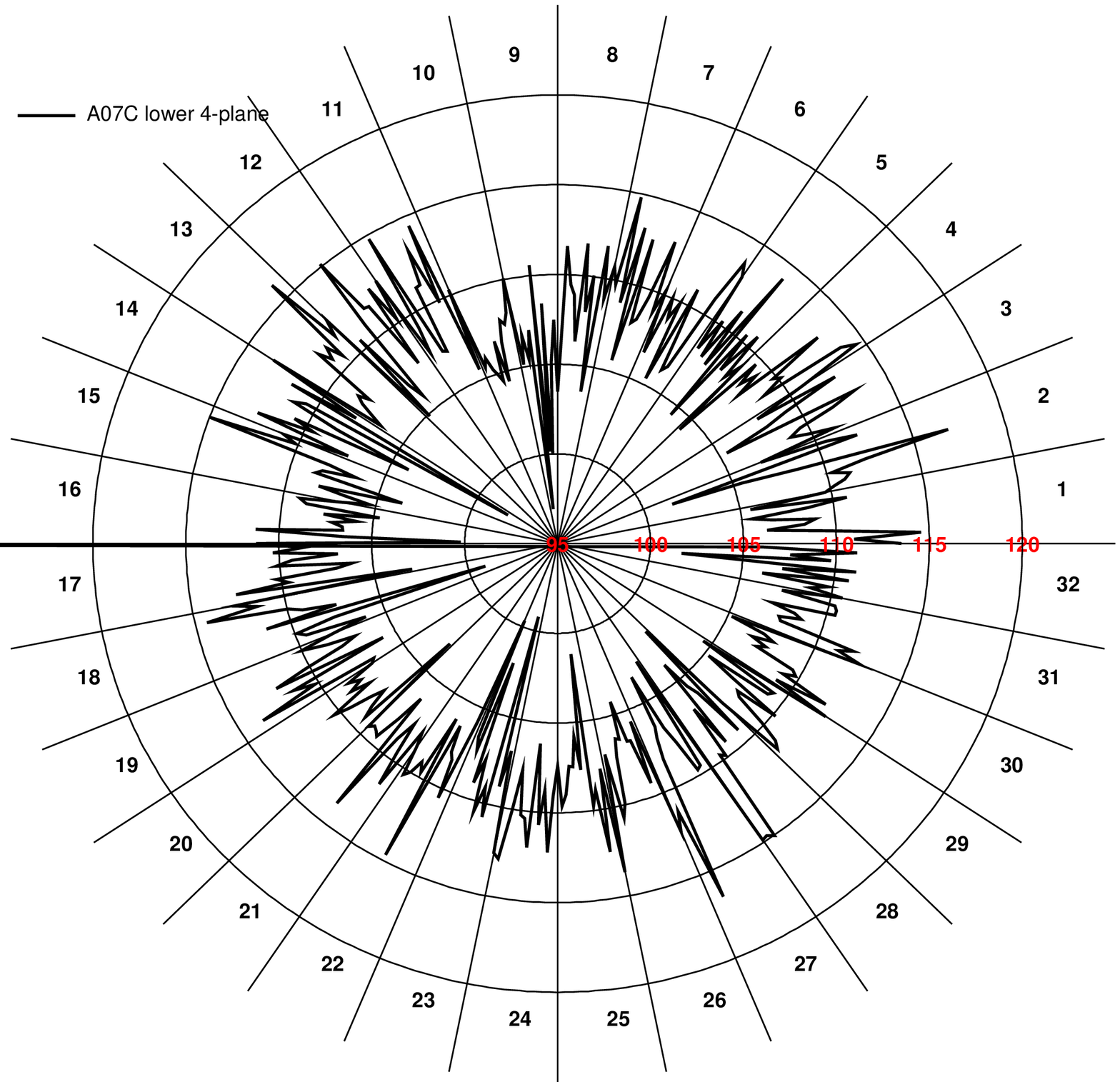}&
\includegraphics[width=9.0cm,angle=0]{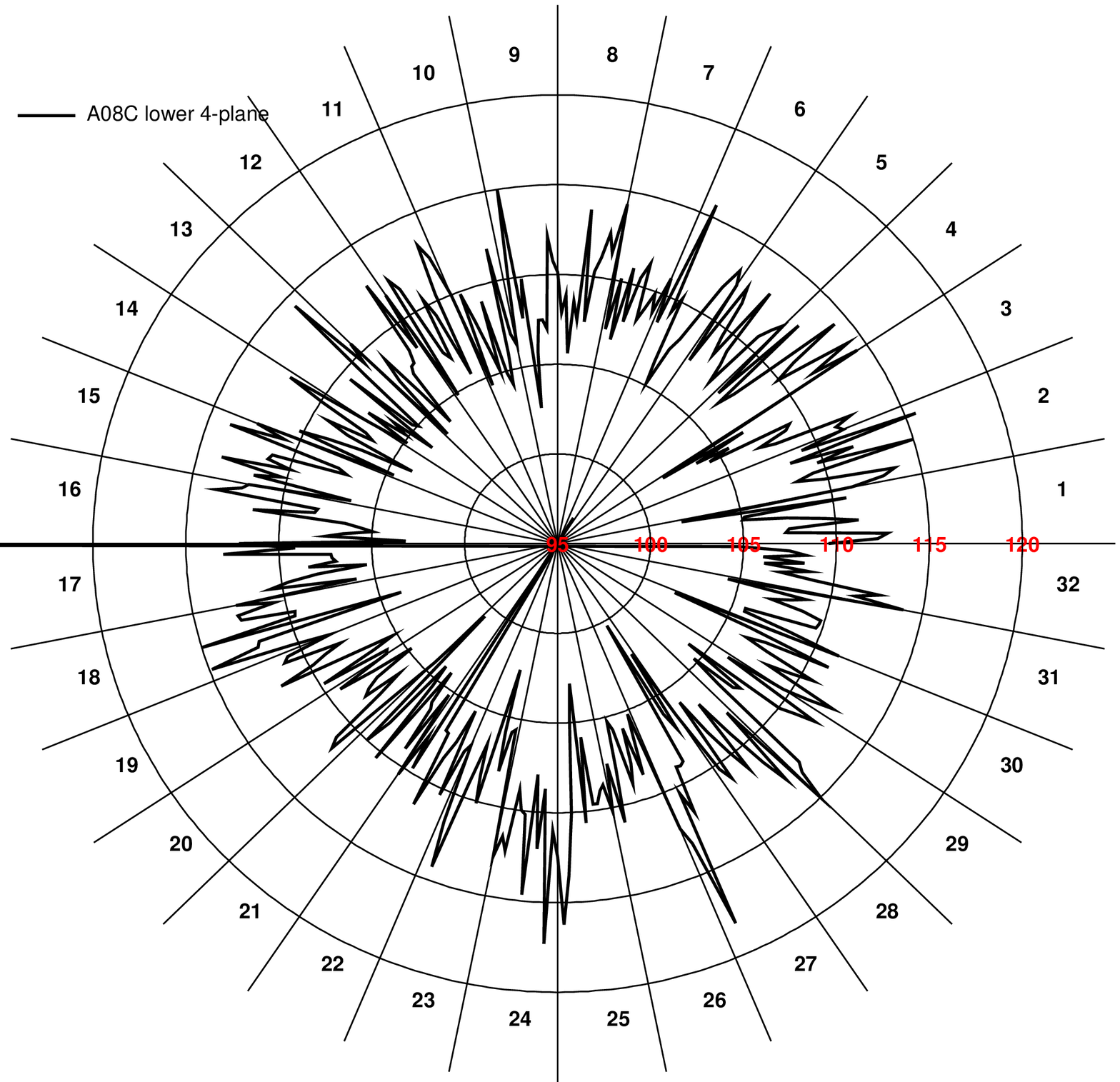}\\
\end{tabular}
\caption{Average per ASDBLR 300 kHz DAC low-threshold for the two upper and lower 4-planes wheels A07C and A08C. }
\label{fig:300kHzA7A8}
\end{center}
\end{figure}

\begin{figure}[ht]
\begin{center}
\begin{tabular}{cc}
\includegraphics[width=9.0cm,angle=0]{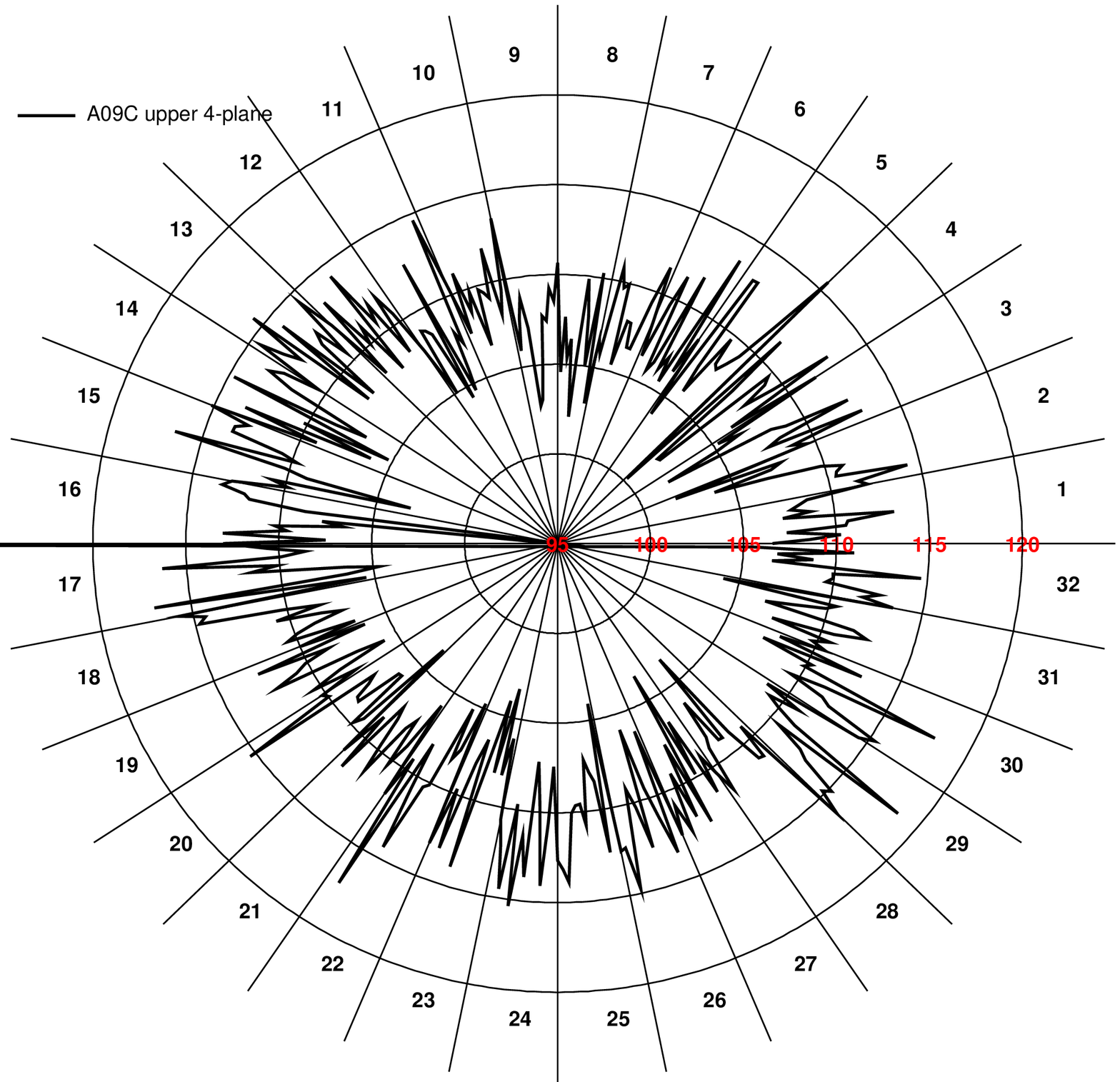}&
\includegraphics[width=9.0cm,angle=0]{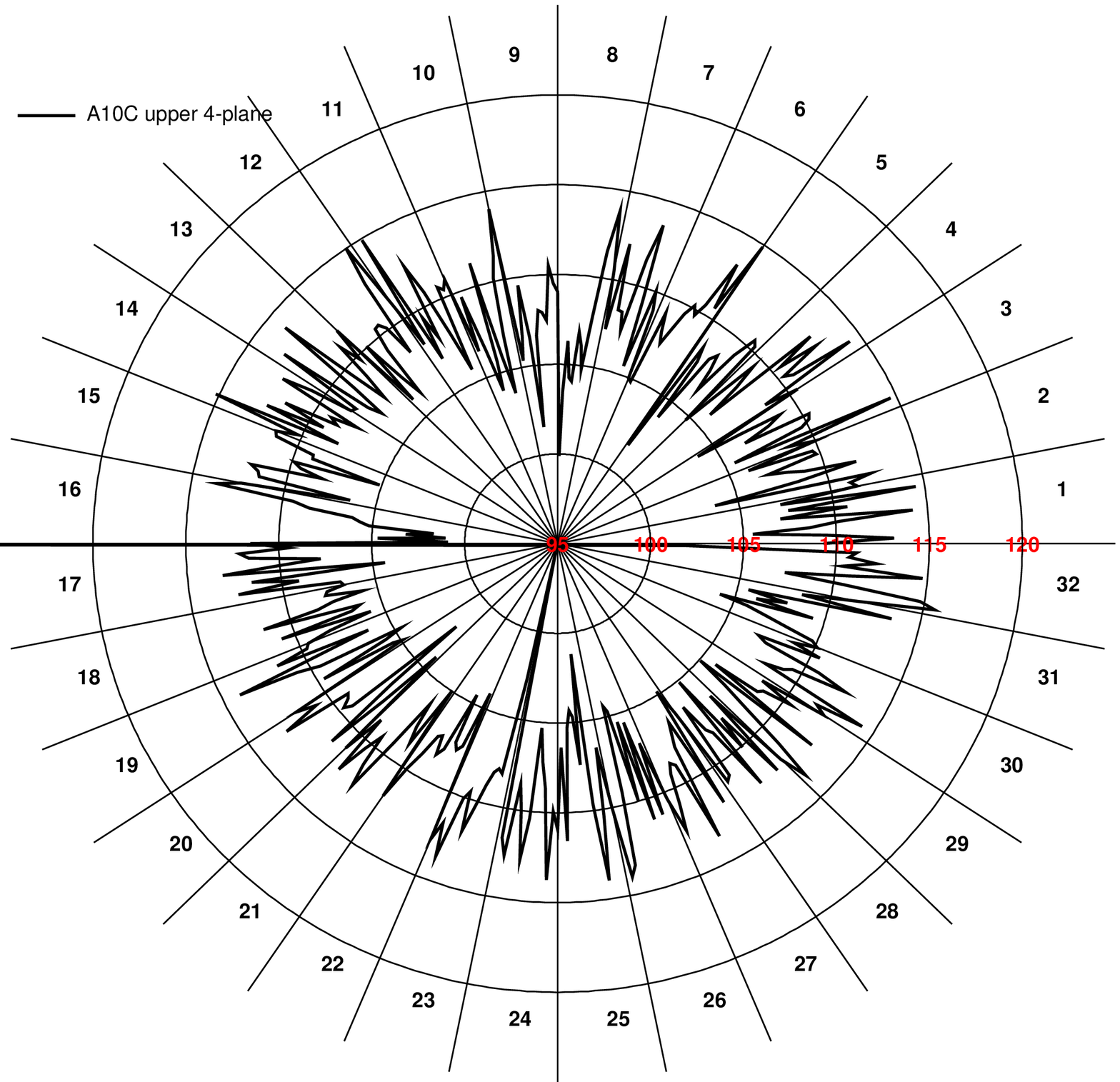}\\

\includegraphics[width=9.0cm,angle=0]{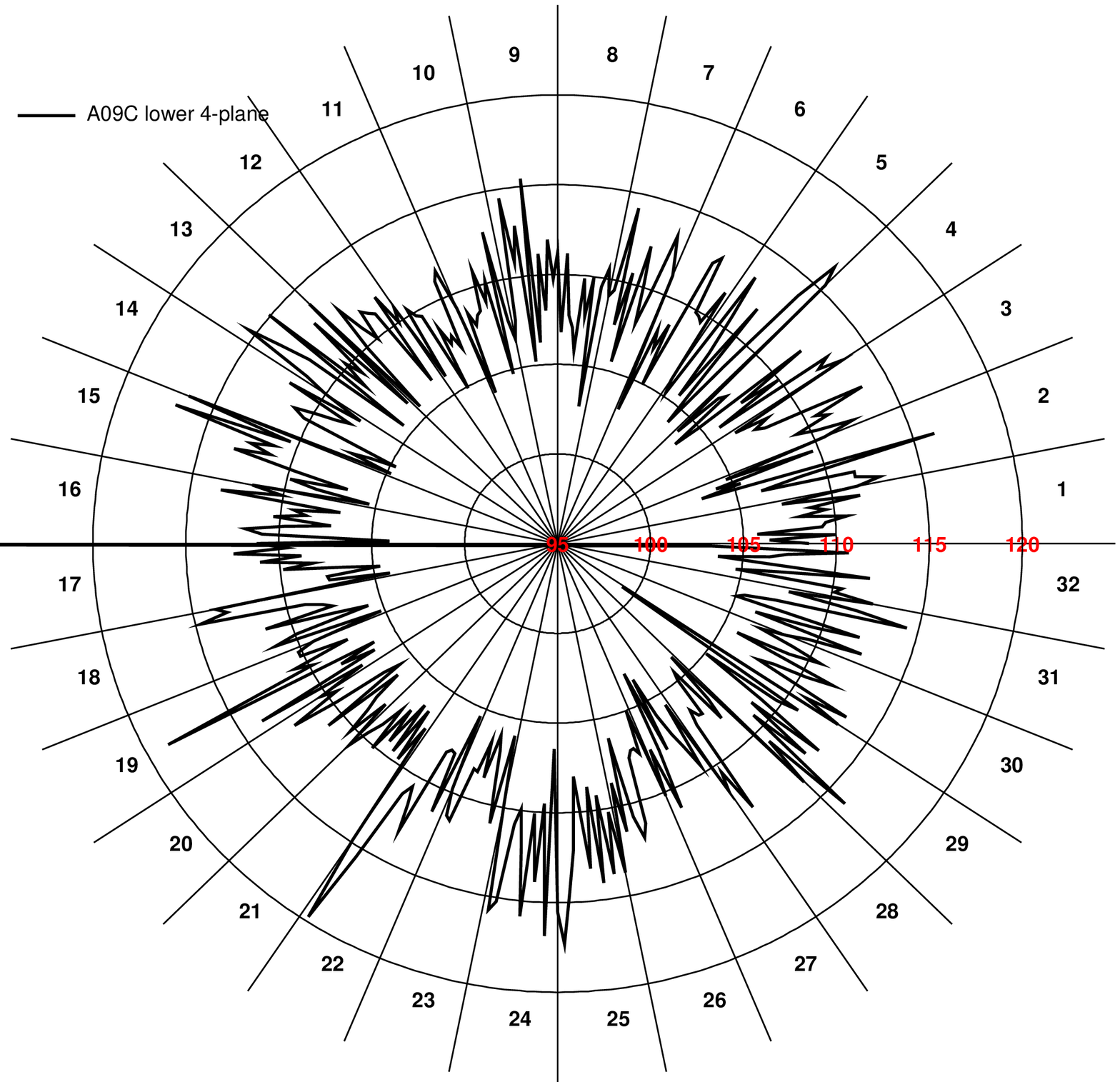}&
\includegraphics[width=9.0cm,angle=0]{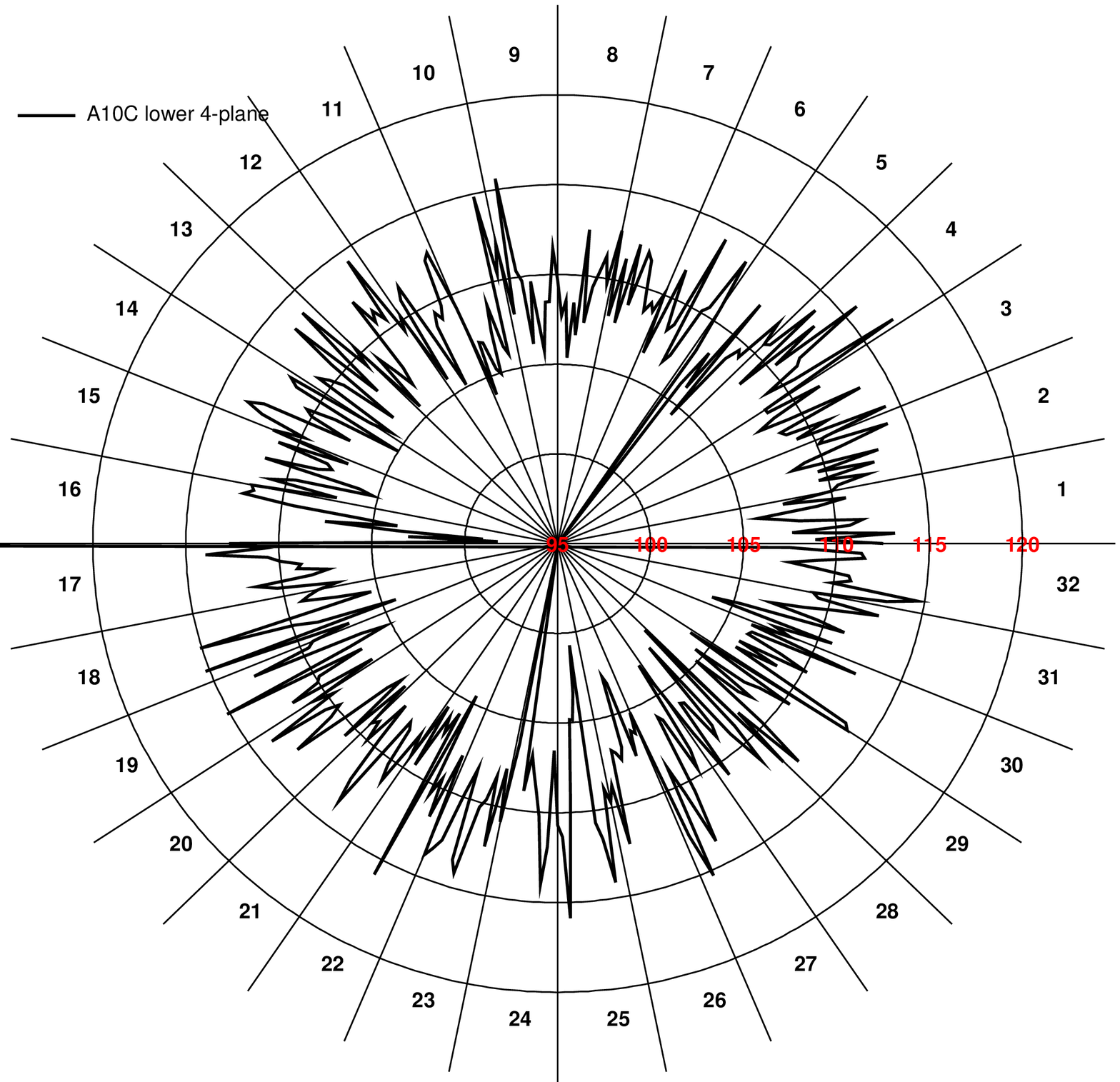}\\
\end{tabular}
\caption{Average per ASDBLR 300 kHz DAC low-threshold for the two upper and lower 4-planes wheels A09C and A10C. }
\label{fig:300kHzA9A10}
\end{center}
\end{figure}

\begin{figure}[ht]
\begin{center}
\begin{tabular}{cc}
\includegraphics[width=9.0cm,angle=0]{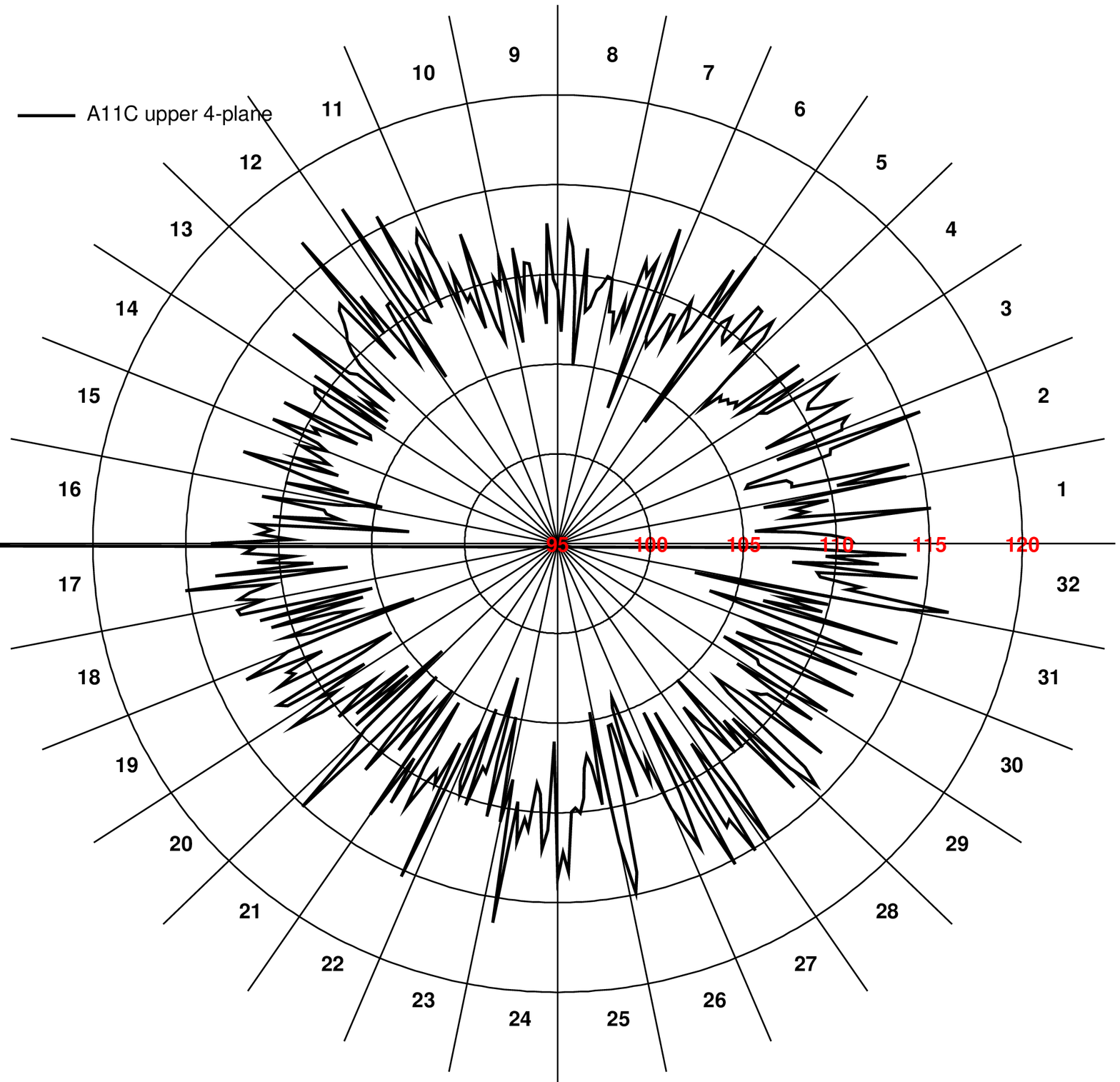}&
\includegraphics[width=9.0cm,angle=0]{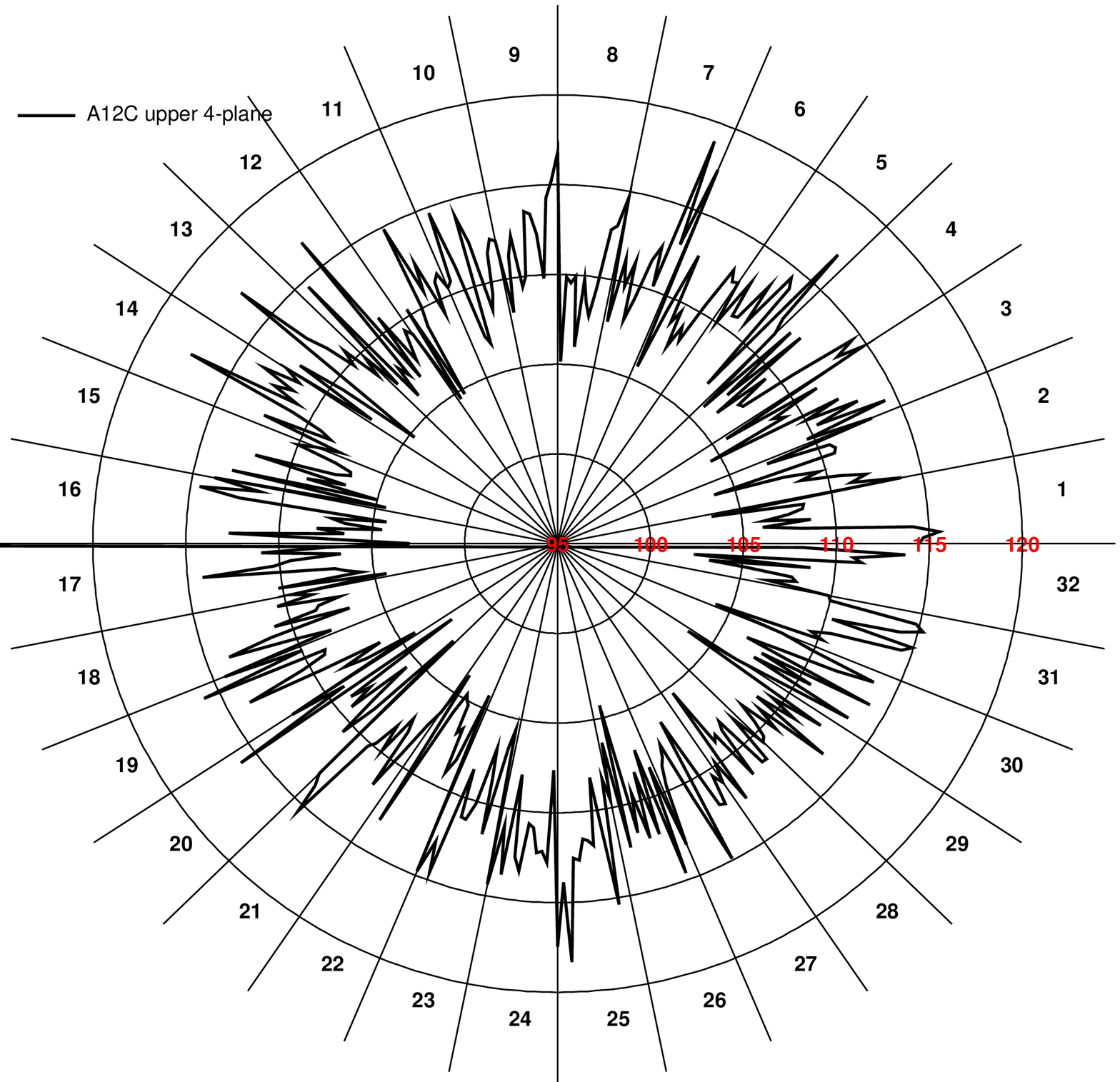}\\

\includegraphics[width=9.0cm,angle=0]{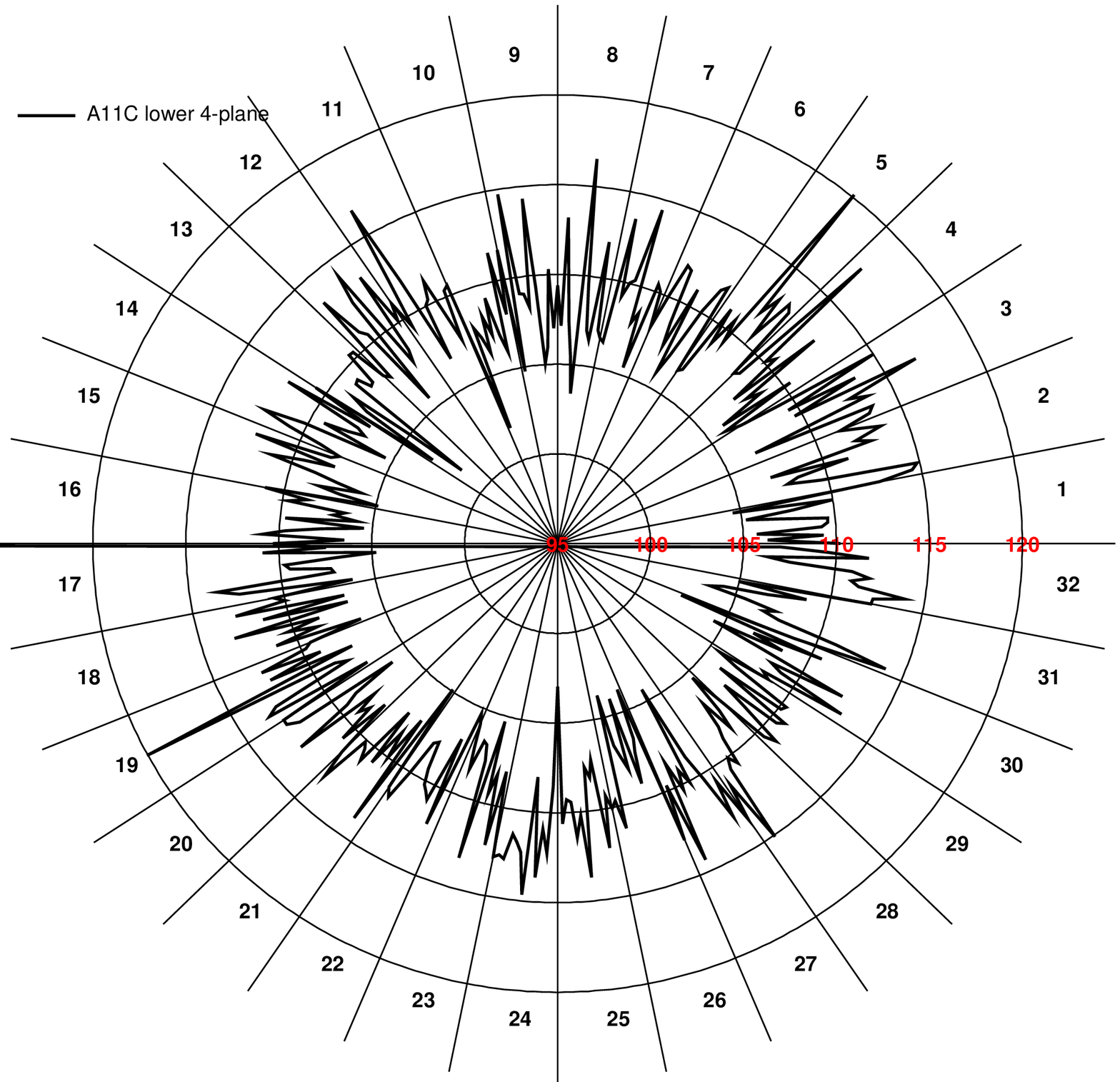}&
\includegraphics[width=9.0cm,angle=0]{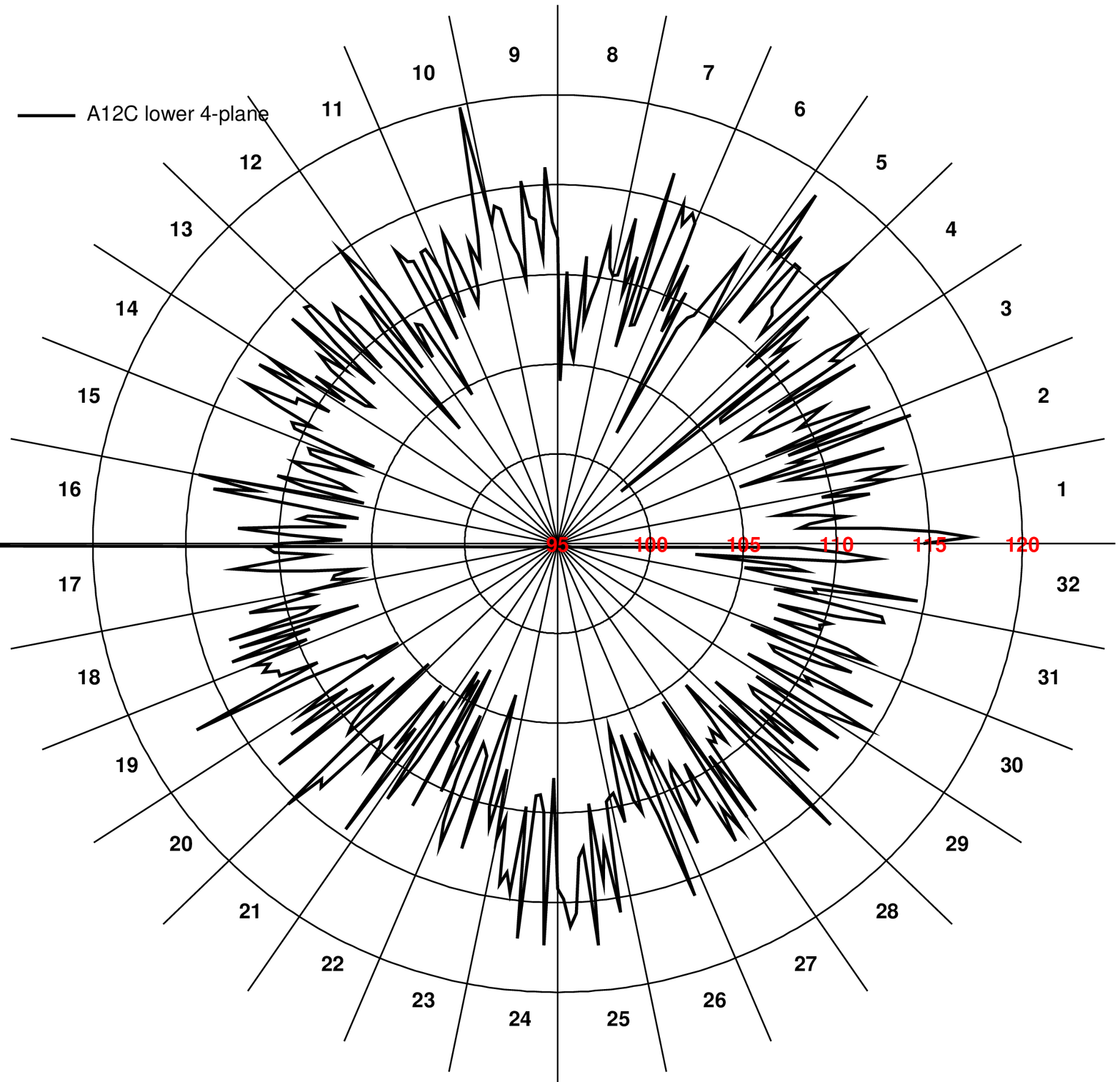}\\
\end{tabular}
\caption{Average per ASDBLR 300 kHz DAC low-threshold for the two upper and lower 4-planes wheels A11C and A12C. }
\label{fig:300kHzA11A12}
\end{center}
\end{figure}

\begin{figure}[ht]
\begin{center}
\begin{tabular}{cc}
\includegraphics[width=9.0cm,angle=0]{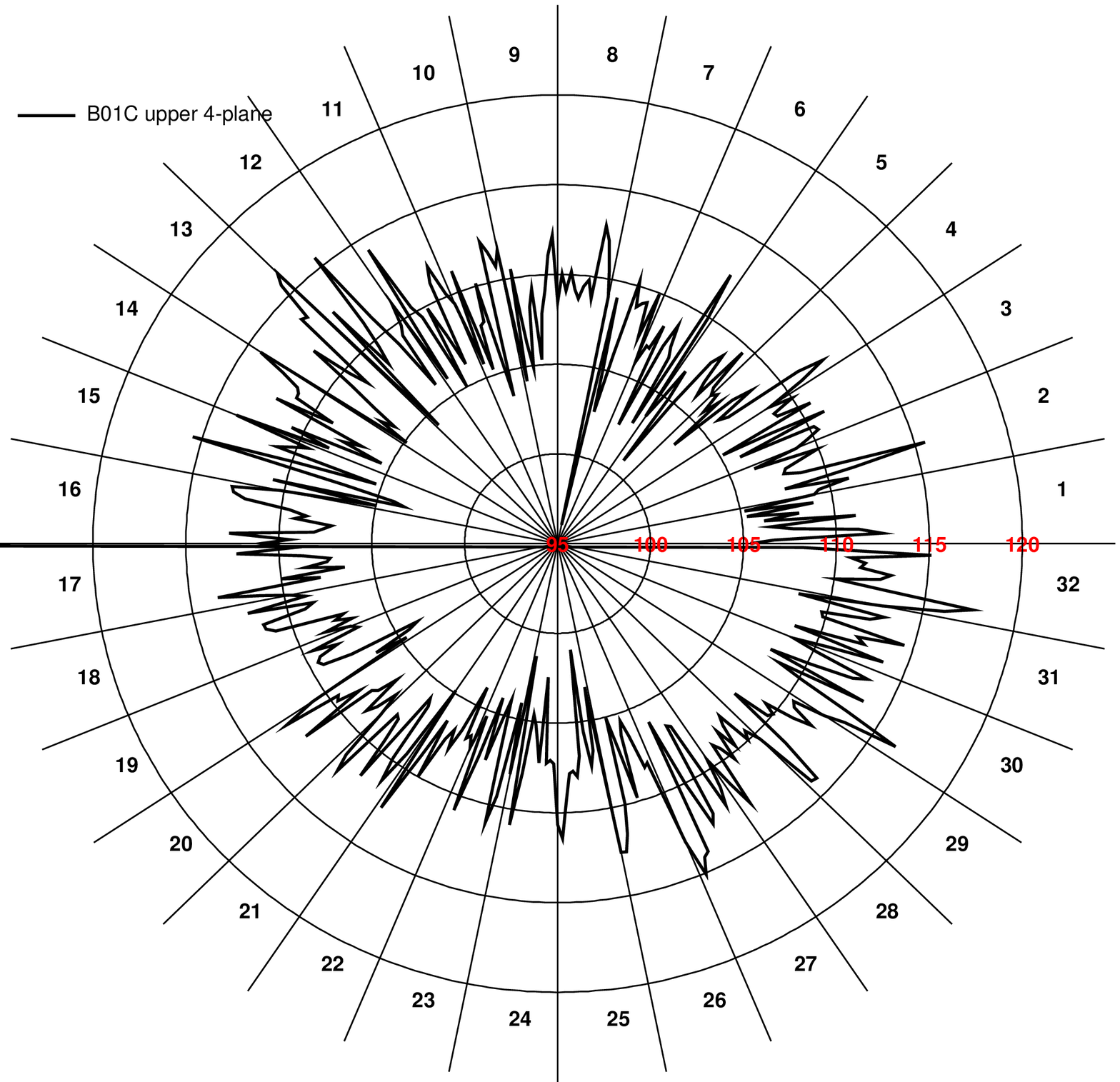}&
\includegraphics[width=9.0cm,angle=0]{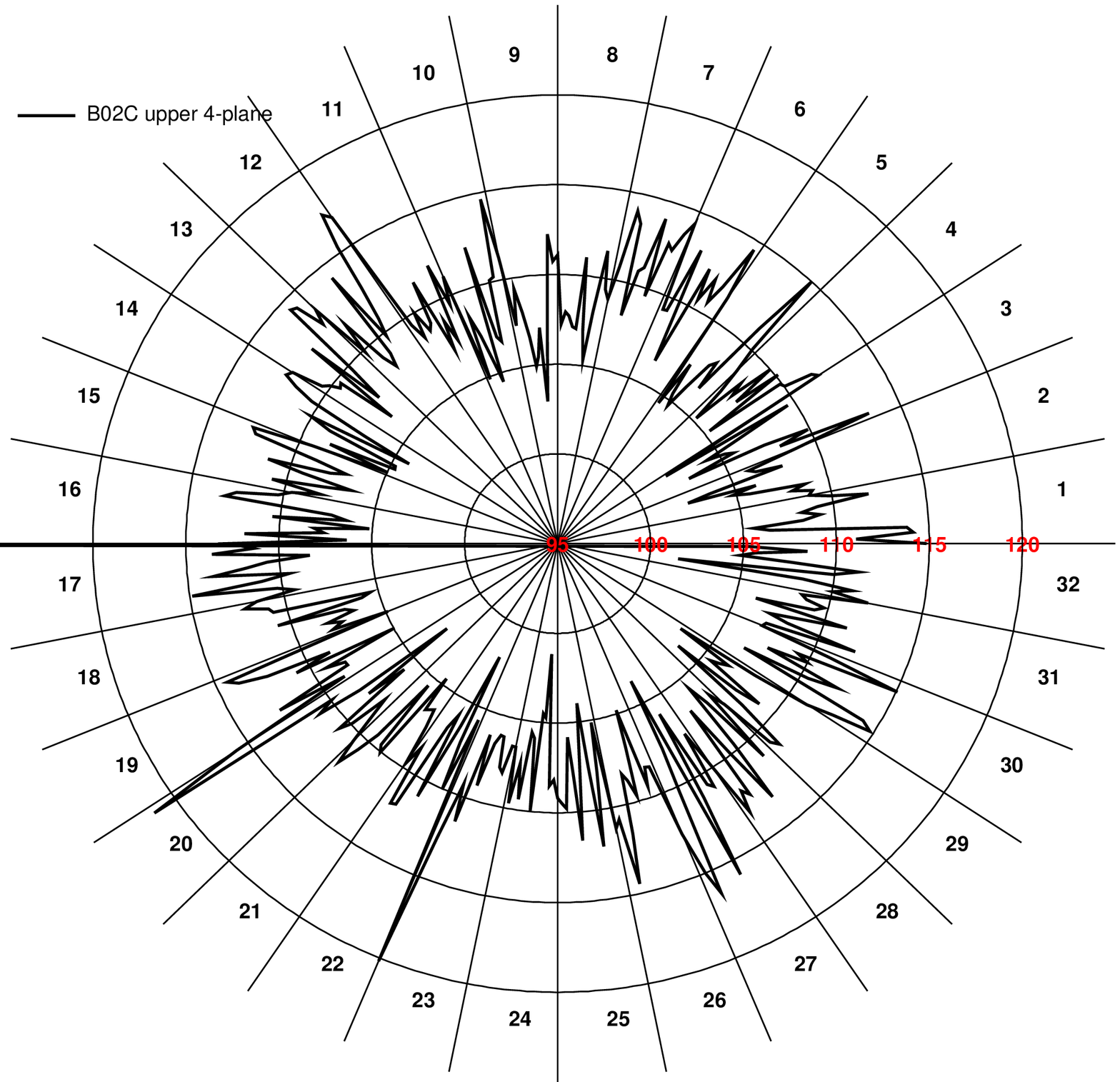}\\

\includegraphics[width=9.0cm,angle=0]{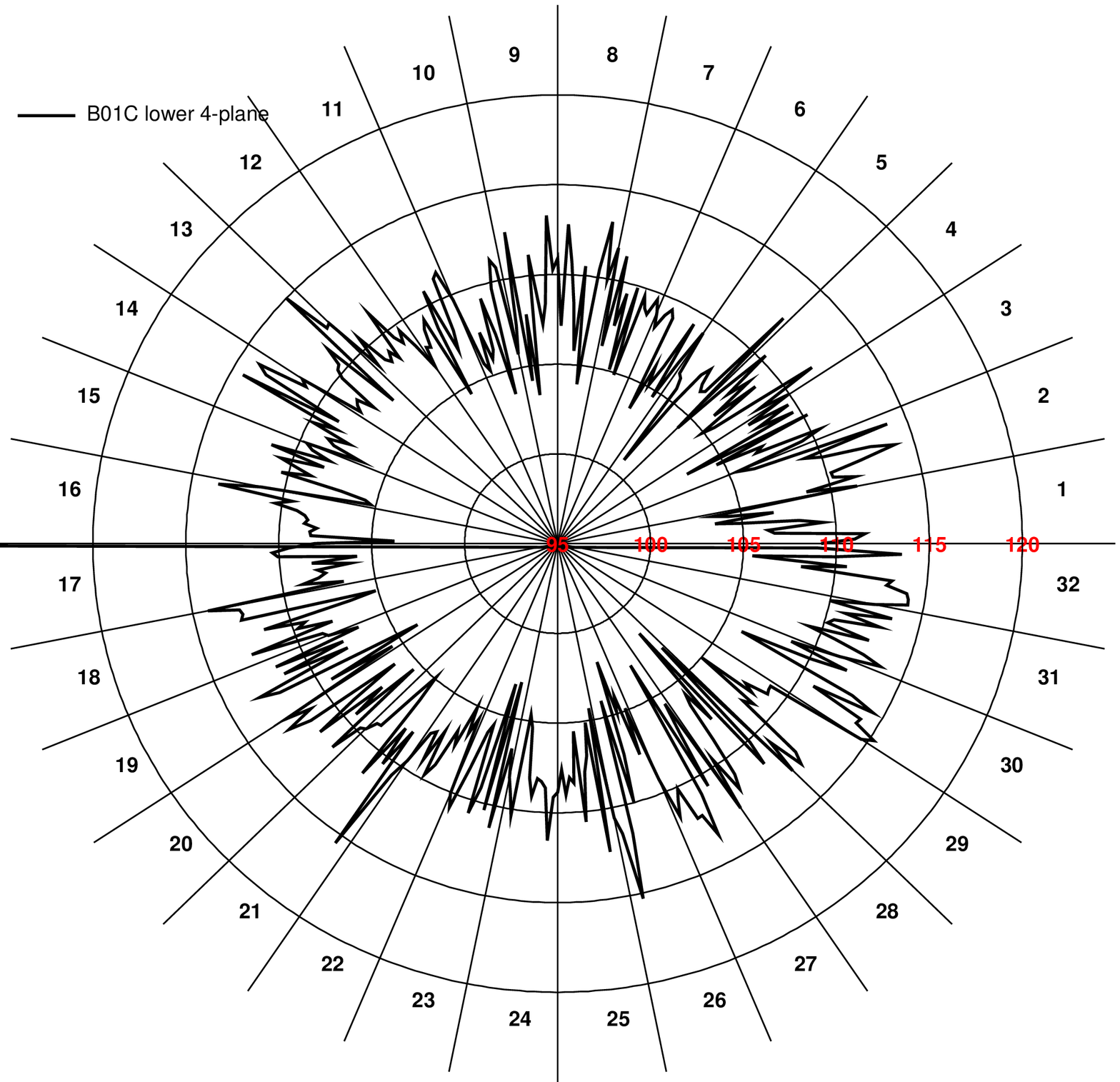}&
\includegraphics[width=9.0cm,angle=0]{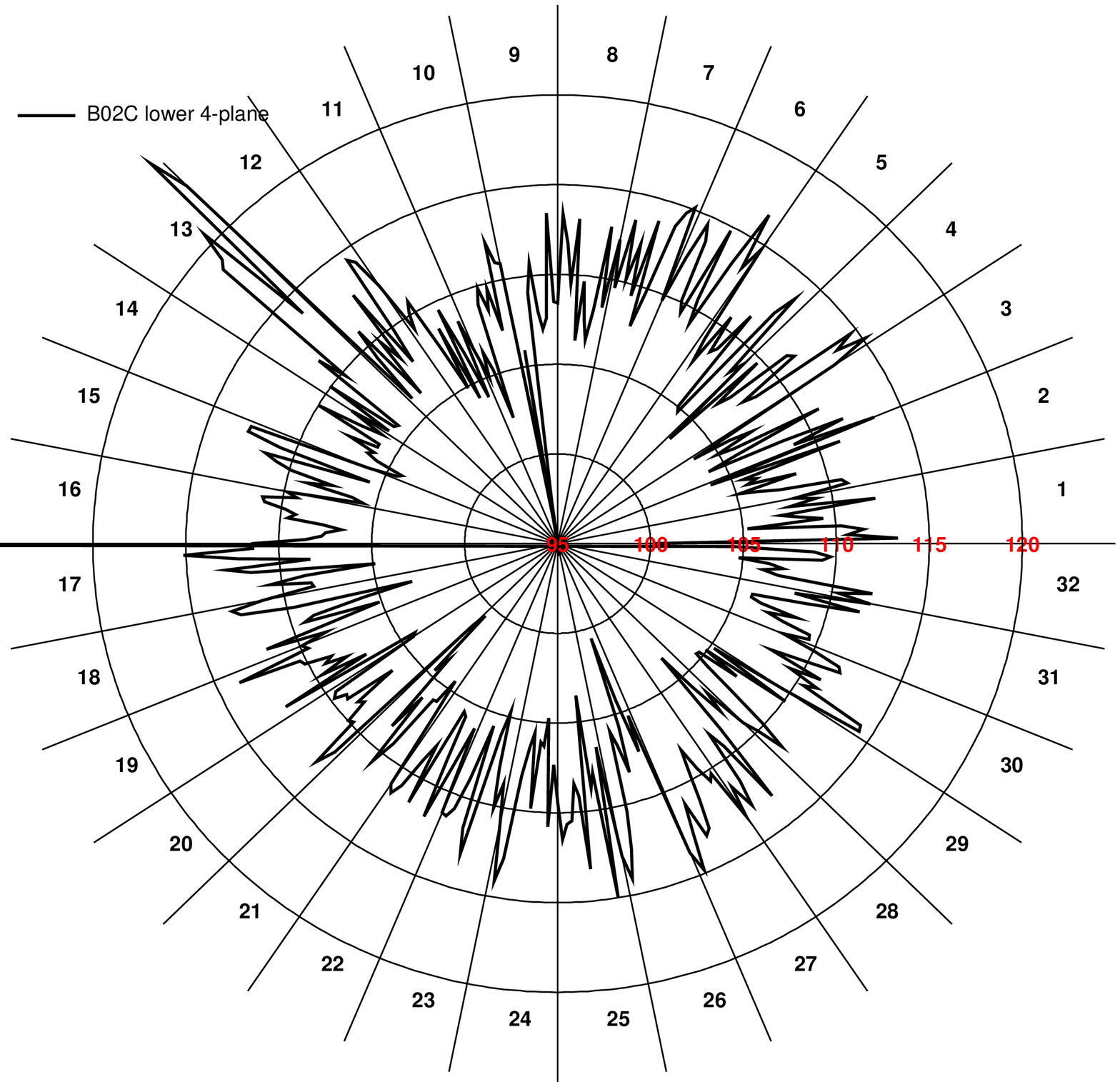}\\
\end{tabular}
\caption{Average per ASDBLR 300 kHz DAC low-threshold for the two upper and lower 4-planes wheels B01C and B02C. }
\label{fig:300kHzB1B2}
\end{center}
\end{figure}

\begin{figure}[ht]
\begin{center}
\begin{tabular}{cc}
\includegraphics[width=9.0cm,angle=0]{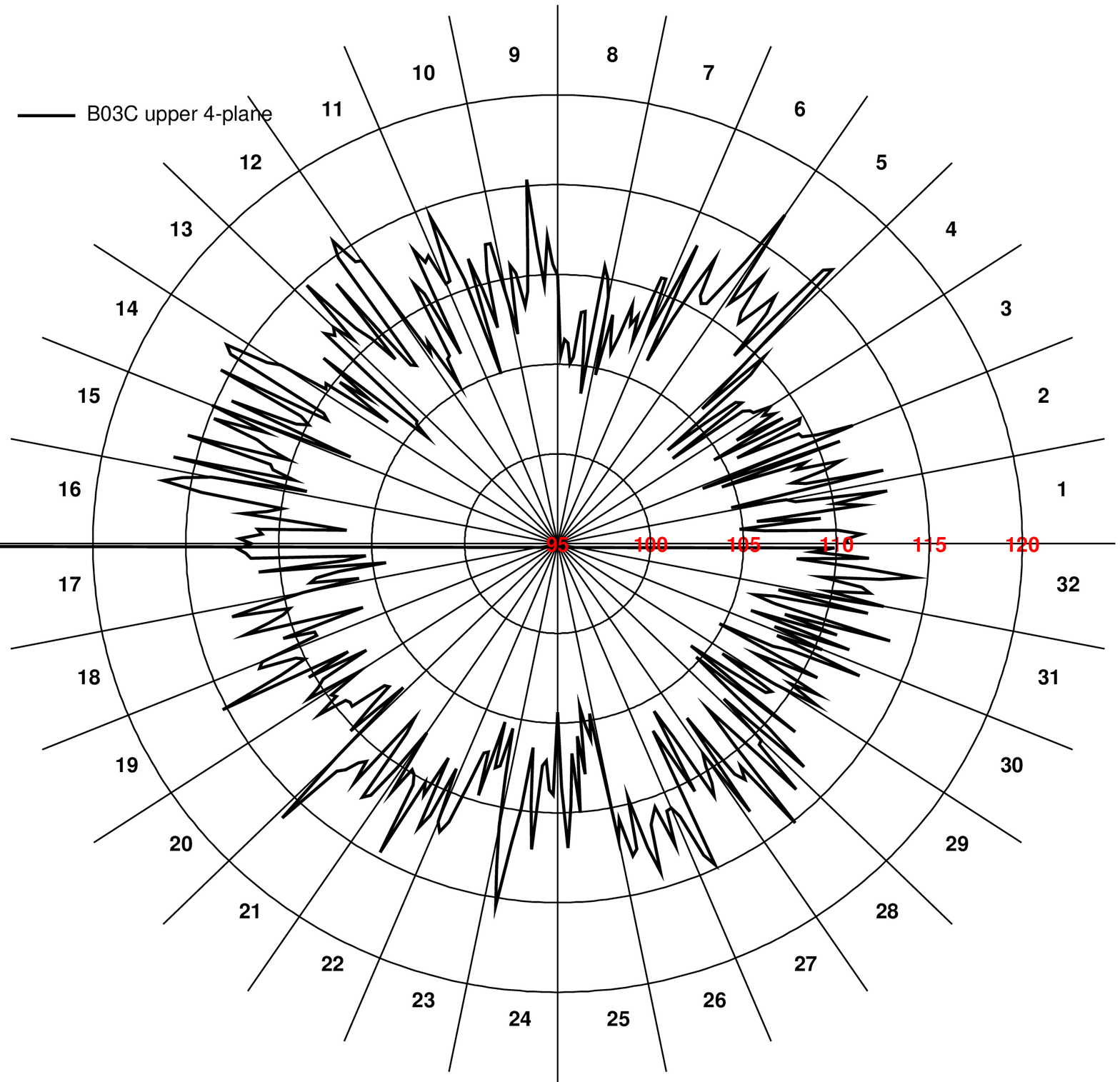}&
\includegraphics[width=9.0cm,angle=0]{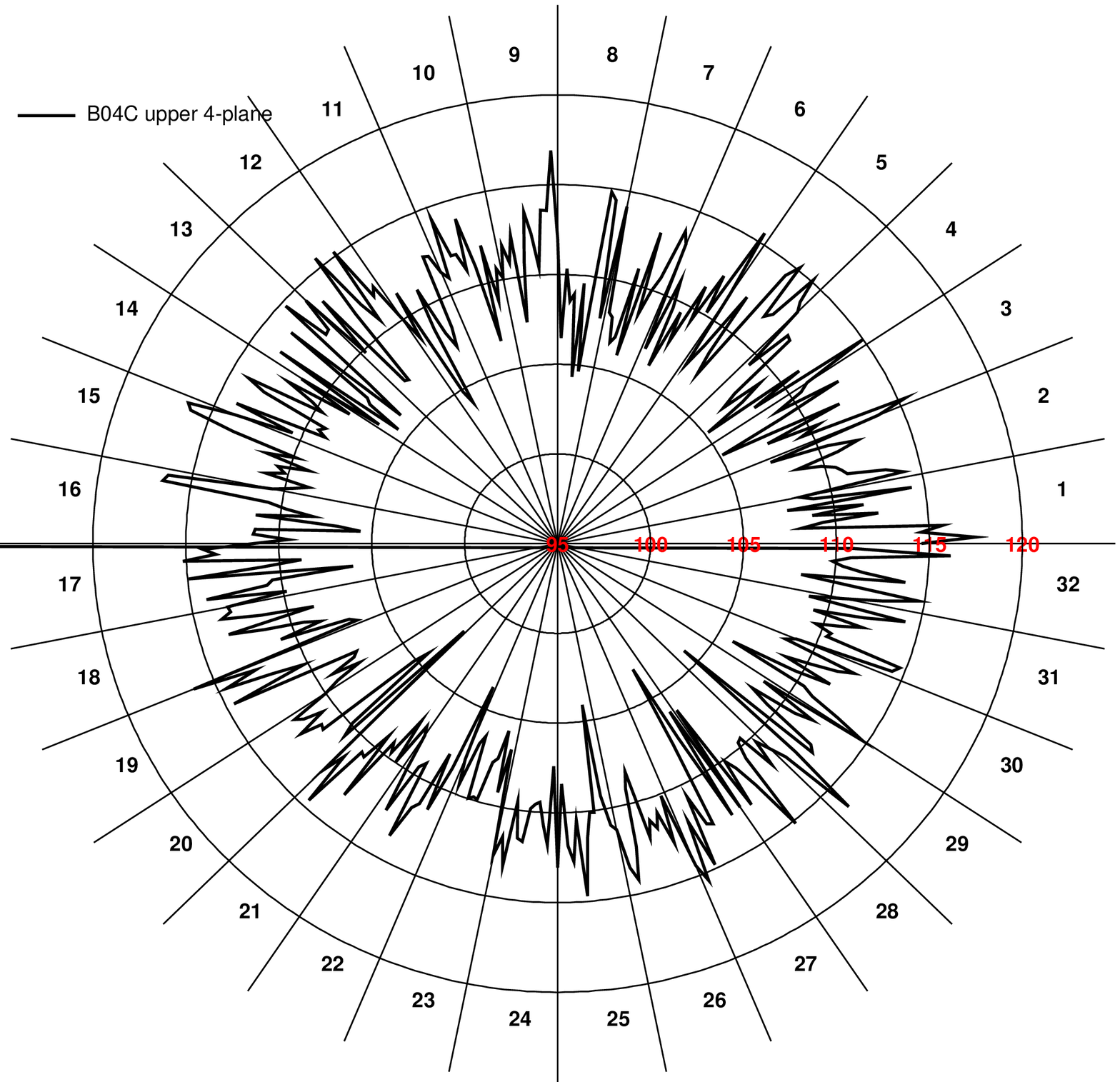}\\

\includegraphics[width=9.0cm,angle=0]{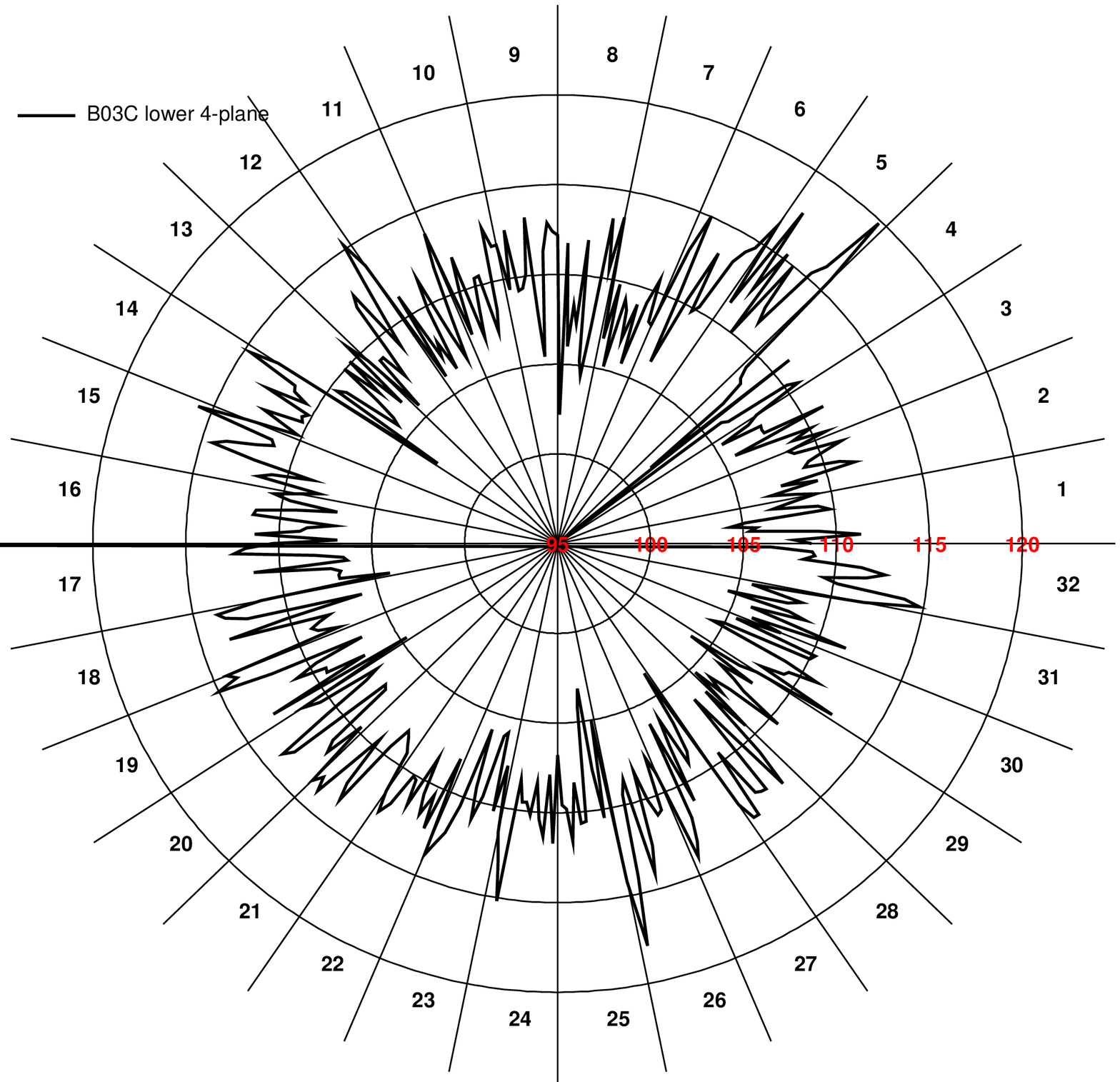}&
\includegraphics[width=9.0cm,angle=0]{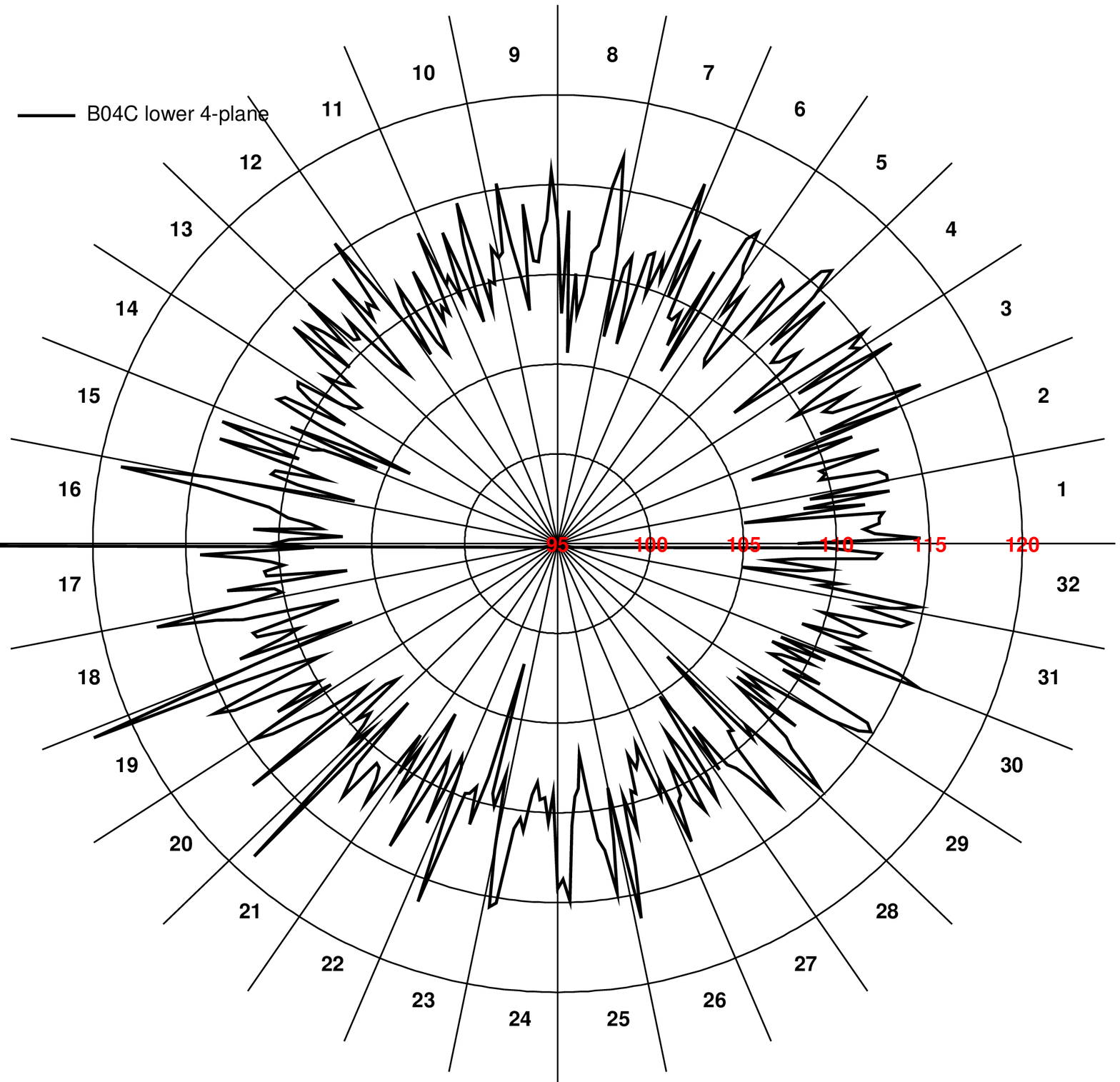}\\
\end{tabular}
\caption{Average per ASDBLR 300 kHz DAC low-threshold for the two upper and lower 4-planes wheels B03C and B04C. }
\label{fig:300kHzB3B4}
\end{center}
\end{figure}

\begin{figure}[ht]
\begin{center}
\begin{tabular}{cc}
\includegraphics[width=9.0cm,angle=0]{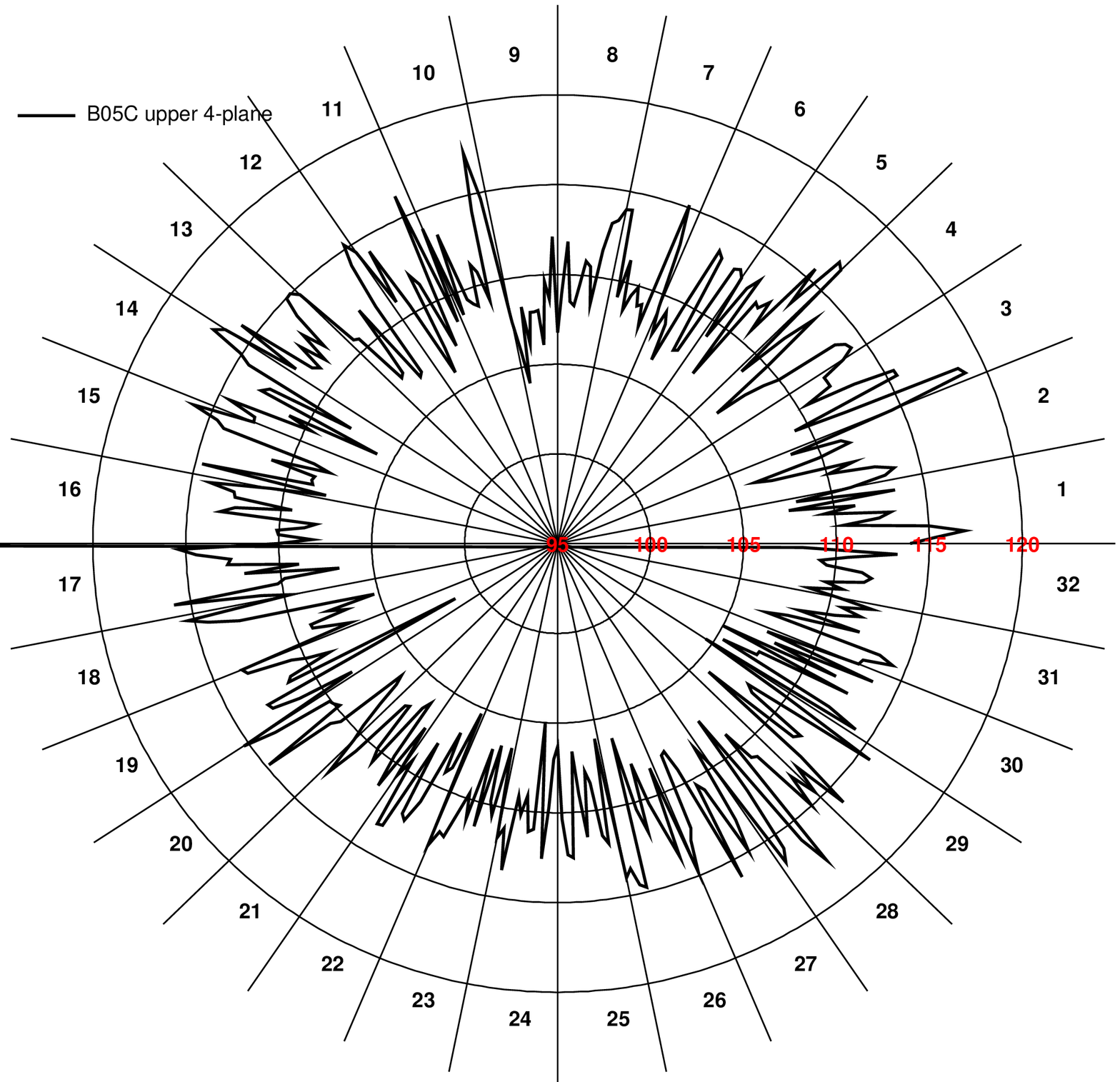}&
\includegraphics[width=9.0cm,angle=0]{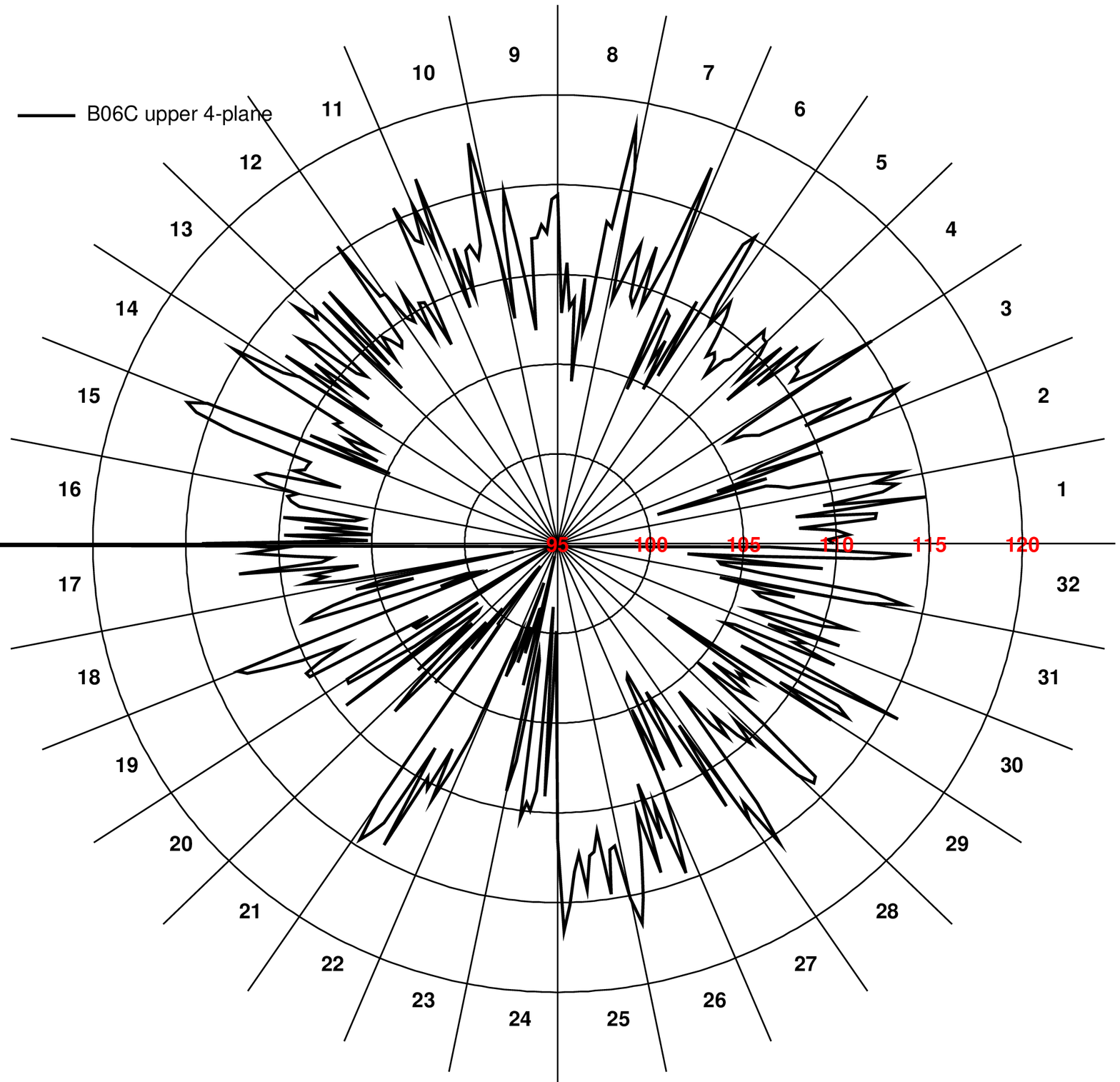}\\

\includegraphics[width=9.0cm,angle=0]{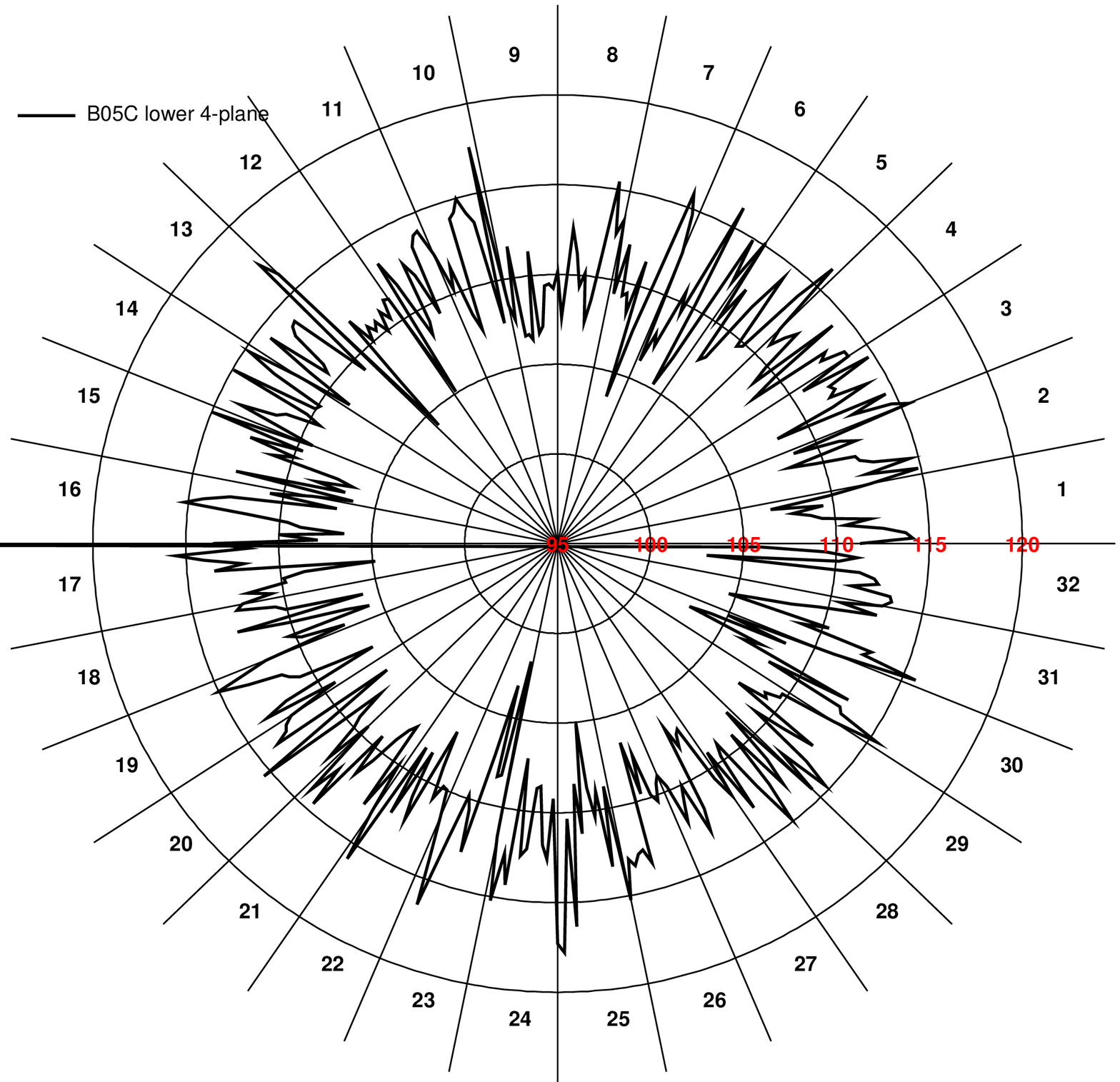}&
\includegraphics[width=9.0cm,angle=0]{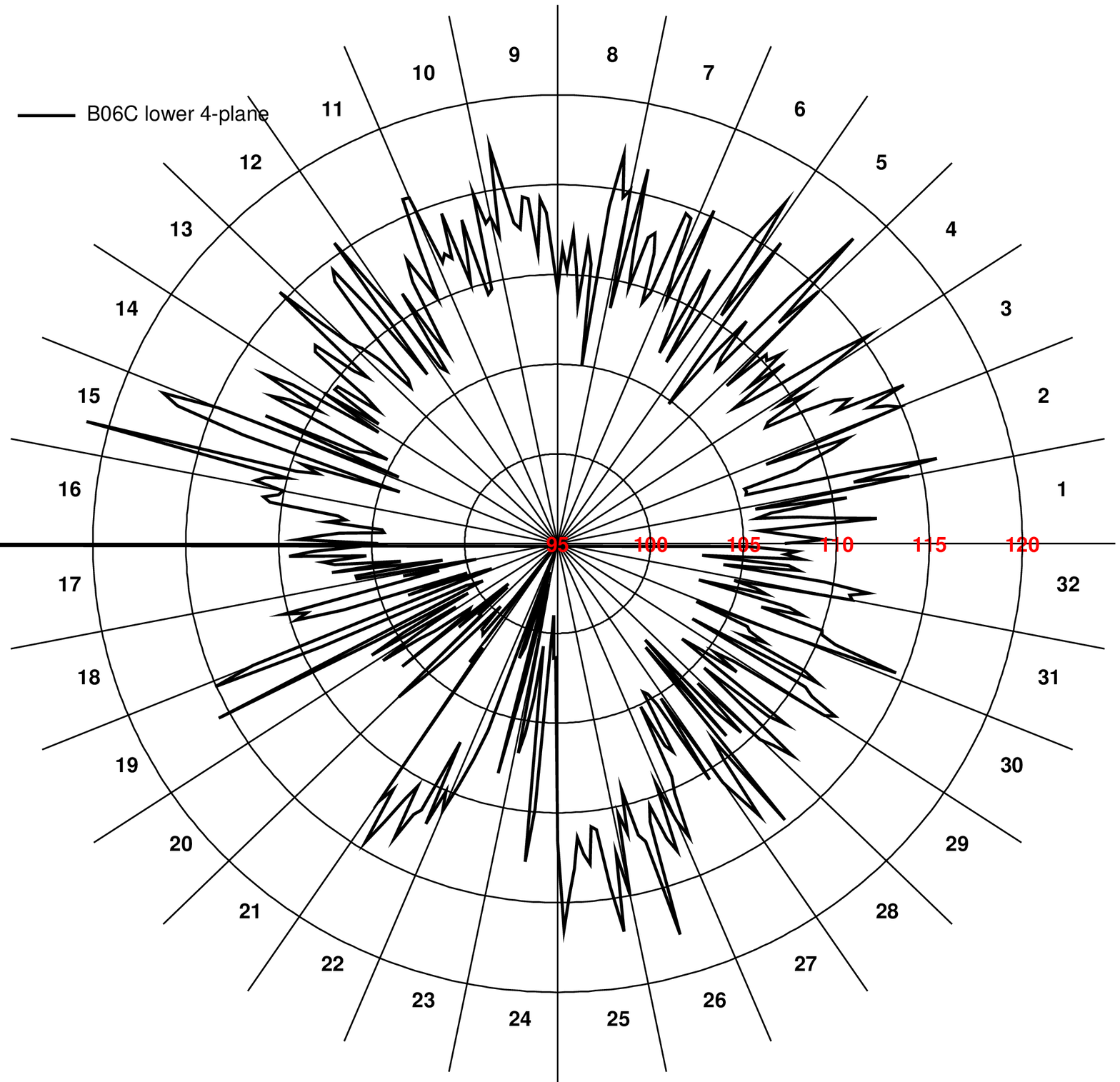}\\
\end{tabular}
\caption{Average per ASDBLR 300 kHz DAC low-threshold for the two upper and lower 4-planes wheels B05C and B06C. }
\label{fig:300kHzB5B6}
\end{center}
\end{figure}

\begin{figure}[ht]
\begin{center}
\begin{tabular}{cc}
\includegraphics[width=9.0cm,angle=0]{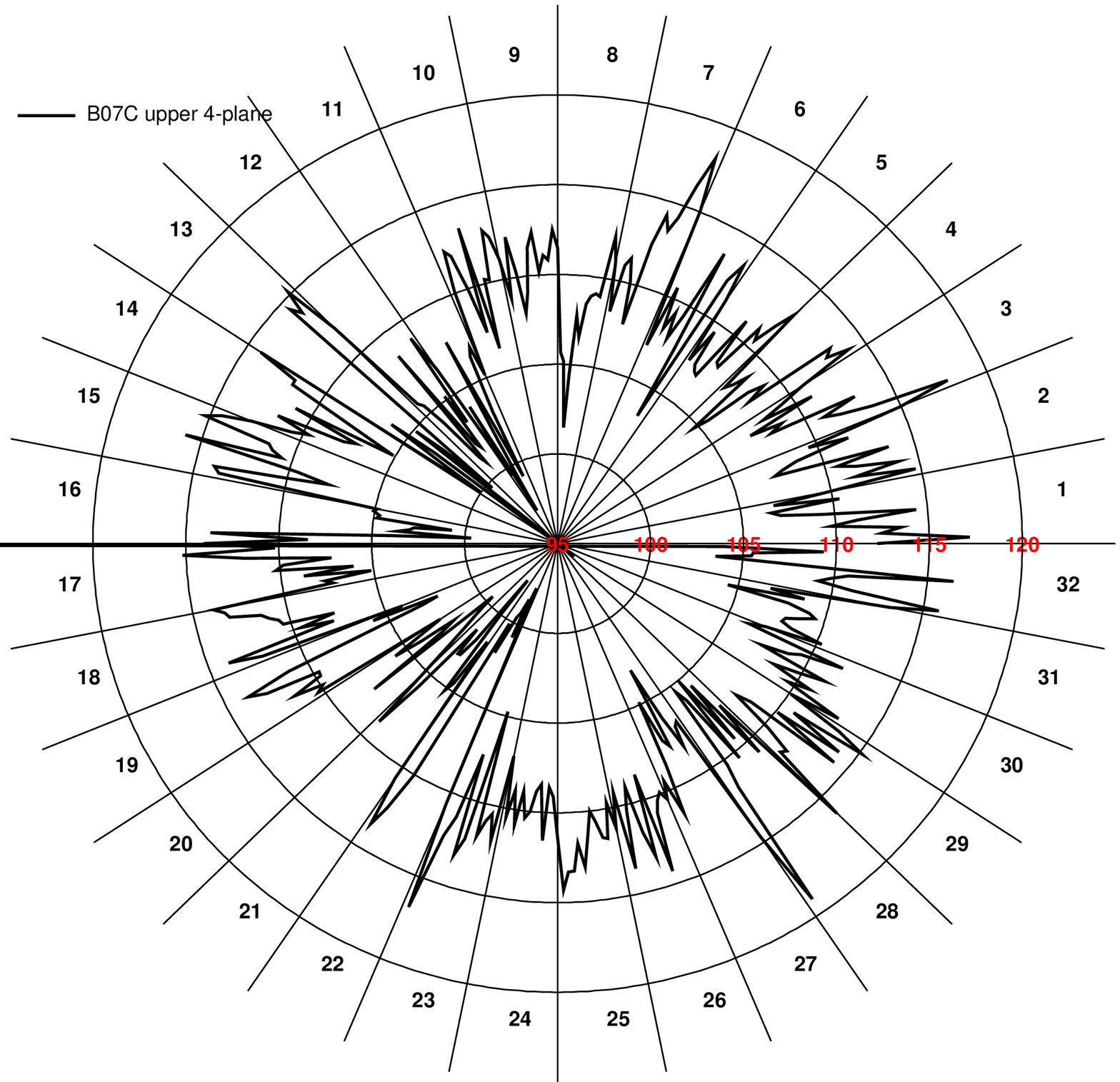}&
\includegraphics[width=9.0cm,angle=0]{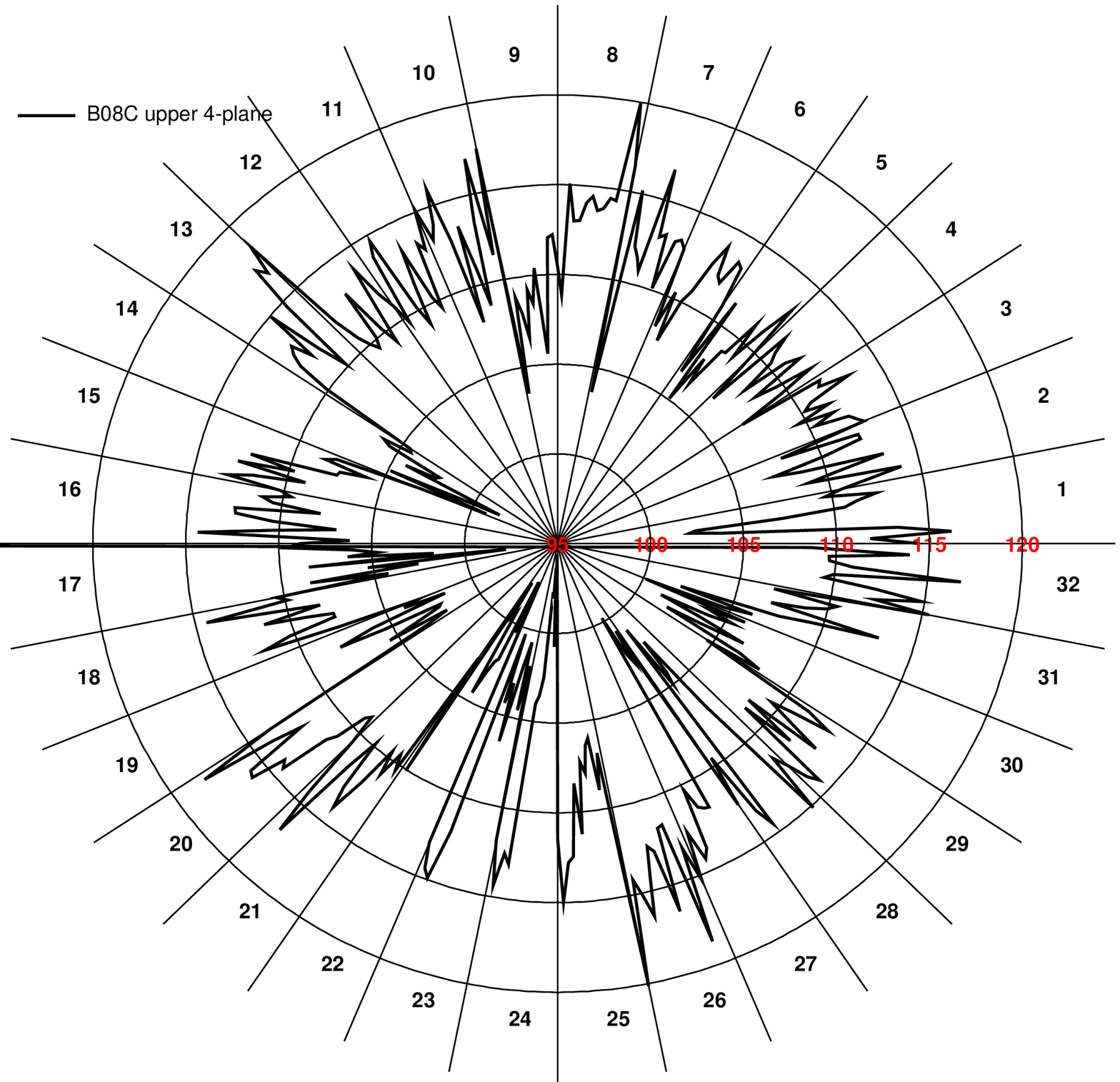}\\

\includegraphics[width=9.0cm,angle=0]{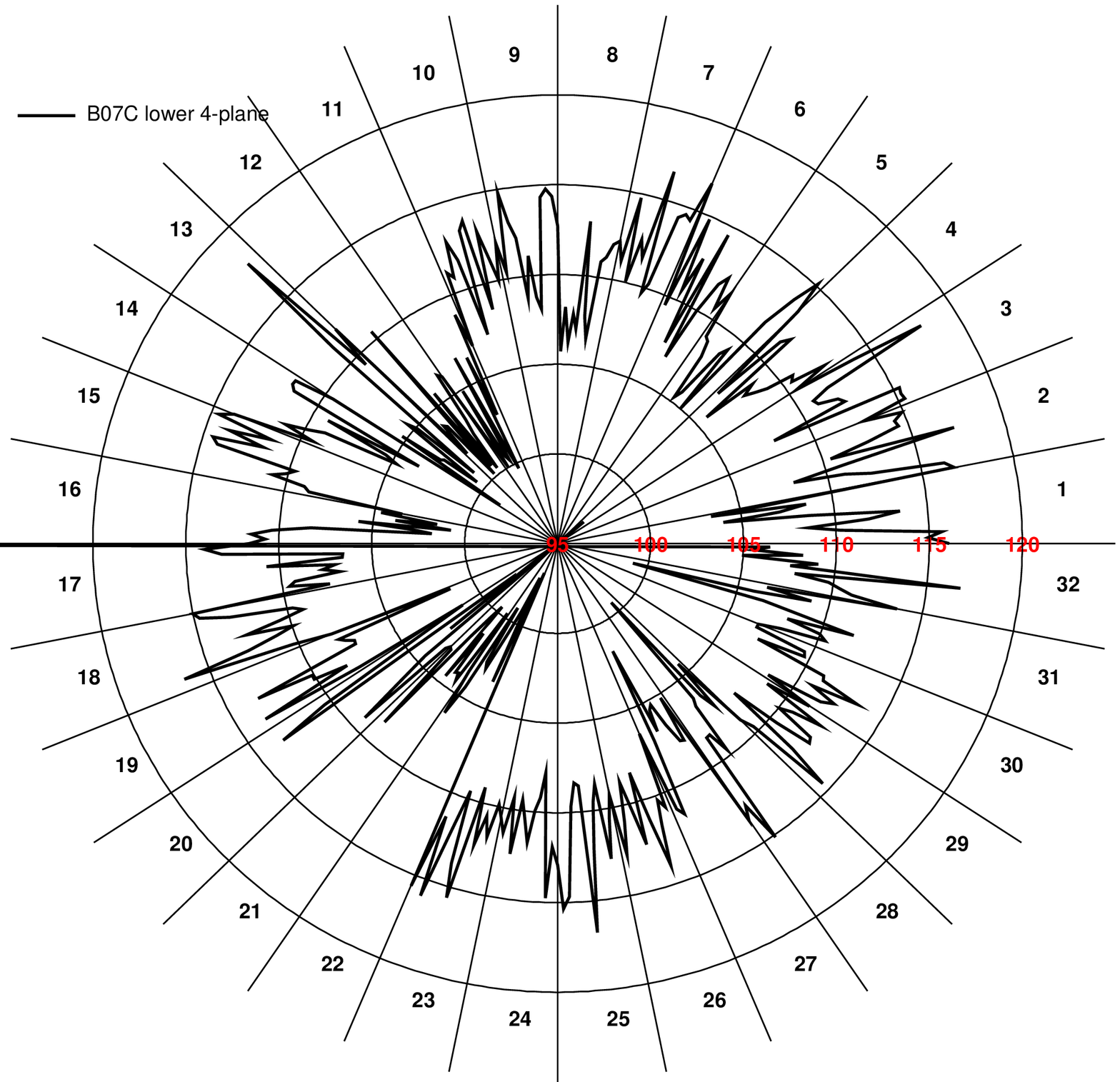}&
\includegraphics[width=9.0cm,angle=0]{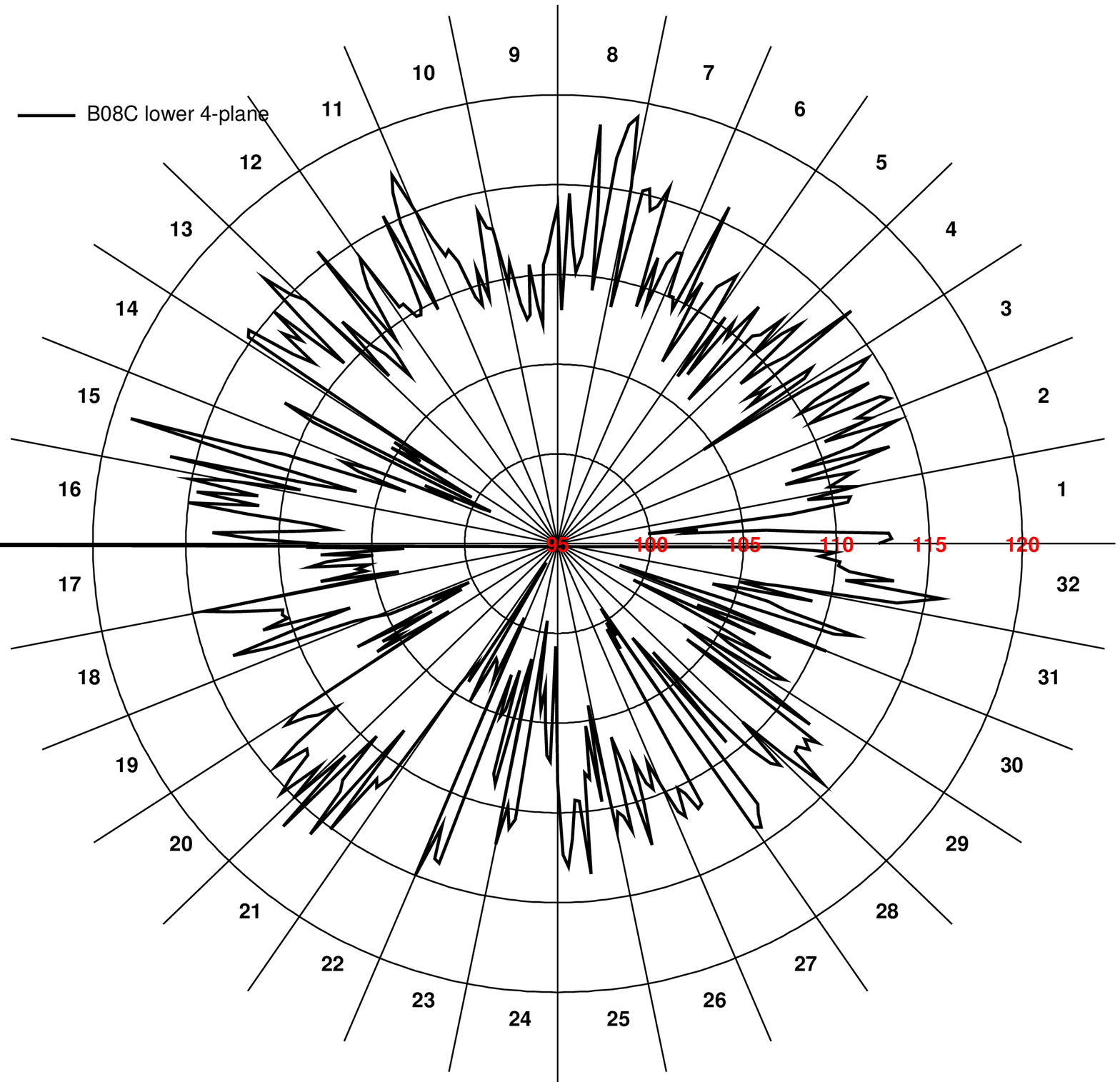}\\
\end{tabular}
\caption{Average per ASDBLR 300 kHz DAC low-threshold for the two upper and lower 4-planes wheels B07C and B08C. }
\label{fig:300kHzB7B8}
\end{center}
\end{figure}

A powerful technique to identify possible problems for the different readout channels of the ATLAS Transition Radiation Tracker has been described. 
Applied to the case of one TRT endcap-detector, it shows that 99.6 \% of the readout channels are perfectly operational. The remaining 0.4 \% are essentially due to metallisation issues of the active WEBs. Moreover, several goals concerning electronics performance and not discussed in this report, have been achieved, like for instance the $\simeq$ 2\% occupancy corresponding to a low-level threshold at about 110-120 DAC counts (see Fig. \ref{fig:300kHzA1A2} to Fig. \ref{fig:300kHzB7B8}).

\section*{Acknowledgements}
We are grateful to D.Froidevaux for a careful reading of the manuscript and for providing fruitful comments on the results.

%


%
\newpage
\begin{landscape}
\begin{figure}[ht]
\begin{center}
\begin{tabular}{ccc}
\epsfig{file=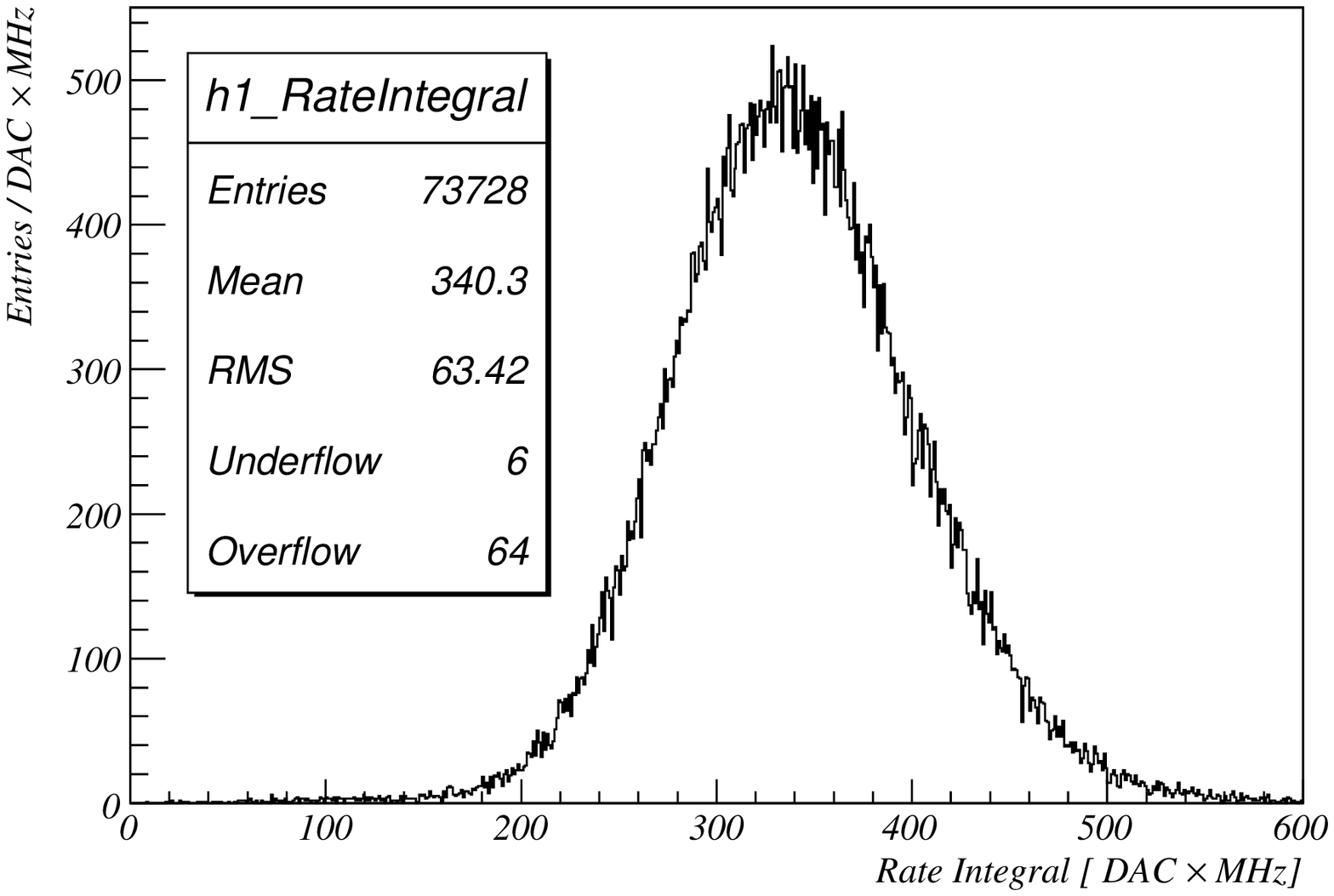,width=0.45\textwidth}  &
\epsfig{file=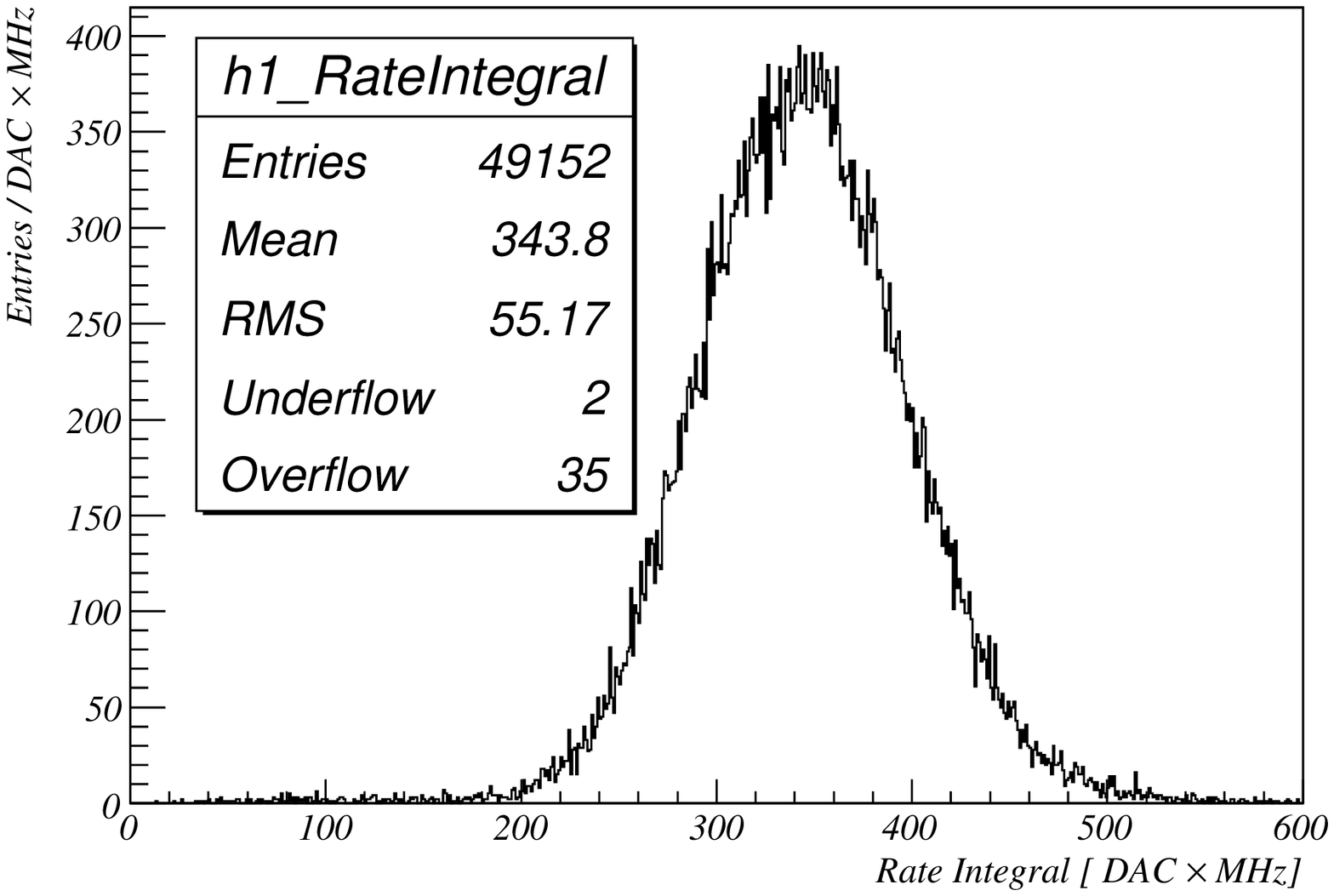,width=0.45\textwidth}  &
\epsfig{file=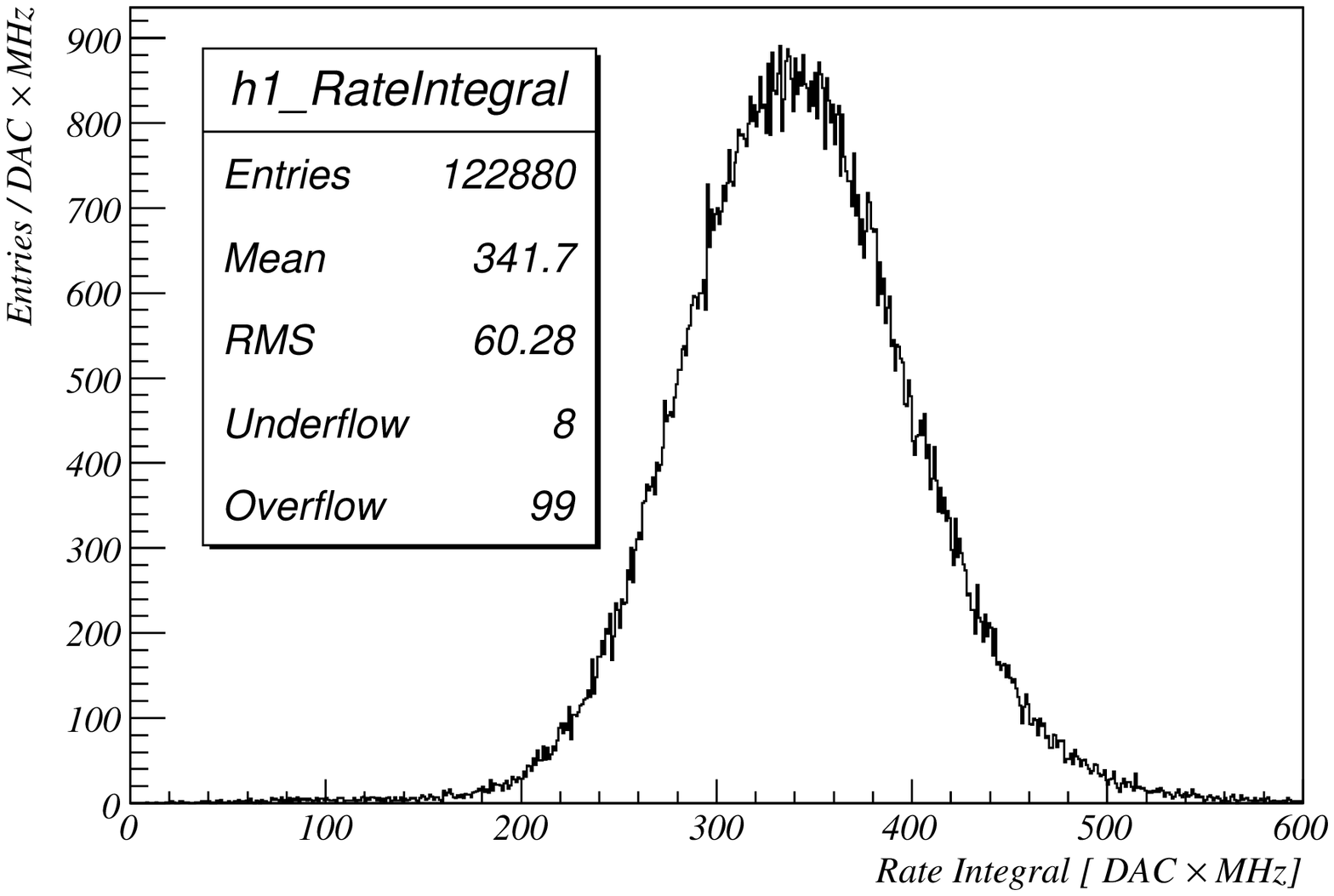,width=0.45\textwidth} \\

\epsfig{file=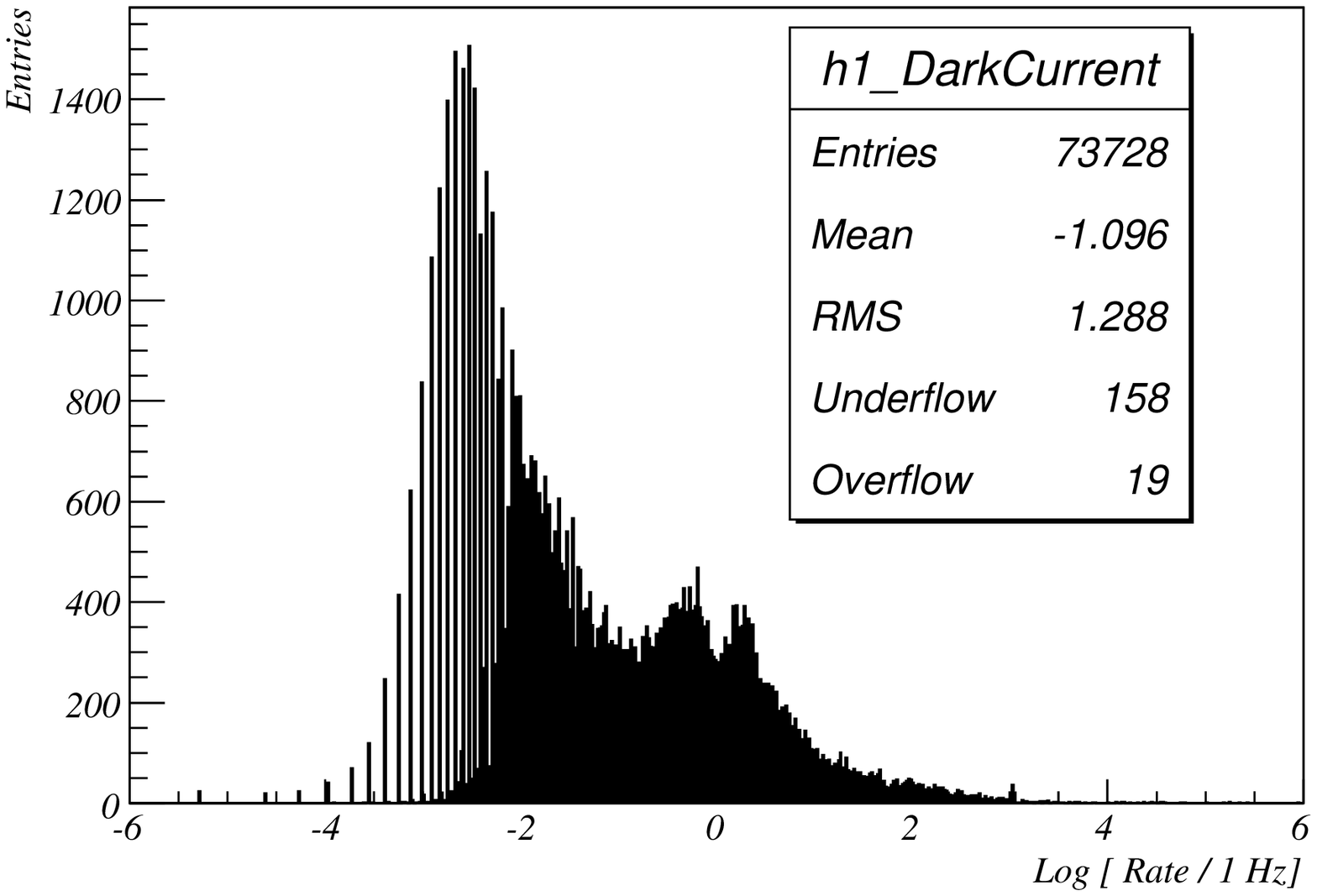,width=0.45\textwidth}  &
\epsfig{file=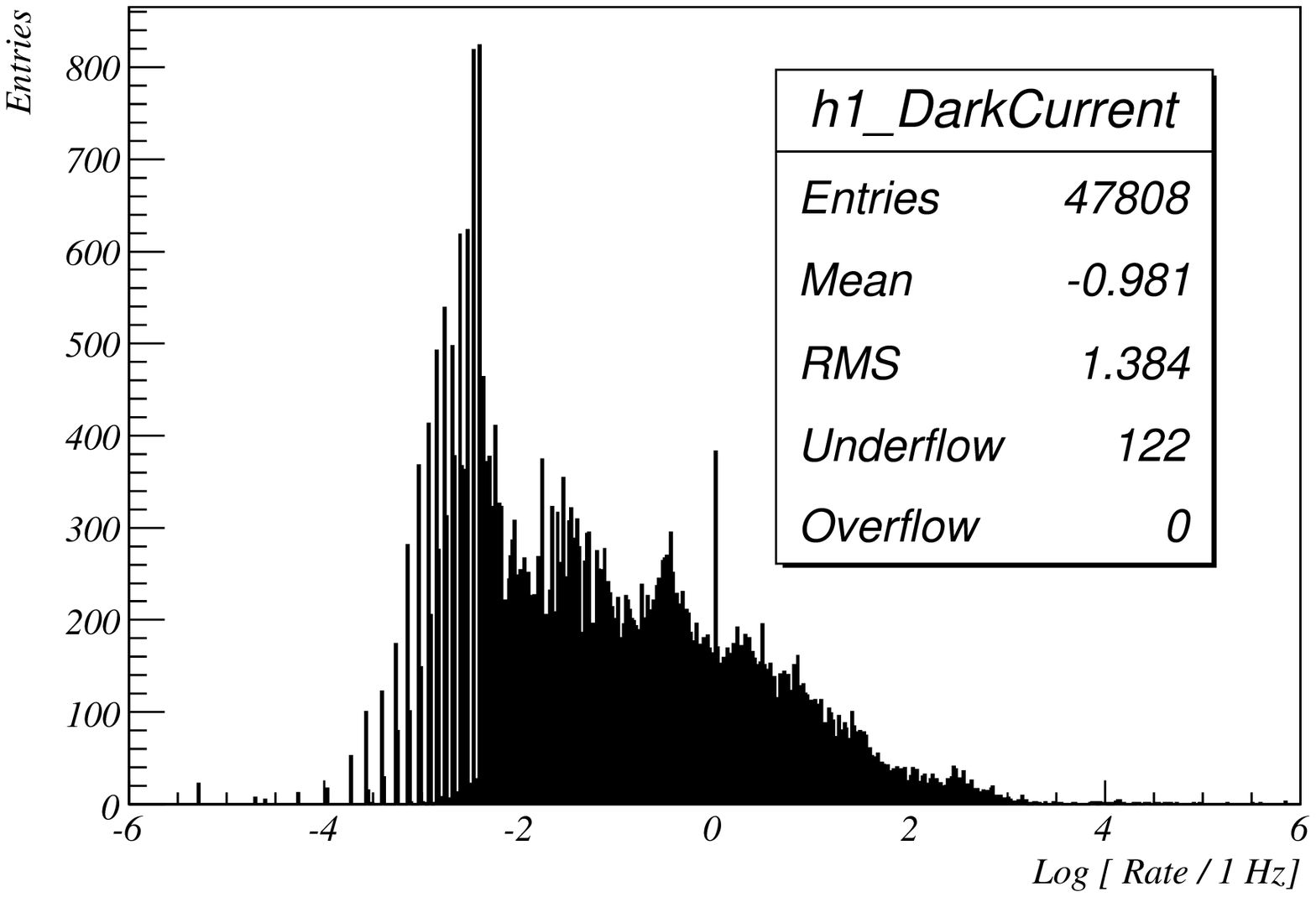,width=0.45\textwidth}  &
\epsfig{file=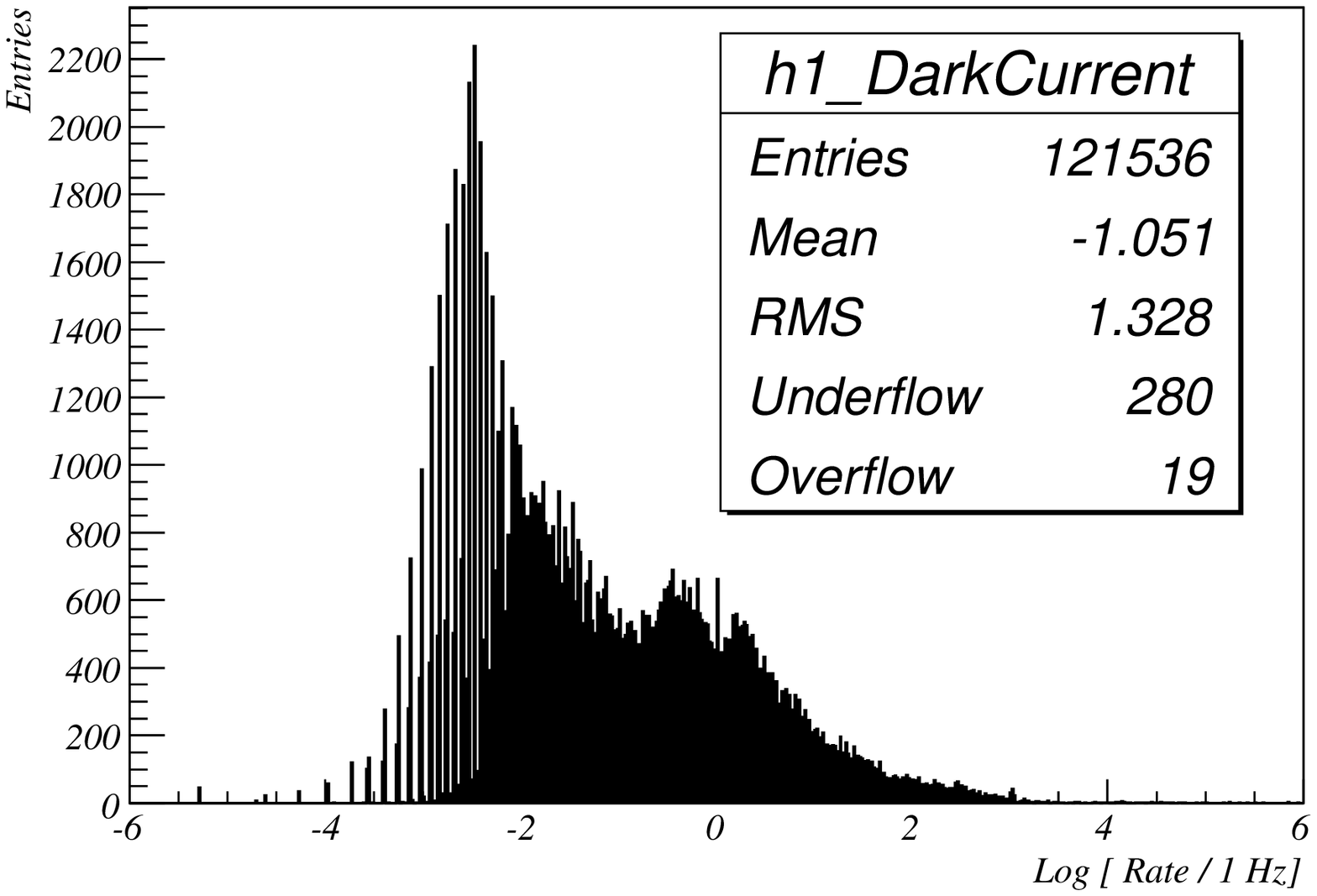,width=0.45\textwidth} \\
A-type &
B-type &
All \\
\end{tabular}
\caption{Noise scan rate integral distribution (upper) and associated accumulation mode dark current rate for both A-Type and B-Type TRT endcap-detectors.}
\label{fig:Channelsrateintegral}
\end{center}
\end{figure}
\end{landscape}

\newpage
\thispagestyle{empty}
\begin{landscape}
\begin{table}
\begin{center}
\begin{tabular}{|l||c|c|c|c|c|c|}
\hline
\hline
Wheel &Straws & AT Losses & AT Repairs & IT Losses & IT Losses $\bigcap$ AT repairs& IT Losses $\bigcap$ bad WEB\\
\hline
\hline
A01C & 6144 & 33& 25& 15&  1& 1\\
\hline
A02C & 6144 & 28& 15& 20& 15&19\\
\hline
A03C & 6144 &  9& 91&  6&  0& 4\\
\hline
A04C & 6144 & 21& 41&  9&  5& 7\\
\hline
A05C & 6144 &  4& 52&  1&  1& 1\\
\hline
A06C & 6144 &  9& 39& 15&  6& 9\\
\hline
A07C & 6144 &  1&  0&  8&  0& 0\\
\hline
A08C & 6144 &  5& 54& 14&  1& 3\\
\hline
A09C & 6144 &  3& 84& 11&  4& 8\\
\hline
A10C & 6144 &  3&203& 21&  3& 16\\
\hline
A11C & 6144 &  6& 11&  1&  1& 1\\
\hline
A12C & 6144 &  2&  9&  6&  1& 1\\
\hline
\hline
Total A-type&73728 &124&624&127&38&70 \\
\hline
\hline
Total A-type(\%)& & 0.17 & 0.85& 0.17& 0.05&0.09 \\
\hline
\hline
B01C & 6144 &16 & 0&  6& 0& 0\\
\hline
B02C & 6144 &26 &46& 20&10&18\\
\hline
B03C & 6144 & 2 & 0& 6&  0& 0\\
\hline
B04C & 6144 &15 & 0& 4&  0& 0\\
\hline
B05C & 6144 &16 & 0& 7&  0& 0\\
\hline
B06C & 6144 & 8 & 4&39&  0& 1\\
\hline
B07C & 6144 &10 &23&24&  0& 2\\
\hline
B08C & 6144 &14 & 2&23&  0& 0\\
\hline
\hline
Total B-type&49152 & 107 & 75& 129& 10&21 \\
\hline
\hline
Total B-type (\%)& & 0.22 & 0.15& 0.26& 0.02&0.04 \\
\hline
\hline
Total &122880 &231 &699 &256 &48 &91 \\
\hline
\hline
Total (\%)& & 0.19 & 0.57& 0.21& 0.04&0.07 \\
\hline
\hline
\end{tabular}
\caption{Correlation between the new non working readout channels and repaired WEBs during the acceptance tests.}
\label{tab:BadChannels}
\end{center}
\end{table}
\end{landscape}

\begin{table}[b!]
\begin{center}
\begin{tabular}{|l||c|c|c|c|c|}
\hline
\hline
Wheel & Capacitance test (\%)& Diode test (\%)&  Acceptance test (\%)& Other (\%)& Total \\
\hline
\hline
A01C  & 12.5 &  2.1 & 68.8 & 16.6 & 48 \\
\hline
A02C  & 20.8 & 12.4 & 58.4 &  8.4 & 48 \\
\hline
A03C  & 13.3 &  0.0 & 60.0 & 26.7 & 15 \\
\hline
A04C  & 26.7 &  0.0 & 70.0 &  3.3 & 30 \\
\hline
A05C  & 20.0 &  0.0 & 80.0 &  0.0 & 5 \\
\hline
A06C  & 33.3 & 16.7 & 37.5 & 12.5 & 24 \\
\hline
A07C  & 22.2 & 22.2 & 11.1 & 44.5 & 9 \\
\hline
A08C  & 15.8 & 21.1 & 26.3 & 36.8 & 19 \\
\hline
A09C  & 28.6 & 42.9 & 21.4 &  7.1 & 14 \\
\hline
A10C  & 25.0 & 37.5 & 12.5 & 25.0 & 24 \\
\hline
A11C  & 14.3 &  0.0 & 85.7 &  0.0 & 7 \\
\hline
A12C  & 13.0 & 12.5 & 25.0 & 37.5 & 8 \\
\hline
\hline
A-Type& 20.5 & 14.0 & 46.3 & 18.2 & 251\\
\hline
\hline
B01C &  4.6 &  9.1 & 72.7 & 13.6 & 22 \\
\hline
B02C & 13.0 &  0.0 & 56.5 & 30.5 & 46 \\
\hline
B03C & 50.0 &  0.0 & 25.0 & 25.0 & 8 \\
\hline
B04C & 10.5 &  5.2 & 79.0 &  5.3 & 19 \\
\hline
B05C & 13.0 &  0.0 & 69.6 & 17.4 & 23 \\
\hline
B06C & 19.1 & 38.3 & 17.0 & 25.6 & 47 \\
\hline
B07C &  0.0 & 52.9 & 29.5 & 17.6 & 34 \\
\hline
B08C &  2.7 & 29.7 & 37.9 & 29.7 & 37 \\
\hline
\hline
B-Type& 14.9 & 16.9 & 48.4 & 20.6 & 236\\
\hline
\hline
\end{tabular}
\caption{Details for the 487 non working channels of the TRT endcap-detector side C integration.}
\label{tab:BadChannels1}
\end{center}
\end{table}

\newpage
\thispagestyle{empty}
\begin{landscape}
\begin{small}
\begin{table}
\begin{center}
\begin{tabular}{|l||ccc|ccc|ccc|ccc|}
\hline
\hline
Wheel & \multicolumn{3}{|c|}{Capacitance}& \multicolumn{3}{|c|}{Diode}& \multicolumn{3}{|c|}{Other}&  \multicolumn{3}{|c|}{All} \\
\hline
\hline
         & bad & Repaired & bad WEB &  bad & Repaired & Bad WEB & bad & Repaired & Bad WEB &  bad & Repaired & Bad WEB \\
\hline
\hline
A01C &   6 & 0 & 0 & 1 & 1 & 1&8 &0 & 0 & 15& 1 &  1 \\
A02C &  10 & 7 & 9 & 6 & 6 & 6&4 &2 & 4 & 20&15 & 19 \\
A03C &   2 & 0 & 1 & 0 & 0 & 0&4 &0 & 3 &  6& 0 &  4 \\
A04C &   8 & 5 & 7 & 0 & 0 & 0&1 &0 & 0 &  9& 5 &  7 \\
A05C &   1 & 1 & 1 & 0 & 0 & 0&0 &0 & 0 &  1& 1 & 10 \\
A06C &   8 & 5 & 7 & 4 & 0 & 0&3 &1 & 2 & 15& 6 &  9 \\
A07C &   2 & 0 & 0 & 2 & 0 & 0&4 &0 & 0 &  8& 0 &  0 \\
A08C &   3 & 1 & 2 & 4 & 0 & 1&7 &0 & 0 & 14& 1 &  3 \\
A09C &   4 & 3 & 4 & 6 & 1 & 3&1 &0 & 1 & 11& 4 &  8 \\
A10C &   6 & 1 & 5 & 9 & 2 & 6&6 &0 & 5 & 21& 3 & 16 \\
A11C &   1 & 1 & 1 & 0 & 0 & 0&0 &0 & 0 &  1& 1 & 10 \\
A12C &   2 & 1 & 1 & 1 & 0 & 0&3 &0 & 0 &  6& 1 &  1 \\
\hline
\hline
A-Type&  53 &25 & 38&33 &10 & 17&41&3 & 15&127&38 & 70\\
\hline
\hline
A-Type (\%)&  &47.2 & 71.7 &33 &30.3 & 51.5&41&7.3 & 36.6&127&29.9 &55.1\\
\hline
\hline
B01C &  1  & 0 & 0& 2 & 0 & 0& 3& 0 & 0& 6& 0 &  0\\
B02C &  6  & 3 & 5& 0 & 0 & 0&14& 7 & 1&20&10 & 18\\
B03C &  4  & 0 & 0& 0 & 0 & 0& 2& 0 & 0& 6& 0 &  0\\
B04C &  2  & 0 & 0& 1 & 0 & 0& 1& 0 & 0& 4& 0 &  0\\
B05C &  3  & 0 & 0& 0 & 0 & 0& 4& 0 & 0& 7& 0 &  0\\
B06C &  9  & 0 & 0&18 & 0 & 1&12& 0 & 0&39& 0 &  1\\
B07C &  0  & 0 & 0&18 & 0 & 2& 6& 0 & 0&24& 0 &  2\\
B08C &  1  & 0 & 0&11 & 0 & 0&11& 0 & 0&23& 0 &  0\\
\hline
\hline
B-Type& 26  & 3 & 5 &50 & 0 & 3 &53& 7 & 13&129&10 & 21\\
\hline
\hline
B-Type (\%)& 26  & 11.5 & 19.2 &50 & 0.0 & 6.0 &53 & 13.2 & 24.5&129& 7.8 & 16.3\\
\hline
\hline
All& 79  &28 & 43 &83 &10 & 20  &94 &10 & 28  & 256 & 48 & 91  \\
\hline
\hline
All (\%)& 79  &35.4 & 54.4&83 &12.1 & 24.1&94 &10.6 & 29.8 & 256 & 18.7 & 35.5 \\
\hline
\hline
\end{tabular}
\caption{Correlation between repaired channels (R) and bad WEBs (W) during acceptance tests and new bad non working channels found after integration}
\label{tab:BadChannels2}
\end{center}
\end{table}
\end{small}
\end{landscape}

\pagebreak

\pagebreak

\pagebreak

\pagebreak

\pagebreak

\pagebreak

\pagebreak

\pagebreak

\pagebreak

\pagebreak

\pagebreak

\end{document}